\documentclass[twocolumn,aps,prd,nofootinbib]{revtex4-1}

\usepackage{epsfig}
\usepackage{dcolumn}
\usepackage{amsmath}
\usepackage{amsfonts}
\usepackage{hhline}
\usepackage{hyperref}

\usepackage[usenames,dvipsnames]{xcolor}

\numberwithin{equation}{section}

\newcommand{\be}{\begin{equation}}
\newcommand{\ee}{\end{equation}}
\newcommand{\bea}{\begin{eqnarray}}
\newcommand{\eea}{\end{eqnarray}}

\newcommand{\vk}{\mathbf{k}}
\newcommand{\vx}{\mathbf{x}}
\newcommand{\vy}{\mathbf{y}}
\newcommand{\ud}{\mathrm{d}}
\newcommand{\ar}{\arrowvert}
\newcommand{\ra}{\rangle}
\newcommand{\la}{\langle}
\newcommand{\da}{\dagger}

\newcommand{\cd}{\! \cdot \!}
\newcommand{\mBF}{\mathcal{B}}

\newcommand{\mP}{\mathcal{P}}
\newcommand{\mD}{\mathcal{D}}

\newcommand{\vpr}{\mathbf{p}_{\rho}}
\newcommand{\vpl}{\mathbf{p}_{\lambda}}
\newcommand{\upa}{\uparrow}
\newcommand{\doa}{\downarrow}

\newcommand{\Nbar}{\bar{N}}

\newcommand{\lbar}{\bar{l}}
\newcommand{\Sbar}{\bar{S}}
\newcommand{\Rbar}{\bar{R}}

\begin{document}

\title{Mapping chiral symmetry breaking in the excited baryon spectrum}

\author{Pedro Bicudo}
\affiliation{CFTP, Departamento de F\'{\i}sica, Instituto Superior T\'ecnico, Avenida Rovisco Pais 1049-001 Lisboa, Portugal}

\author{Marco Cardoso }
\affiliation{CFTP, Departamento de F\'{\i}sica, Instituto Superior T\'ecnico, Avenida Rovisco Pais 1049-001 Lisboa, Portugal}

\author{Felipe J. Llanes-Estrada}
\affiliation{Departamento de F\'{\i}sica Te\'orica I, Universidad Complutense de Madrid, 28040 Madrid, Spain}

\author{Tim Van Cauteren}
\affiliation{Departamento de F\'{\i}sica Te\'orica I, Universidad Complutense de Madrid, 28040 Madrid, Spain}

\begin{abstract}
We study the conjectured ``Insensitivity to Chiral Symmetry Breaking'' in the highly excited light baryon spectrum. While the experimental spectrum is being measured at JLab and CBELSA/TAPS, this insensitivity remains to be computed theoretically in detail.
As the only existing option to have both confinement, highly excited states and chiral symmetry, we adopt the truncated Coulomb gauge formulation of QCD,
considering a linearly confining Coulomb term. Adopting a systematic and numerically intensive variational treatment up to 12 harmonic oscillator shells we are able to access several angular and radial excitations. 
We compute both the excited spectra of $I=1/2$ and $I=3/2$ baryons, up to large spin $J=13/2$, and study in detail the proposed chiral multiplets.
While the static-light and light-light spectra clearly show chiral symmetry restoration high in the spectrum, the realization of chiral symmetry is more complicated in the baryon spectrum than earlier expected.
\end{abstract}

\maketitle

\twocolumngrid

\section{Introduction}

In a short letter \cite{Bicudo:2009cr} and several communications \cite{LlanesEstrada:2008mh,LlanesEstrada:2009sb,VanCauteren:2009vm,LlanesEstrada:2010pb,LlanesEstrada:2011jd} we proposed a variational approach to the  truncated Coulomb gauge formulation of QCD,
considering a linearly confining term, 
to gain novel insights into the baryon spectrum, particularly into the standing conjecture of Insensitivity to Chiral Symmetry Breaking (IChSB) in the high spectrum.

After our first result computed with Monte Carlo integration \cite{Bicudo:2009cr}, we continued to refine our numerical technique, and opted for Gauss integration, in order to analyse with sufficient detail the splittings in the proposed chiral multiplets. We now provide  full detail to this research line and comprehensively report our methods and findings.

The fundamental motivation to embark on this investigation of possible chiral multiplets in the baryon spectrum is driven by works initiated by de Tar, Kunihiro, Cohen and Glozman,
\cite{Detar:1988kn,Cohen:1996zz,Cohen:1996sb,Glozman:1999tk,Jido:1999hd,Cohen:2001gb,Swanson:2003ec,Cohen:2005am,Glozman:2005tq,Wagenbrunn:2006cs,Glozman:2007ek}

in which a chiral degeneracy, or chiral symmetry restoration, was suggested to occur in the excited baryon spectrum. Detailed examination of the arguments therein require to address the excited spectrum up to several excitations while breaking chiral symmetry in a controlled manner. 

We have found particularly interesting~\cite{Bicudo:2009cr} the study of Yrast baryons (that, with the same usage of the Swedish word for ``dizziest'' as in nuclear physics, are the lowest mass baryons for each angular momentum $J$) as they can unambiguously be identified for each parity. This is an advantage over the case of radial-like excitations:
there it is less clear which state of positive parity should be compared to which state of negative parity.

So far, large angular excitations have not been investigated in lattice gauge field theory studies, and thus model studies are in order. 
The model computations for simpler hadron spectra, such as light-light mesons and static-light mesons, already theoretically support the concept \cite{Bicudo:1998bz,LlanesEstrada:1999uh,Wagenbrunn:2007ie,Sazonov:2014qla,Bicudo:2015mjf}.
The agreement~\cite{Nefediev:2008dv} is that, in addition to parity doubling brought about by the symmetry, the pion couplings to various ${N^*}$, ${\Delta^*}$, etc. resonances become decreasingly small with the excitation quantum number (due to smaller wavefunction overlaps between the boosted decay-pion, and incoming and outgoing ${N^*}$'s).

In this work we employ the only model for baryons that can presently test the concept.
In essence, one needs a \emph{confining model} where high excitations,
e.g. angular, 
can be studied. 
At the same time, we need a model where \emph{chiral symmetry breaking is spontaneously, not explicitly, broken}. 
As the only existing option to have both confinement, highly excited states and chiral symmetry, we adopt the truncated Coulomb gauge formulation of QCD. Moreover we consider the case of a linearly confining Coulomb term, necessary for linear Regge trajectories.
The model Hamiltonian is given in equation \ref{Hmodel} below, but suffice it to say here that
the interaction vertex is chiral symmetric, entailing the spinor combination $U^\dagger U$
that corresponds to the non-relativistic reduction of the chiral-symmetry preserving, QED-like vertex $\bar{\psi} \gamma^\mu  \psi$. We will compute the mass spectra of positive parity and negative parity
excited baryons, 
$M_J^+$ and $M_J^-$, for both isospin $I=1/2$ and $I=3/2$, and compare them looking for signs of chiral symmetry 
restoration. The parameters of interest are the differences between these four energies.

The truncated Coulomb gauge approach is equivalent to a field theory version of chiral invariant quark models
 \cite{LeYaouanc:1984ntu,Amer:1983qa,LeYaouanc:1983huv,LeYaouanc:1983it,Amer:1982fg,Bicudo:1989sh,Bicudo:1989si,Bicudo:1989sj,LlanesEstrada:2001kr,TorresRincon:2010fu,LlanesEstrada:2010bs}.
 Because the model includes linear confinement one can examine an excited spectrum with infinitely many excitations. We do not necessarily claim they should be found in experiment, as we do not treat open decay channels~\cite{Coito:2011qn} in this work, given the already very large task at hand.
Second, because of the field theory formulation, one can employ the renowned BCS approximation to spontaneously break chiral symmetry by a $^3P_0$ Cooper-pair condensate \cite{Bicudo:1989sh,Bicudo:1989si,Bicudo:1989sj} and thus control the sensitivity of the high spectrum to this breaking thanks to the dynamical quark mass.

As additional motivation, recent efforts at CBELSA/TAPS ~\cite{Thoma,Wilson:2015uoa,Sokhoyan:2015fra}, CLAS and CLAS12 at Jefferson Lab ~\cite{Crede:2013sze,Glazier:2015cpa,Crede:2016wqu} and by theBonn-Gatchina partial wave analysis group \cite{Anisovich:2004zz,Anisovich:2006bc,Anisovich:2011fc,Denisenko:2016ugz} show good prospects for progress in the excited baryon spectrum. Some of the resonances that we address, particularly for high angular momentum, are of experimental interest.
 
Surprisingly, the spectrum of excited baryons, well addressed within massive-quark models~
\cite{Loring:2001kv,Loring:2001kx,Loring:2001ky,Plessas:2015mpa,Day:2013rza,Theussl:2000sj}, remains to be studied in depth with a chirally invariant and linearly confining quark model, although many studies have addressed the approach for the vacuum and mesons~
\cite{Adler:1984ri,Bicudo:1995kq,LlanesEstrada:1999uh,Wagenbrunn:2007ie}.
Here we provide the first detailed study
of the baryon spectrum with such chirally invariant approach and a linearly rising potential.

\begin{figure}
\includegraphics*[width=0.9\columnwidth]{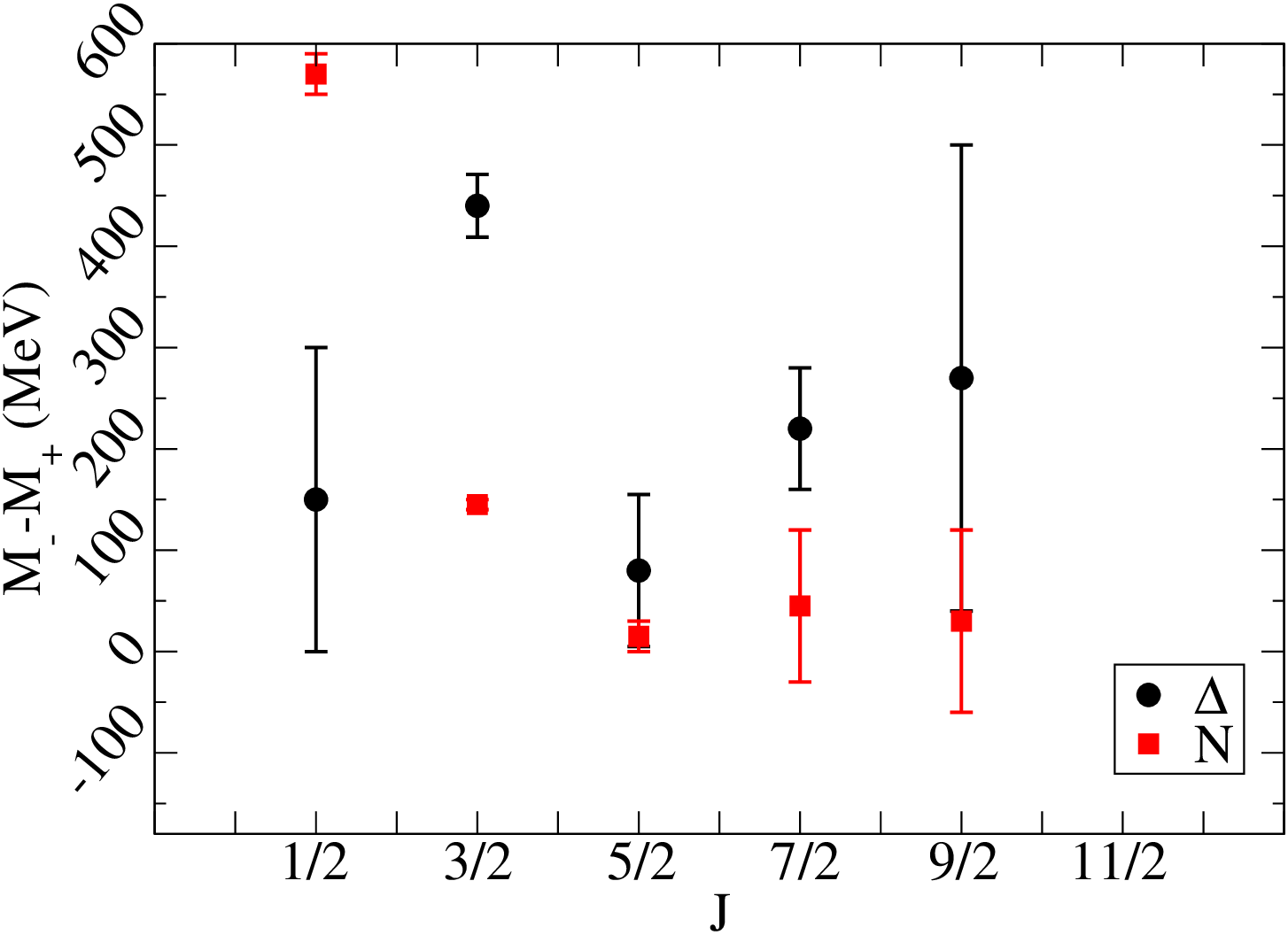}
\caption{\label{fig:expt}  Experimental parity splittings in the confirmed} $N$ and $\Delta$ Yrast baryons as function of angular momentum~\cite{Agashe:2014kda}.
\end{figure}

Before we report our computations, it is worth recalling the current experimental situation \cite{Agashe:2014kda} in what concerns Yrast-baryon parity splittings as function of angular momentum. Fig. ~\ref{fig:expt} shows the difference of $N^+$ and $N^-$ masses (or parity splitting) as well as that of $\Delta^+$ and $\Delta^-$ as function of $J$.
It is apparent from this figure that the nucleon splittings do seem smaller on average for larger $J$ consistently with the theory prediction, while the $\Delta$ splittings do not show clear evidence of decrease; they are at best constant, and the experimental uncertainty is large.

Unlike recent nuclear computations that also use the variational principle extensively to address the few-body problem~\cite{Vary:2015dda,Dikmen:2015tla},
the major difficulty that we here face is not the diagonalization of huge matrices, but rather the computation of the matrix elements  of the Hamiltonian. As will be shown below in section~\ref{sec:basis}, there is a lot of bookkeeping in the construction of the basis, but not massive computer time.
Each matrix element however is a complicated integral that, without the
simplifications brought about by the non-relativistic limit (which, while being adequate in nuclear physics, is not suited for quarks inside hadrons), 
poses a significant computational challenge. Thus, we are 
only able to reach a
finite shell number in the 3-body problem.

 In exchange, our Hamiltonian matrix is relatively small and its diagonalization  straightforward.

In the next sections the reader will find an \emph{expos\'e} of why chiral symmetry has been expected to be manifest in the high hadron spectrum~(Sec.~\ref{sec:insensitivity});
a short discussion of why tracking the possibly decreasing Yrast parity splittings in the high spectrum is interesting to access the running quark mass function (Sec.~\ref{sec:runningmass}); a brief discussion of the model field-theory Hamiltonian and its variational treatment (Sec.~\ref{sec:model});
then the detailed construction of the three-quark variational wavefunction basis~(Sec.~\ref{sec:basis}); the report of our numerical results~(Sec.~\ref{sec:numerics}); 
and a discussion thereof of the 
mechanisms
that may be slowing the convergence to
the expected parity doubling in the high baryon spectrum by breaking chiral symmetry, which deserves future investigation. (Sec.~\ref{sec:conclusion}). A small appendix (App.~\ref{sec:trocas}) is dedicated to a somewhat technical computation of exchange matrix elements needed for the wavefunction overlaps that appear in anti-symmetrizing and orthonormalizing the variational basis.

\section{Insensitivity to chiral symmetry breaking in the high hadron spectrum}
\label{sec:insensitivity}

Here, in the first subsection~\ref{subsec:quartet} we quickly revise the argument suggesting parity doubling in the high spectrum, and in the next, subsection~\ref{subsec:expansion}, we show one could conceivably use this possible doubling, if experimentally found, to make a statement about the running quark mass.

\subsection{Three-Quark Chiral (nondegenerate) Quartet and 
its two (possibly degenerate) doublets}\label{subsec:quartet}
Baryon valence wavefunctions~\footnote{We will follow standard notation and use the greek indices $\lambda$, $\mu$, $\nu\dots$ for spin, latin $k$, $p$ and $q$ for momenta, $f$ for flavor, $a$ for isospin, $c$ for color, and generically $i$, $j$, $k$ for excitations with the same quantum numbers (in baryons) or if we wish to be unspecific (in quarks).} with three quarks
$\sum_{ijk} F_{ijk}\  B^\da_i B^\da_j B^\da_k \ar \Omega \ra$
naturally group into reducible, nondegenerate quartets high in the spectrum, with two states having positive parity and two states having negative parity, and this quartet is split into two doublets. 
We frame the discussion within standard Hamiltonian techniques without invoking yet any particular model; in section~\ref{sec:model} below we will be more specific and introduce the actual Hamiltonian that will be treated numerically.

Following the concise overview of Jaffe, Pirjol and Scardicchio~\cite{Jaffe:2005sq},
let us recall the possible representations of chiral symmetry in the hadron spectrum.
In Wigner-Weyl mode, the commutator of the chiral charge $Q^a_5$
with a positive parity  baryon field is a sum over negative parity baryons and viceversa
\begin{eqnarray}  \label{linearrealization} \nonumber 
\left[Q_5^a,\sigma^+_i\right]  &=& \Theta_{ij}^a \sigma^-_j 
\\ 
 \left[ Q_{5}^{a},\sigma^{-}_{i}\right]  &=&  \Theta_{ij}^a \sigma^+_j \  .
\end{eqnarray}
Wigner's theorem then guarantees that if $[Q^a_5,H]=0$ 
(as is the case for the $\gamma^0 \gamma^0$ Coulomb-type kernel in section~\ref{sec:model} below and more generally in Chromodynamics), and the vacuum state is unique, 
$Q^a_5\ar \Omega \ra=0$, then $\sigma^+$ and $\sigma^-$ are degenerate.

However, if the symmetry is spontaneously broken (or hidden), a Goldstone
boson appears, the pion, and chiral symmetry is realized
non-linearly. Because the pion has negative parity,
\be
[Q_5^a,\sigma^\pm_i]= v_0(\pi^2) \epsilon_{abc}\pi^c \Theta_{ij}^b \sigma^\pm_j
\ee
where $v_0$ is a scalar function (e.g. $v_0=1$) of the pion field.

We now descend to the quark level, in the spirit of Nefediev, Ribeiro
and Szczepaniak~\cite{Nefediev:2008dv}.
The chiral charge is the spatial integral of the zeroth component of the axial current,
\be \label{chiralchargeX}
Q_5^a = \int d^3x \Psi^\da(x) \gamma_5 \frac{\tau^a}{2} \Psi(x)\ .
\ee
We transform the fields to momentum space in a Bogoliubov-rotated basis
\be \label{FieldExpansion}
\Psi(\vx) = \int \frac{\ud^3 k}{(2\pi)^3}
e^{i\mathbf{k}\cdot\vx} \sum_{\lambda i}
\left[U_{k\lambda}B_{k\lambda i} +
  V_{-k\lambda}D^+_{-k\lambda i} \right]
\ee 
in terms of the spinors
\bea \label{spinors}
U_{k\lambda} = \frac{1}{\sqrt{2}} \left[ \begin{array}{c}
    \sqrt{1+\sin{\phi_k}}\chi_\lambda \\ \sqrt{1-\sin{\phi_k}}
    \mathbf{\sigma}\cdot\hat{k} \chi_\lambda\end{array} \right]  \\
V_{-k\lambda} = \frac{1}{\sqrt{2}} \left[ \begin{array}{c}
    -\sqrt{1-\sin{\phi_k}} \mathbf{\sigma}\cdot\hat{k}i\sigma_2 \chi_\lambda  \\
    \sqrt{1+\sin{\phi_k}}i\sigma_2 \chi_\lambda  \end{array} \right]\; .
\eea
written as functions  of the chiral angle with $\sin\phi(k) = m(k)/\sqrt{m(k)^2+k^2}$
(the $k$ dependence would be absent of this parametrization for free spinors: these are more general and useful in the interacting theory).

There are four possible terms after normal ordering the charge in Eq.~(\ref{chiralchargeX}), $B^\da B$, $B^\da D^\da$, $D^\da D$, $DB$. The last two vanish when the chiral charge acts on an initial baryon state made of three quarks
$\ar \sigma \ra = \sum F B^\da B^\da B^\da \ar \Omega \ra$
since the antiquark destruction operator $B$  acts directly on the vacuum
and gives zero. After spinor contractions,
\begin{eqnarray} \label{EffectQ5}
Q^5_a \ar \sigma \ra = \int \frac{d^3k}{(2\pi)^3} \sum_{\lambda 
\lambda ' f f'c} \left(  \frac{\tau^a}{2} \right)_{ff'}\\ \nonumber
\left(\cos \phi(k) ({\bf \sigma}\cd{\bf \hat{k}})_{\lambda \lambda'}
B^\da_{k\lambda f c} B_{k\lambda'f'c} + \right. \\\left. \nonumber
 \sin \phi(k) (i\sigma_2)_{\lambda \lambda'}
B^{\da}_{k\lambda f c}D^\da_{-k \lambda'f'c}
\right) \ar \sigma \ra \ .
\end{eqnarray}
The first term conserves the quark number and therefore remains within the same variational subspace. The second however creates a quark-antiquark pair on the baryon wavefunction. It is well known that the wavefunction associated with this pair, $\sin \phi(k) i\sigma_2$ precisely
corresponds to a pion creator operator in RPA approximation (and more generally it is an
adequate pion interpolating field due to its quantum numbers). 
Therefore this second term is
responsible for the non-linear realization of the symmetry: if chiral
symmetry is spontaneously broken, $m(k)\not = 0$, thus $\sin \phi(k) \not = 0$ and under a chiral rotation a baryon is mapped to a baryon plus a pion $\sigma\to \pi \sigma$.

Now, if the wavefunction $F(k_i,k_j,k_l)$ has its support for very large values of the momenta, then $m(k)\simeq 0$ there, since the running quark mass is large precisely for small values of the momentum (see Fig. ~\ref{fig:gap} below). This means that there is a chance that chiral symmetry is realized in Wigner-Weyl mode in the high spectrum.
This would happen because the second term in Eq.~(\ref{EffectQ5}) is now small, so chiral symmetry is realized linearly in the number of quarks by the first term of that Eq.~(\ref{EffectQ5}) that  returns Eq.~(\ref{linearrealization}).

Once chiral symmetry becomes realized linearly in the Wigner-Weyl
mode, degenerate baryon multiplets appear. The three-quark wavefunctions
group in quartets that can be immediately found by successively applying
the charge $Q^a_5$ to a given three-quark wavefunction. 

Employing only one quark index $i$, for the case of
one flavor, and showing the parity of the states, we can very transparently
list the four linearly independent vectors for the quartet
~\footnote{$F$ is antisymmetric in its three indices because the fermion $B$ operators anticommute.},
\begin{eqnarray} \label{quartetstates}
\ar \sigma_1^P \ra &=& \sum F^P_{ijk} B^\da_i B^\da_j B^\da_k \ar \Omega \ra \ ,
\\ \nonumber
\ar \sigma_2^{-P} \ra &=& \sum F^P_{ijk} \left({\bf \sigma}\cd
{\bf \hat{k}}_i \frac{\tau^a}{2} B^\da  \right)_i
 B^\da_j B^\da_k \ar \Omega \ra \ ,
 \\ \nonumber
\ar \sigma_3^{P} \ra &=& \sum F^P_{ijk} \left({\bf \sigma}\cd
{\bf \hat{k}}_i \frac{\tau^a}{2} B^\da \right)_i
\left({\bf \sigma}\cd {\bf \hat{k}}_j \frac{\tau^b}{2} B^\da \right)_j B^\da_k \ar \Omega \ra \ ,
\\ \nonumber
\ar \sigma_3^{-P} \ra &=& \sum F^P_{ijk} \left({\bf \sigma}\cd
{\bf \hat{k}}_i \frac{\tau^a}{2} B^\da \right)_i
\left({\bf \sigma}\cd {\bf \hat{k}}_j \frac{\tau^b}{2} B^\da \right)_j 
\\ \nonumber
&&
\qquad \qquad \left({\bf \sigma}\cd {\bf \hat{k}}_k \frac{\tau^c}{2} B^\da \right)_k \ar \Omega \ra\ ;
\end{eqnarray}
the sequence arises by acting over the three quarks
of $\ar \sigma_1^{P} \ra$ in the first line. The baryon charge is the sum of three components, 
one for each quark $Q_5^a= {q_5^a}_1 + {q_5^a}_2+{q_5^a}_3$, 
each of them
reversing the spin projection over the quark  momentum axis with the matrix 
${\bf \sigma}\cd
{\bf  \hat{k}}_i$. Repeated application of a charge over the same quark does
not change the parity since $\left( {\bf \sigma}\cdot
{\bf  \hat{k}}_i\right)^2= I$. Therefore, successive application generates
new terms by rotating different quarks.

The total angular momentum $J$ of the four states coincides because the chiral charge is a 
pseudoscalar operator, so that $[J,Q^a_5]=0$.  Each member of this chiral quartet is itself a multiplet under isospin transformations (e.g. a proton-neutron doublet).

Moreover, the isospin Pauli matrices $\tau^a$ produce tensor isospin operators who may mix the Nucleon $I=1/2$ spectrum with the Delta $I=3/2$ spectrum \cite{Cohen:1996zz,Cohen:1996sb,Glozman:1999tk,Cohen:2001gb}. Thus we are interested in studying possible multiplets with the Yrast and Yrare (groundstate and first excitation) with parity $+$ and $-$ and isospin $1/2$ and $3/2$.

\subsection{Series expansion in $m(k)/k$}\label{subsec:expansion}

We now try to track the size of the chiral-symmetry breaking effect of the
dynamical quark mass in the high spectrum.
The quark mass appears both in the quark spinors and in
the QCD Hamiltonian. As we wish to track what happens for small quark mass
at high momenta, we pursue a series expansion of the spinors and Hamiltonian in the
parameter $m(k)/k$.

The spinors are expanded as
\begin{eqnarray}\label{spinorexp}
 U_{k\lambda} = & &\frac{1}{2E(k)} \left[ \begin{array}{c} \sqrt{E(k) +
m(k)}\ \ \  \chi_\lambda \\ \sqrt{E(k) - m(k)} \mathbf{\sigma} \cdot \hat{k}\ \ \  \chi_\lambda
\end{array} \right] \\
\overrightarrow{k \to \infty} & & \frac{1}{\sqrt{2}} \left[ \begin{array}{c}
\chi_\lambda \\ \mathbf{\sigma} \cdot \hat{k}\ \ \  \chi_\lambda \end{array} \right] +
\frac{1}{2\sqrt{2}} \frac{m(k)}{k} \left[ \begin{array}{c} \chi_\lambda \\
-\mathbf{\sigma} \cdot \hat{k}\ \ \  \chi_\lambda \end{array} \right] \, , \nonumber
\end{eqnarray}
with $E(k) = \sqrt{k^2 + m(k)^2}$. Therein we have worked up to the leading chiral-symmetry breaking term which must be $O(m(k)/k)$.

When expanding the matrix elements of the QCD Hamiltonian~\cite{Christ:1980ku} 
restricted to the Hilbert space of highly excited resonances where $\langle k \rangle$ is large,  the zeroth order term in the $m(k)/k$ expansion is chirally invariant, and the first order term 
involves nonchiral, spin-dependent potentials in the quark-quark interaction, 
\begin{eqnarray}
 \langle n_1 | H^{QCD} | n_2 \rangle \simeq \nonumber \\   \langle n_1 |
H^{QCD}_\chi | n_2\rangle + \langle n_1 | \frac{m(k)}{k} H^{QCD\ '}_\chi |
n_2 \rangle \dots \label{QCDexp}
\end{eqnarray}
 Importantly, the lower spinor component of the first-order term has the opposite sign of the lower component of the chiral invariant term preceding it in the second line of Eq.~(\ref{spinorexp}).

With massless quarks, we have a vanishing commutator~\cite{Nefediev:2008dv} $\la i \ar [Q_5^a,H_{QCD}] \ar j\ra=0$ in the \emph{perturbative} basis;
but chiral noninvariant mass terms appear in the strong interactions due to dynamical mass generation, with chiral symmetry spontaneously broken by the ground state, 
$Q^a_5 | 0 \rangle \neq 0$,
a large quark mass in the propagator, pseudo-Goldstone bosons
and the loss of parity-degeneracy in ground-state baryons. It would appear that in the hadron basis $\la n_1 \ar [Q_5^a,H_{QCD}] \ar n_2\ra\neq 0$.

To discuss this situation, we write the chiral charge once more, 
now in terms of $m(k)$ instead of $\sin \phi (k)$,
\begin{eqnarray} \label{chiralcharge} \\ \nonumber
Q_5^a  = \int \frac{d^3k}{(2\pi)^3} \sum_{\lambda
\lambda ' f f'c} \left(  \frac{\tau^a}{2} \right)_{ff'}
{ k \over \sqrt{ k^2 + m^2(k)}}
 \\ \nonumber 
\times \left[ \phantom{\frac{(a)}{b}}\!\!\!\!\!\!\!\! ({\bf \sigma}\cdot{\bf
\hat{k}})_{\lambda \lambda'}  	
\left( B^\dagger_{k \lambda f c} B_{k \lambda' f' c} + D^\dagger_{-k \lambda' f'
c}
D_{-k \lambda f c} \right) + \right. \\  \nonumber \left.
{ m(k) \over k} (i\sigma_2)_{\lambda \lambda'} \
\left( B^{\dagger}_{k\lambda f c} D^\dagger_{-k\lambda'f'c}+
B_{k \lambda' f' c} D_{-k \lambda f c}
\right) \right] \ .
\end{eqnarray}
The first term between the square brackets once more consists of quark and antiquark
number operators flipping spin and parity. The observation is that for $m(k) \ll k$, it dominates over the
second term which creates or annihilates a pion (realizing chiral symmetry nonlinearly).
Thus, in the high-lying spectrum, 
\be \label{neardeg}
\la n_1 \ar [Q_5^a,H_{QCD}] \ar n_2\ra\simeq 0 \ \ \ \ {\rm for\   large}\ n_1,n_2
\ee  and each parity doublet appears to become degenerate when $m(k)$ vanishes. Moreover, the mass splitting between partners is a direct measure of $m(k)$.

\section{The running quark mass and the 
$\Delta \ , \ N$ spectra
}\label{sec:runningmass}

Eq.~(\ref{neardeg}) permits to link the mass splitting $|M^{P=+} - M^{P=-}|$ in a parity doublet to the running quark mass.  We concentrate on the  lowest lying $N$ and $\Delta$ parity doublets for increasing spin $j$ (the Yrast baryons) and reason as follows.
\begin{enumerate}
 \item Baryon masses fall on Regge trajectories: 
\be \label{Regge} j = \alpha_0 + \alpha {M^{\pm}}^2
    \, \stackrel{_{j\to\infty}}{\longrightarrow} \, \alpha {M^{\pm}}^2
\ee
 so that we can track the gross behavior of baryon masses with angular momentum.
\item The relativistic virial theorem~\cite{Lucha:1989jf} in a few body system at fixed particle number, 
\be \label{Lucha:1989jf}
\langle k \rangle \to c_2 M^\pm \to \frac{c_2}{\sqrt{\alpha}} \sqrt{j}\ , 
\ee
shows how the average momentum grows high in the spectrum. (Here $c_2$ is an unspecified constant).
 \item The chirally invariant term ($\langle n|H_\chi^{QCD}|n \rangle$) of Eq.~(\ref{QCDexp}) cancels out
in the splitting $\Delta M$ so that for small $m(k)/k$,
 \begin{displaymath}
  |M^+-M^-| \ll M^\pm
 \end{displaymath}
 and
 \bea 
      |M^+-M^-| \to \,
      \langle \frac{m(k)}{k} {H_\chi^{QCD}}' \rangle \, \\ \nonumber \to \, c_3 \frac{m(k)}{k}
      \langle {H_\chi^{QCD}}'\rangle
 \eea
 \item In the first nonvanishing term in the splitting, $H_\chi^{QCD}$, 
the spin-orbit $\mathbf{L}_i \cdot \mathbf{S}_i$ term transforms the angular momentum in the centrifugal barrier term from $\mathbf{L}_i^2$ to the chirally invariant
    $\mathbf{L}_i^2 + 2\mathbf{L}_i \cdot \mathbf{S}_i = \mathbf{J}_i^2 -
    \frac{3}{4}$. (Because of  the sign difference in the
    helicity-dependent term $\sim - \boldsymbol{\mathbf{\sigma}} \cdot
      \hat{\mathbf k}$ in the spinor, the spin-orbit term in
    ${H^{QCD}_\chi}'$ adds to the mass difference $\Delta M$, instead
    of cancelling out as for $H^{QCD}_\chi$). Thus, the centrifugal
barrier scales like the mass of the state itself $M^\pm$ for high $j$, as per Eq.~(\ref{Regge}). 
The spin-orbit term on the other hand $\mathbf{L}_i \cdot \mathbf{S}_i \sim J_i$ scales with one less power of $j$, and so the term producing the parity splitting actually decreases
    \begin{equation}
     \langle {H^{QCD}_\chi}' \rangle \to c_5 \frac{M^\pm}{ j} \to
    \frac{c_5}{\sqrt{\alpha}} \sqrt{\frac{1}{j}} \, . \label{eq:tensor}
\end{equation}
\end{enumerate}
Combining these four arguments, we obtain 
\begin{equation} 
 |M^+-M^-| \to \frac{c_3 c_5}{c_2
    \sqrt{\alpha}} m(\langle k \rangle) j^{-1} \; .
\end{equation}
An experimental extraction can  fit the exponent $-i$ of $j$ in
the splitting 
\begin{equation} \label{key1}
|M^+ - M^-| \propto j^{-i} \ .
\end{equation}
Then an experimental extraction of  the power-law behaviour of the
running quark mass in the region of chiral symmetry breaking is given by 
\begin{equation} \label{key2}
m(k) \propto k^{-2i+2}\ . 
\end{equation}
Returning to Fig. ~\ref{fig:expt}, we see that 
the experimentally known splittings of lowest-lying $N$ and $\Delta$ resonances 
 are insufficient to derive the exponent $i$. 

Knowledge of the masses of the parity
doublets for spins $j>9/2$ would greatly enhance this (typically, only the natural
parity mass is known above this, and all partners are missing from the tables).

The beauty of this application of the would-be doublets is that it directly maps experimental data to an underlying QCD property, here the exponent of the running quark mass. The numerical value of the mass itself, being gauge dependent, is not accessible. But the exponent of its power-law running at mid-momentum, related to chiral symmetry breaking, could well be, as other symmetry-related quantities are (for example, the quark spin is accessible in deeply inelastic scattering). If the exponent is indeed gauge-independent, then we should be able to access it in lattice gauge theory, and indeed one of us has already found some indications in heavy-light mesons~\cite{Bicudo:2015mjf}.

It is interesting to note that a whole different way to access the running mass function from experiment is through the study of electromagnetic form factors~\cite{Bicudo:1998qb,Cloet:2013gva} where, due to Ward identities, the hadron wavefunctions are related to the running quark mass, so that certain integrated moments thereof can be probed.

\section{Model Hamiltonian} \label{sec:model}
Our so far generic arguments would make it compelling to look for the parity doublets in the high baryon spectrum. Yet we wish to have a complete model computation at hand to thoroughly explore the physics. After an intense effort, we have now carried the program out and report it in this article. 

We employ a well-known Hamiltonian density~\cite{LeYaouanc:1984ntu,Amer:1983qa,LeYaouanc:1983huv,LeYaouanc:1983it,Amer:1982fg,Bicudo:1989sh,Bicudo:1989si,Bicudo:1989sj,Bicudo:1998bz} that respects all global symmetries of QCD, and particularly chiral symmetry which is only broken spontaneously. 

We adopt the truncated Coulomb gauge formulation of QCD, also equivalent to 

 the field-theory upgrade of the Cornell model,
\begin{multline} 
\mathcal{H}(\vx) = \Psi^\dagger(\vx) (-i\mathbf{\alpha}\cdot\nabla + m
\beta) \Psi \label{Hmodel} \\
- \frac{1}{2} \int \ud^3 y \Psi^\dagger(\vx) T^a \Psi(\vx)
V(\vx-\vy) \Psi^\dagger(\vy) T^a \Psi(\vy)\ ;
\end{multline}
here $T^a$ is Gell-Mann's color matrix, and the field operator is
expanded as in Eq.~(\ref{FieldExpansion}) above to obtain the momentum-space
representation.  
Moreover we consider the case of a linearly confining Coulomb term, necessary for linear Regge trajectories.

This model has been extensively discussed in previous literature~\cite{Bicudo:1991kz}, 
so we recapitulate only the minimal set of features that we need.

\subsection{Gap equation and numerical solution}
\begin{figure}
\includegraphics[angle=90,width=0.9\columnwidth]{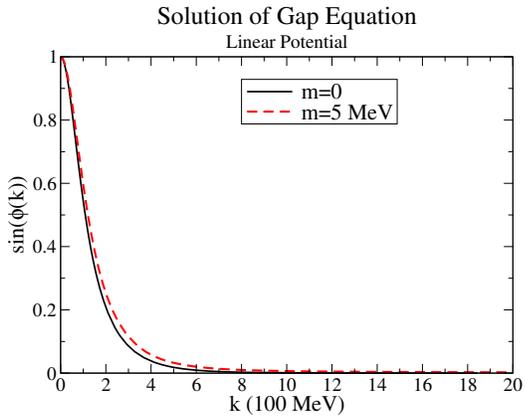}
\caption{Solution of the gap equation, $\sin \phi(k)$, for the linear
potential  with Coulomb-gauge type $\gamma_0\gamma_0$ vector coupling.
The current quark mass is fixed at $m_u=5 \ MeV$ at a high scale. 
\label{fig:gap}}
\end{figure}

Minimizing the vacuum expectation value of the Hamiltonian  in the BCS
approximation yields a gap equation for $\sin \phi(k)$ 
(the sine and the cosine are shortened to $s_k$ and $c_k$ as needed). This has been
solved for the linear potential in the past \cite{Adler:1984ri,Bicudo:1995kq,LlanesEstrada:2001kr}
and is known to have a tower of symmetry breaking solutions. The ground-state
solution is  a monotonous function of $k$ and has been obtained along
standard methods  \cite{Bicudo:2003cy,Bicudo:2010qp} by solving the non-linear integral
equation on a computer. It is shown in figure \ref{fig:gap}.
The chiral angle enters later computation through the three-quark 
matrix elements of the Hamiltonian in Eq.~(\ref{modelHelement}).

The BCS gap function for the linear potential, resulting solely from the Hamiltonian density in Eq.~(\ref{Hmodel}), yields a small constituent quark mass of order  $100\ MeV$,
already parametrized in excellent approximation by the ansatz
\cite{Bicudo:2010qp}
\be
 {\cal A}_3 (p) = m_0 + {1 \over c_0 + c_2 p^2 + c_4 p^4} \ , 
\ee
with the vacuum minimization parameters,
$c_0=6.01623, \, c_2=23.2517, \, c_4=12.0965$ in dimensionless units of the meson string tension ${4 \over 3}\sigma=1$,
for a bare mass of $m_0 = 0.01 \sqrt{{4 \over 3}\sigma} \simeq 4\, \mbox{MeV}$
A more complete model for the truncated Coulomb gauge potential, with more parameters, would be necessary to get a larger constituent quark mass.
As a sanity check we have also employ a constant quark mass $m(k) \to m(0)$.
In Ref. ~\cite{Bicudo:2009cr}
the mass gap function generated in Coulomb gauge SU(2) lattices~\cite{Langfeld:2004qs}, as well as in Landau gauge SU(3) lattices~\cite{Bowman:2005vx} was also used. 
There are of course differences of detail from using different masses, but not of principle.

High in the spectrum, various relativistic splittings are suppressed and  
 the onset of chiral restoration could be masked by other effects. Therefore we need a benchmark calculation to know the quantitative amount of insensitivity to chiral symmetry breaking.

The litmus test for whether the parity partners are becoming degenerate due to chiral symmetry will be passed if the energy difference between doublers is significantly smaller for the chiral theory.

Moreover, we will use our same computer code but fix $m(k)=M$ to  a constant (like in the constituent  model). This means $\sin \phi(k)=1$ in that benchmark calculation, that will be reported below in Fig. ~\ref{fig:constqm} next to other systematic analysis.
This allows to estimate whether the running mass can be extracted from the spectrum, as discussed in sub-section \ref{subsec:expansion}.

\subsection{Exposing parity doubling within the model}

The complete three-quark baryon wavefunction given below in Eqs.~(\ref{baryondef}) and (\ref{baryondef2}),
can be shortened to 
\be
|\sigma \rangle = \frac{\epsilon^{ijk}}{\sqrt{6}}
F_\sigma^{\lambda_1\lambda_2\lambda_3}(\vk_1,\vk_2,\vk_3)
B_{\vk_1\lambda_1i}^\dagger B_{\vk_2\lambda_2j}^\dagger B_{\vk_3\lambda_3k}^\dagger |
\Omega \rangle \ .
\ee
 The momentum part of the wavefunction is
\bea
F_\sigma^{\lambda_1\lambda_2\lambda_3}(\vk_1,\vk_2,\vk_3) = \\ \nonumber
e^{i\mathbf{K}\cdot\mathbf{R}} F_\sigma^{\lambda_1\lambda_2\lambda_3}
(\rho(\vk_1,\vk_2,\vk_3), \lambda(\vk_1,\vk_2,\vk_3))
\eea
with $\mathbf{K}$ and $\mathbf{R}$ the total momentum and center of mass coordinate,
and $\rho$, $\lambda$ are the Jacobi coordinates (see subsec.~\ref{subsec:Jacobi} below).

The resulting matrix elements of the Hamiltonian in Eq.~(\ref{Hmodel}) 
are then~\cite{LlanesEstrada:2000fa}
\begin{widetext}
\begin{multline} \label{modelHelement}
\langle\sigma|H|\sigma'\rangle = \delta^3(\mathbf{R}-\mathbf{R}') \, 3 \int \frac{\ud^3
  k_1}{(2\pi)^3} \frac{\ud^3 k_2}{(2\pi)^3} \left[
  F_\sigma^{\lambda_1\lambda_2\lambda_3} (\vk_1,\vk_2) \right]^*
  \Bigg\{ F_{\sigma'}^{\lambda_1\lambda_2\lambda_3} (\vk_1,\vk_2)
  \left(m s_{k_1} + |\vk_1| c_{k_1} \right) - \frac{2}{3} \int
  \frac{\ud^3 q}{(2\pi)^3} V(|\mathbf{q}|) \phantom{\Bigg\}} \\
  \phantom{\Bigg\{} \times \Big[
  F_{\sigma'}^{\lambda_1\lambda_2\lambda_3} (\vk_1,\vk_2) \left(
  s_{k_1}s_{k_1+q} + c_{k_1}c_{k_1+q} \hat{\mathbf{k}_1}\cdot\widehat{\mathbf{k}_1+\mathbf{q}}
  \right) - F_{\sigma'}^{\mu_1\mu_2\lambda_3}
  (\vk_1+\mathbf{q},\vk_2-\mathbf{q}) \left(U^+_{\vk_1\lambda_1}
  U_{\vk_1+\mathbf{q}\mu_1} \right) \left(U^+_{\vk_2\lambda_2}
  U_{\vk_2-\mathbf{q}\mu_2} \right) \Big] \Bigg\} \; .
\end{multline}
\end{widetext}
Here, the arguments of the wavefunctions $F$ are the momenta of first
and second quark, and should be transformed into Jacobi-momenta in the
rest frame ($\mathbf{K}=\mathbf{0}$). Furthermore, sums over all spin
quantum numbers are assumed. $V(|\mathbf{q}|)
= \frac{-8\pi\sigma}{q^4}$ is the non-regularized linear string potential
in momentum space. It is seen that one has to compute 9-dimensional
integrals although the integral over $\mathbf{q}$ can be performed
efficiently. Also, the linear infrared divergence of the $\mathbf{q}$
integral (due to the $q^{-4}$ dependence of the potential) becomes a
logarithmic divergence because of the cancellation between the 1-body
and the 2-body potential terms in the limit $q\to0$.

Let us now see how the mechanism for insensitivity to chiral symmetry breaking
discussed in generic terms heretofore is at work for the
simplified Hamiltonian in Eq.~(\ref{Hmodel}) above. That the kinetic energy 
is invariant under the spin rotation caused by the $\sigma\cd \hat{\bf k}$ in Eq.~(\ref{chiralcharge}) is obvious. Indeed, if the first state of the quartet in Eq.~(\ref{quartetstates}) has as wavefunction
\be
\ar 1 \ra = F^{s_1s_2s_3}({\bf k}_1,{\bf k}_2)\ ,
\ee
its kinetic and self-energies in Eq.~(\ref{modelHelement}) are proportional to  the simple overlap
\be \label{compareKE1}
F^{s_1s_2s_3\ \ \da} ({\bf k}_1,{\bf k}_2)
F^{s_1s_2s_3}({\bf k}_1,{\bf k}_2)
\ee
and if we take as the second element of the quartet that with only
one spin rotation, we have
\be
\ar 2 \ra = \left( \sigma\cd \hat{\bf k}_1 \right)_{s_1 s'_1}
F^{s'_1s_2s_3}({\bf k}_1,{\bf k}_2)\ ,
\ee
and its kinetic energy and self-energy are then proportional to
($\sigma$ being Hermitian)
\be \label{compareKE2}
F^{s'_1s_2s_3\ \ \da} ({\bf k}_1,{\bf k}_2)
\left( \sigma\cd \hat{\bf k}_1 \right)_{s'_1 s_1}
\left( \sigma\cd \hat{\bf k}_1 \right)_{s_1 s^{,,}_1 }
F^{s^{,,}_1s_2s_3}({\bf k}_1,{\bf k}_2)
\ee
and since $ \sigma\cd \hat{\bf k}_1\sigma\cd \hat{\bf k}_1={\mathbb{I}}$ 
the terms in (\ref{compareKE1}) and (\ref{compareKE2})  are identical.

Moving on to the potential energy in  Eq.~(\ref{modelHelement}), degeneracy requires the following equality to hold (irrelevant parts of $\la V\ra $ are omitted)
\begin{widetext}
\bea
F^{\lambda_1\lambda_2\lambda_3\ \ \da} ({\bf k}_1,{\bf k}_2)
\left(U^+_{\vk_1\lambda_1} U_{\vk_1+\mathbf{q}\mu_1} \right)
F^{\mu_1\mu_2\lambda_3}({\bf k}_1+{\bf q},{\bf k}_2-\bf{q})= ?
\\ \nonumber
F^{\lambda'_1\lambda_2\lambda_3\ \ \da} ({\bf k}_1,{\bf k}_2)
\left( \sigma\cd \hat{\bf k}_1 \right)_{\lambda'_1 \lambda_1}
\left(U^+_{\vk_1\lambda_1} U_{\vk_1+\mathbf{q}\mu_1} \right) 
\left( \sigma\cd (\widehat{{\bf k}_1+{\bf q}}) \right)_{\mu_1 \mu'_1}
F^{\mu'_1\mu_2\lambda_3}({\bf k}_1+{\bf q},{\bf k}_2-\bf{q})
\eea
Substituting the spinors in Eq.~(\ref{modelHelement}) this reduces to whether the following equality is or not satisfied
\bea \label{condicion}
\sqrt{1+s_{k_1}}  \sqrt{1+s_{k_1+q}} \delta_{\lambda_1\mu_1}+
\sqrt{1-s_{k_1}}  \sqrt{1-s_{k_1+q}}
\left(\!\!  \sigma\cd \hat{\bf k}_1 
\sigma\cd (\widehat{{\bf k}_1+{\bf q}})\!\!  \right)_{\lambda_1 \mu_1}
 = ?
 \\ \nonumber
\left( \sigma\cd \hat{\bf k}_1\right)_{s'_1 s_1}\!\! \left(
\sqrt{1+s_{k_1}}  \sqrt{1+s_{k_1+q}} \delta_{s_1\lambda_1}+
\sqrt{1-s_{k_1}}  \sqrt{1-s_{k_1+q}}
\left(\!\!  \sigma\cd \hat{\bf k}_1
\sigma\cd (\widehat{{\bf k}_1+{\bf q}})\!\!  \right)\right)_{s_1 \lambda_1} \!\! 
\left(\!\!  \sigma\cd (\widehat{{\bf k}_1+{\bf q}})\!\!  \right)_{\lambda_1 
\lambda'_1} .
\eea
\end{widetext}
The equality does not generally hold and the two states $\ar 1\ra$, 
$\ar 2\ra$ are not degenerate if chiral symmetry is broken.
But high in the spectrum where the $F$ wavefunctions have support, $\sin \phi(k)$ is effectively zero. Substituting $s_k=0$ in Eq.~(\ref{condicion}) and using $\sigma 
\hat{{\bf k}} \sigma\hat{{\bf k}}=\mathbb{I}$, it is satisfied. 
We tentatively conclude from this analysis 
that the Hamiltonian in eq. (\ref{Hmodel} makes the 
two states degenerate when $\sin \phi(k)\to 0$ (but see later in Sec.~\ref{sec:conclusion}).

The degeneracy would clearly be absent if we employed a chiral-symmetry violating 
interaction with vertex $\bar{U}U=U^\da \gamma_0 U$ instead of the vector-coupled quarks in Eq.~(\ref{Hmodel}). In the Pauli-Dirac 
representation that we employ, 
\be
\gamma_0=\left(
\begin{tabular}{cc}
$\mathbb{I}$ & 0 \\
0 &$ - \mathbb{I}$
\end{tabular}
\right)
\ee
and hence 
\bea
\left( \sigma\cd \hat{\bf k}_1\right)_{s_1 s'_1}
(U^\da \gamma_0 U)_{s'_1\lambda_1}
\left( \sigma\cd (\widehat{{\bf k}_1+{\bf q}}) \right)_{\lambda_1 
\lambda'_1}
\\ \nonumber
\to  - (U^\da \gamma_0 U)_{s'_1\lambda_1}
\eea
when $\sin\phi(k)\to 0$, and there is no degeneracy. This demonstrates 
explicitly the cancellation in the chiral symmetric case, and the lack 
of cancellation in the chiral non-symmetric.

Beyond our simple model Hamiltonian, examination of the exact QCD one formulated in Coulomb gauge~\cite{Christ:1980ku,Szczepaniak:2001rg} immediately 
reveals that the differences lay in the gauge part but not in the quark-spin structure, and though the mass function $\sin \phi (k)$ may be chosen to involve more complicated correlations than the BCS ansatz, chiral symmetry appears so far to be manifested in Wigner mode high in the spectrum.

To conclude this subsection, let us comment on the additional known consequence of  possible insensitivity to chiral symmetry breaking in the high spectrum
 (a dynamical effect due to hadrons decoupling from the condensed fermions) 
This is that all resonances
decouple from their $N\pi\dots\pi$ decay channels, they have to decay through
intermediate excited states (or if this is not possible, be very narrow)
That is, $g_{N^{**}N^*\pi}\to 0$ for all possible couplings involving pions.
This is easy to see in BCS approximation~\cite{Nefediev:2008dv}: the coupling will be
proportional to a wavefunction overlap, formally 
\be \label{coupling}
g_{N^**N^*\pi} \propto \int F^{N^{**}} {\mathcal B}_{N^{*}} F^{N^{*}} {\mathcal B}_{\pi}\sin\phi(k)
\ee
where the pion has been taken as an exact Goldstone boson with wavefunction 
$F^{\pi}\propto \sin\phi(k)$ and $\mathcal B$ are boost operators (as the outgoing wavefunctions correspond to moving particles). Both the boost operators and the very excited
resonance wavefunction $F^{N^{**}}$ in Eq.~(\ref{coupling}) select large values of the quark momentum $k$. This entails that $\sin\phi (k) \to 0$, and thus $g_{N^{**}N^*\pi}\to 0$ too.

\section{Variational wavefunction basis} \label{sec:basis}

The three-fermion problem is a classic of quantum mechanics (e.g. the triton nucleus),
and extensive work has been carried out in the quark model for baryons ~\cite{Loring:2001kv,Loring:2001kx,Loring:2001ky,Plessas:2015mpa,Day:2013rza,Theussl:2000sj,Isgur:1978wd}
We build on well-established results and put together a practical
package of wavefunctions that can also be used in future applications. 
The construction of this basis and its numeric implementation is by far the most time consuming part of this project and we report it with some detail for reproducibility. 

We can divide the construction of the wavefunction basis in several steps
that are conceptually well-separated and given in the following subsections.
In essence, we start with the three-dimensional harmonic oscillator
for the intrinsic coordinates in momentum space, 
whose wavefunctions we denote by $\phi$. The spins need to be combined
to yield states labelled by the conserved $J$, $m_J$.
Then we antisymmetrize respect to
the three sets of quark quantum numbers, to yield a new set of three-body
wavefunctions $\Phi$ that are orthonormal and antisymmetric. This is done with the
help of the Moshinsky-Brody-Ribeiro-van Beveren (MBRB)
coefficients~\cite{Moshinsky:1959,Brody:1967,Moshinsky:1969,Ribeiro:1978gx,vanBeveren:1982sj}.

\subsection{Momentum space wavefunction and Jacobi coordinates} \label{subsec:Jacobi}
As the first step, we employ a standard variational approximation to the vacuum, the BCS state
$\ar \Omega \ra$. A baryon, in lowest order in the Fock-space expansion is a three-quark
excitation thereof,
\begin{widetext}
\be \label{baryondef}
\ar {\mathcal{B}} \ra = \sum_{c,s,f} \prod \int \frac{d^3p_i}{(2\pi)^3}
\frac{\epsilon^{c_1 c_2 c_3}}{\sqrt 6} F_{\mathcal{B}}^{s,f}({\bf p}_1,
{\bf p}_2,{\bf p}_3) B^\da_{c_1,s_1,f_1,{\bf p}_1} 
B^\da_{c_2,s_2,f_2,{\bf p}_2} B^\da_{c_3,s_3,f_3,{\bf p}_3} \ar \Omega \ra \ .
\ee
\end{widetext}

We could use one-particle Cartesian momenta ${\bf p}_1$ and ${\bf p}_2$, while 
enforcing ${\bf p}_3=-{\bf p}_1-{\bf p}_2$. However the possibility
to use the algebraic MBRB coefficients, that
reduce the basis wavefunction overlap without need for any numerical 
integration, prompted us to use the Jacobi coordinates.
These however are not the standard non-relativistic Jacobi coordinates that
depend on the particle mass, since the relativistic kinetic energy is not quadratic
and does not separate anyway. Because of symmetry considerations we choose instead
\begin{equation}
\left(
\begin{array}{c}
{\bf p}_\rho \\
{\bf p}_\lambda \\
{\bf p}_R \\
\end{array}
\right) 
=
\left(
\begin{array}{ccc}
{1  \over \sqrt 2 }& - { 1 \over \sqrt2 } & 0 \\
{1  \over \sqrt 6 }& { 1 \over \sqrt 6 } & - { 2 \over \sqrt 6 } \\
{1  \over \sqrt3 }&  { 1 \over \sqrt3 } & { 1 \over \sqrt3 } \\
\end{array}
\right)
\left(
\begin{array}{c}
{\bf p}_1 \\
{\bf p}_2 \\
{\bf p}_3 \\
\end{array}
\right) \ .
\end{equation}

The color wavefunction $\frac{\epsilon^{c_1 c_2 c_3}}{\sqrt 6}$ is normalized to $1$ taking into account all possible contractions 
of the fields in the $\la {\mathcal{B}}\ar {\mathcal{B}} \ra$ overlap.
Since we construct $F$ to be totally symmetric under exchange of any two 
quarks, we do not sum over permutations of the $B$'s in the orthonormalization of the
basis (to avoid double-counting).

In terms of the Jacobi coordinates, and in the center of mass
${\bf P}_R=0$ the wavefunction becomes 
\be \label{baryondef2}
F_{\mathcal{B}}^{s,f}({\bf p}_1,{\bf p}_2,{\bf p}_3)=
F_{\mathcal{B}}^{s,f}({\bf p}_\rho,{\bf p}_\lambda) 
\ee

\subsection{Single-particle Harmonic Oscillator wavefunctions}
For atomic physics systems, or weakly bound systems such as the 
deuteron, an adequate basis of functions is the Laguerre Polynomial basis, decaying
at large distances $e^{- c \ r}$. For particles in a box the Bessel
functions are maybe optimal. In the present case of a linear confining
potential, a practical basis is the eigenbasis of the harmonic oscillator (HO).
This is because many analytical results exist to simplify intermediate computations, and since
the wavefunctions are confining, variational convergence for the linear potential
should be straightforward. The more natural Airy basis (that would
diagonalize the non-relativistic Schr\"odinger equation with a linear 
potential) looses its advantage for relativistic kinetic energies, as 
necessary with a running mass function, and is more cumbersome to use.

Working in momentum space, we Fourier transform the HO functions, which depend on a parameter 
$\alpha$ with dimensions of length, that scales 
with the size of the ground state. 
If we used only one or two basis elements, it would have to be interpreted as a variational parameter to be varied until a minimum was found. However, since we will use an extended basis
and diagonalize the Hamiltonian in it, $\alpha$ can be left fixed at
any value (of course, an unreasonable value thereof would lead to very
poor convergence in the number of wavefunctions, so it is logic to take it
of order $\alpha \simeq 1/\Lambda_{\rm QCD}$). For each of the two ${\bf p}_\rho$, ${\bf p}_\lambda$, we have
\bea
\varphi_{nlm}^\alpha(\mathbf r) & = & i^{2n+l} \varphi_{nl}^\alpha (r) Y_{lm} 
\left(\hat r \right)
\nonumber \\
& = & i^{2n+l} \alpha^{- 3/2} \varphi_{nl}\left(\alpha^{-1} \, r \right) Y_{lm} \left(\hat r \right)
\nonumber \\
{ F.T. \atop \longrightarrow } \varphi_{nlm}^\alpha \left(\mathbf p 
\right)
& = & \left(2 \pi \alpha \right)^{3/2}  \varphi_{nl}
\left(\alpha \, p \right) Y_{lm} \left(\hat p \right) \ .
\label{eq:HO_wf}
\eea
There being two independent three-dimensional variables 
${\bf p}_\rho$, ${\bf p}_\lambda$, we take the product of two such 
H. O. functions, so that the unsymmetrized basis diagonalizes two
independent harmonic oscillators. We take their two constants $\alpha_\rho=
\alpha_\lambda=\alpha$ since this is useful to simplify the symmetrization of the wavefunctions on the variational
basis. As just argued, the choice of $\alpha$ is only relevant for convergence speed.
Doing so, the wavefunction space 
splits in "shells" of the sum of harmonic oscillators, each shell
being characterized by a shell-quantum number
\be
N=(2n_\rho + l_\rho + 2 n_\lambda+ l_\lambda)\; .
\label{eq:HO_energy}
\ee

It is useful to note that all states in the same shell have the same 
parity $P=(-1)^N=(-1)^{l_\rho+l_\lambda}$. Therefore, to study a baryon
sector of given parity, only half the shells need to be kept (the three
quark intrinsic parities being positive).

For even parity, a shell with quantum number $N$ has 
$\frac{(N+2)(N+3)(N+4)}{24}$ independent degenerate (for the HO) 
spin-multiplets of fixed $n_\rho$, $n_\lambda$, $l_\rho$, $l_\lambda$. Within each 
multiplet there are eight states corresponding to the two possible 
values of the spins $s_1$, $s_2$, $s_3$.

For odd parity, the corresponding number of multiplets is rather
$\frac{(N+1)(N+3)(N+5)}{24}$.
Thus, if we variationally limit the basis up to $N_{\rm max}$ shells,
the number of multiplets (hence storage space in the computer, for example)
grows asymptotically as $N_{\rm max}^4$.

Within a certain HO shell $N$, one can assign an \textit{in-shell}
quantum number or index (\textit{e.g.} $l$ or $k$) to the different
combinations of $(n_\rho,n_\lambda,l_\rho,l_\lambda)$ that make up $N$
according to Eq.~(\ref{eq:HO_energy}). The one-to-one correspondence
between the in-shell index $l$ and the set of orbital quantum numbers
is illustrated in Table~\ref{tab:orb_qns}. Instead of writing the
whole set of orbital quantum numbers, it is easier to refer to an
$(N,l)$-pair.

\begin{table}[t!]
\begin{ruledtabular}
\begin{tabular}{c|c|c c c c}
$ N $ & $ l $ & $ n_\rho $ & $ n_\lambda$  & $ l_\rho $ &$  l_\lambda $ \\
\hline
0 & 0 & 0 & 0 & 0 & 0 \\
\hline
1 & 0 & 0 & 0 & 0 & 1 \\
  & 1 & 0 & 0 & 1 & 0 \\
\hline
2 & 0 & 0 & 0 & 0 & 2 \\
  & 1 & 0 & 0 & 1 & 1 \\
  & 2 & 0 & 0 & 2 & 0 \\
  & 3 & 0 & 1 & 0 & 0 \\
  & 4 & 1 & 0 & 0 & 0 \\
\hline
3 & 0 & 0 & 0 & 0 & 3 \\
  & 1 & 0 & 0 & 1 & 2 \\
  & 2 & 0 & 0 & 2 & 1 \\
  & 3 & 0 & 0 & 3 & 0 \\
  & 4 & 0 & 1 & 0 & 1 \\
  & 5 & 0 & 1 & 1 & 0 \\
  & 6 & 1 & 0 & 0 & 1 \\
  & 7 & 1 & 0 & 1 & 0 \\
\hline
4 & 0 & 0 & 0 & 0 & 4 \\
  & 1 & 0 & 0 & 1 & 3 \\
  & 2 & 0 & 0 & 2 & 2 \\
  & 3 & 0 & 0 & 3 & 1 \\
  & 4 & 0 & 0 & 4 & 0 \\
  & 5 & 0 & 1 & 0 & 2 \\
  & 6 & 0 & 1 & 1 & 1 \\
  & 7 & 0 & 1 & 2 & 0 \\
  & 8 & 0 & 2 & 0 & 0 \\
  & 9 & 1 & 0 & 0 & 2 \\
  & 10 & 1 & 0 & 1 & 1 \\
  & 11 & 1 & 0 & 2 & 0 \\
  & 12 & 1 & 1 & 0 & 0 \\
  & 13 & 2 & 0 & 0 & 0 \\
\hline
$ \cdots $ &$ \cdots $ &$ \cdots $ &$ \cdots $ &$ \cdots $ &$ \cdots $ \\ 
\end{tabular}

\end{ruledtabular}
\caption{\label{tab:orb_qns} 
The in-shell index $l$ within a certain oscillator shell $N$
corresponds to a set of orbital quantum numbers
$(n_\rho,n_\lambda,l_\rho,l_\lambda)$ as illustrated. There are degeneracies {\it e.g.} $(2l_\rho+1)(2l_\lambda+1)$ that will lift in subsequent steps.}
\end{table}

\subsection{Spin coupling and spin index convention}
The wavefunctions of the variational basis include three $1/2$-spins, 
and the two orbital angular momenta, to yield
the total angular momentum, recoupling with the help of 
Clebsch-Gordan coefficients, for which we use bra-ket notation
$ \la l_1 m_1 l_2 m_2 \ar l m_l \ra $. 
(The isospin coupling will barely be discussed but to construct $I=1/2$, $3/2$ the procedure is the same, though simpler because of the orbital angular momentum complication here.)

The basis functions are made for a certain total spin
$J$ and spin projection $M_J$. The basis is cut at a certain maximum
value for the harmonic oscillator shell-quantum number $N_{\rm max}$. 
Ideally one would like to take the limit $N_{\rm max}$ to make the variational
treatment arbitrarily accurate, but this is not possible with finite computer power. In practice we have been able to reach $N_{\rm max}=12$ or less.

The starting set of basis functions with spin is simply chosen to be factorized
\begin{multline}
\mBF^{\alpha}_{N,l,m_\rho,m_\lambda,R_i} (\vpr,\vpl,S_i) =
\\ \varphi^\alpha_{n^{N,l}_\rho,l^{N,l}_\rho,m_\rho}(\vpr) \,
\varphi^\alpha_{n^{N,l}_\lambda,l^{N,l}_\lambda,m_\lambda}(\vpl) \,
\chi_{R_i}(S_i) \; . \label{eq:BF}
\end{multline}
Here, the $\varphi$-functions are the harmonic oscillator
wavefunctions from Eq.~(\ref{eq:HO_wf}). The spin-dependent part of
the basis function, $\chi_{R_i}(S_i)$, is basically a kronecker-delta
$\delta_{R_i S_i}$. The spin indices $R_i$ and $S_i$ are a shorthand
notation for the three uncoupled quark spins according to
Table~\ref{tab:spin_index}.

\begin{table}[hbt]
\begin{ruledtabular}
\begin{tabular}{c|c c c}
$ S_i $ & $ s_1 $ & $ s_2 $ & $ s_3 $ \\
\hline
$ 0 $ & $ \upa $ & $ \upa $ & $ \upa $ \\
$ 1 $ & $ \upa $ & $ \upa $ & $ \doa $ \\
$ 2 $ & $ \upa $ & $ \doa $ & $ \upa $ \\
$ 3 $ & $ \upa $ & $ \doa $ & $ \doa $ \\
$ 4 $ & $ \doa $ & $ \upa $ & $ \upa $ \\
$ 5 $ & $ \doa $ & $ \upa $ & $ \doa $ \\
$ 6 $ & $ \doa $ & $ \doa $ & $ \upa $ \\
$ 7 $ & $ \doa $ & $ \doa $ & $ \doa $ \\
\end{tabular}
\end{ruledtabular}
\caption{\label{tab:spin_index} 
Labelling of the three uncoupled quark spins $(s_1,s_2,s_3)$ with a spin index $S_i$ for
indexation in the computer.}
\end{table}

The set of basis functions~(\ref{eq:BF}) is  to be recoupled to total $J$ and  projection
$M_J$. This is accomplished by constructing an array $\phi$ of
coefficients which consist of a sum of products of Clebsch-Gordan
coefficients that couple the quark spins and orbital quantum numbers
to a fixed $(J,M_J)$. This changes the set of uncoupled quantum numbers
\be
n_\rho^{N,l}, l_\rho^{N,l}, m_\rho, n_\lambda^{N,l}, l_\lambda^{N,l}, m_\lambda,
S_1, s_1, S_2, s_2, S_3, s_3 \, ,
\ee
by the set of coupled quantum numbers
\be
n_\rho^{\Nbar,\lbar}, l_\rho^{\Nbar,\lbar}, n_\lambda^{\Nbar,\lbar},
l_\lambda^{\Nbar,\lbar}, L^{\Nbar}, S_1, S_2, S_3, S, S_{12}, J, M_J \, 
\ee
The barred ordering scheme for the coupled orbital quantum numbers is
illustrated in Table~\ref{tab:barred_orb_qns}, and we use a bar to distinguish the indices specific to the coupled basis. The barred ordering scheme for
the coupled spin quantum numbers is given in Table~\ref{tab:barred_spin_index}.

The basis functions of fixed $(J,M_J)$ are given by
\begin{widetext}
\be
\phi^{(J,M_J),\Nbar,\lbar,\Rbar_i} (\vpr, \vpl, S_i) =
\sum^{l^{\Nbar,\lbar}_\rho}_{m_\rho =
  -l^{\Nbar,\lbar}_\rho} \sum^{l^{\Nbar,\lbar}_\lambda}_{m_\lambda =
-l^{\Nbar,\lbar}_\lambda}
\sum^{7}_{R_i = 0} 
\phi^{(J,M_J)[\Nbar][\lbar][\Rbar_i]}_{[m_\rho][m_\lambda][R_i]}
\; \mBF^\alpha_{N^{\Nbar},l^{\Nbar,\lbar},m_\rho,m_\lambda,R_i} (\vpr, \vpl,
S_i) \;
, \label{eq:unsym_BF}
\ee
where the array $\phi$ of coefficients is in turn
\begin{multline}
\phi^{(J,M_J)[\Nbar][\lbar][\Rbar_i]}_{[m_\rho][m_\lambda][R_i]} =
\sum_{M_S=-S^{\Rbar_i}}^{S^{\Rbar_i}} \sum_{M_L=-L^{\Nbar}}^{L^{\Nbar}}
\sum_{M_{S_{12}}=-S_{12}^{\Rbar_i}}^{S_{12}^{\Rbar_i}}
\la l^{\Nbar,\lbar}_\rho \, m_\rho \, l^{\Nbar,\lbar}_\lambda \, m_\lambda \ar
L^{\Nbar} \, M_L \ra 
\la L^{\Nbar} \, M_L \, S^{\Rbar_i} \, M_S \ar J \, M_J \ra \\
\times \la \frac{1}{2} \, s_1^{R_i} \, \frac{1}{2} \, s_2^{R_i} \ar
S_{12}^{\Rbar_i} \, M_{S_{12}} \ra
\la S_{12}^{\Rbar_i} \, M_{S_{12}} \, \frac{1}{2} \, s_3^{R_i} \ar S^{\Rbar_i}
\, M_S \ra \; .
\label{eq:unsym_coef}
\end{multline}
\end{widetext}

\begin{table}[t!]
\begin{ruledtabular}
\begin{tabular}{c|c|c|c c c c|c c}
$ \Nbar $ & $ L $ & $ \lbar $ & $ n_\rho $ & $ n_\lambda $ & $ l_\rho $ & $ l_\lambda $ & $ N  $ & $ l $ \\
\hline \hline
0 & 0 & 0 & 0 & 0 & 0 & 0 & 0 & 0\\
\hline \hline
1 & 1 & 0 & 0 & 0 & 0 & 1 & 1 & 0\\
  &   & 1 & 0 & 0 & 1 & 0 &   & 1\\
\hline \hline
2 & 0 & 0 & 0 & 0 & 1 & 1 & 2 & 1\\
  &   & 1 & 0 & 1 & 0 & 0 &   & 3\\
  &   & 2 & 1 & 0 & 0 & 0 &   & 4\\
\hline
3 & 1 & 0 & 0 & 0 & 1 & 1 & 2 & 1\\
\hline
4 & 2 & 0 & 0 & 0 & 0 & 2 & 2 & 0\\
  & 2 & 1 & 0 & 0 & 1 & 1 &   & 1\\
  & 2 & 2 & 0 & 0 & 2 & 0 &   & 2\\
\hline \hline
5 & 1 & 0 & 0 & 0 & 1 & 2 & 3 & 1\\
  &   & 1 & 0 & 0 & 2 & 1 &   & 2\\
  &   & 2 & 0 & 1 & 0 & 1 &   & 4\\
  &   & 3 & 0 & 1 & 1 & 0 &   & 5\\
  &   & 4 & 1 & 0 & 0 & 1 &   & 6\\
  &   & 5 & 1 & 0 & 1 & 0 &   & 7\\
\hline
6 & 2 & 0 & 0 & 0 & 1 & 2 & 3 & 1\\
  &   & 1 & 0 & 0 & 2 & 1 &   & 2\\
\hline
7 & 3 & 0 & 0 & 0 & 0 & 3 & 3 & 0\\
  &   & 1 & 0 & 0 & 1 & 2 &   & 1\\
  &   & 2 & 0 & 0 & 2 & 1 &   & 2\\
  &   & 3 & 0 & 0 & 3 & 0 &   & 3\\
\hline \hline
8 & 0 & 0 & 0 & 0 & 2 & 2 & 4 & 2\\
  &   & 1 & 0 & 1 & 1 & 1 &   & 6\\
  &   & 2 & 1 & 0 & 1 & 1 &   & 10\\
  &   & 3 & 0 & 2 & 0 & 0 &   & 8\\
  &   & 4 & 1 & 1 & 0 & 0 &   & 12\\
  &   & 5 & 2 & 0 & 0 & 0 &   & 13\\
\hline
9 & 1 & 0 & 0 & 0 & 2 & 2 & 4 & 2\\
 $ \cdots $ & $ \cdots $ & $ \cdots $ & $ \cdots $ & $ \cdots $ & $ \cdots $ & $ \cdots $ & $ \cdots $ & $ \cdots $ \\
\end{tabular}
\end{ruledtabular}
\caption{\label{tab:barred_orb_qns} 
The barred shell number $\Nbar$ with its barred in-shell index $\lbar$
correspond to a set of orbital quantum numbers
$(L,n_\rho,n_\lambda,l_\rho,l_\lambda)$ as illustrated. The unbarred shell
number and in-shell index corresponding to the same set of orbital quantum
numbers are also given. Remark that whenever an unbarred set $(N,l)$ appears
more than once in the last two columns, this means that the Clebsch-Gordan
coefficients corresponding to $\langle L l_\rho l_\lambda M_L | l_\rho l_\lambda
m_\rho m_\lambda \rangle$ are different from 1.}
\end{table}

\begin{table}[hbt]
\begin{ruledtabular}
\begin{tabular}{|c|c c c|}
$ \Sbar_i $ & $ S $ & $ S_{12} $ & $ M_S $ \\
\hline
$ 0 $ & $ 1/2 $ & $ 0 $ & $ -1/2 $ \\
$ 1 $ & $ 1/2 $ & $ 0 $ & $ +1/2 $ \\
$ 2 $ & $ 1/2 $ & $ 1 $ & $ -1/2 $ \\
$ 3 $ & $ 1/2 $ & $ 1 $ & $ +1/2 $ \\
$ 4 $ & $ 3/2 $ & $ 1 $ & $ -3/2 $ \\
$ 5 $ & $ 3/2 $ & $ 1 $ & $ -1/2 $ \\
$ 6 $ & $ 3/2 $ & $ 1 $ & $ +1/2 $ \\
$ 7 $ & $ 3/2 $ & $ 1 $ & $ +3/2 $ \\
\end{tabular}
\end{ruledtabular}
\caption{\label{tab:barred_spin_index} 
This defines the one-to-one correspondence between barred spin index $\Sbar_i$ for the computer code and the three coupled quark spin quantum numbers $(S,M_S,S_{12})$.}
\end{table}

The distinction between upper and lower indices in $\phi$ in
Eq.~(\ref{eq:unsym_coef})) is generically useful when handling arrays of such indices related to basis functions.
The upper indices ($\Nbar,\lbar,\Rbar_i$) are not
summed over in order to evaluate a (basis) function, they are tags. 
Then the lower ones, such as ($m_\rho,m_\lambda,R_i$), are summed over.

In the end, these coefficients $\phi$ describe a unitary transformation between the uncoupled
and the coupled basis.

\subsection{Wavefunction symmetry}
We need to have a symmetrized momentum-spin-isospin wavefunction since fermion antisymmetrization is taken care of by the color degree of freedom. 
A way to formulate symmetry is by enforcing that the functions of the basis be
eigenfunctions of the exchange operators, $P^{12}$, $P^{13}$, $P^{23}$, $P^{123}$ and $P^{132}$.

All these exchange operators can be written as functions of $P^{12}$ and $P^{23}$
\bea
P^{123}&=& P^{23}P^{12}
\nonumber \\
P^{132} &=& P^{12}P^{23}
\nonumber \\
P^{13\phantom{2}}&=& P^{12}P^{23}P^{12}
\label{eq:symmetrizer}
\eea
so that if a state is an even eigenstate of both $P^{12}$ and $P^{23}$, it is automatically symmetric for any permutation of three quarks (antisymmetric when accounting for color). So, we only need to consider the effect of these two operators on the basis.
Our strategy  is then to obtain the eigenfunctions of these symmetry operators.

The fixed--$(J,M_J)$ basis defined in
Eqs.~(\ref{eq:BF},\ref{eq:unsym_coef}) is still
\textit{unsymmetrized}, so the individual functions are not necessarily eigenfunctions of
the operators $P_{ij}$ and they will be recombined. 

But notice that since that is an eigenbasis of the
harmonic oscillator, and the $P_{ij}$ commute with that
Hamiltonian, as also with total spin and orbital angular momentum, the
symmetrizer will not mix certain quantum numbers, and in particular
only states within the same shell $N$ will be combined.

Thus symmetrization only needs to mix wavefunctions within
a HO shell of fixed $N=(2n_\rho + l_\rho + 2 n_\lambda+ l_\lambda)$,
$L$, (and thus $\Nbar$ which only depends on these two is not mixed either in the symmetrization/orthonormalization process)
and $S$, $J, m_J$. 
The next step in our computer treatment is then to construct
orthonormal sets of basis functions for each HO shell $N$, spin $J$ and a certain spin
projection $M_J$.

In summary, we seek within each shell the eigenfunctions $\Phi$ of $P_{12}$ and $P_{23}$ for which
\be
P_{12} \Phi = P_{23} \Phi = \Phi \, .
\ee
Since $P_{ij}$ are Hermitian, the basis so generated is automatically orthogonal.

The action of the symmetrizers on the spin and flavor indices amounts to a simple exchange.
As for the momenta, we need the representation of the permutation 
group on the Jacobi coordinates. Let
\be\left(
\begin{tabular}{c} 
${\bf p}_\rho'$ \\ $ {\bf p}'_\lambda$ 
\end{tabular} \right) = P_{\sigma}  
 \left(
\begin{tabular}{c} 
${\bf p}_\rho$\\  ${\bf p}_\lambda$ 
\end{tabular} \right) 
\ee
then

\bea
{\mathbb{I}} &=& \left( 
\begin{tabular}{cc}
$1$ & $0$ \\ $0$ & $1$
\end{tabular} 
\right)
\\ \nonumber
P^{12} &=& \left( 
\begin{tabular}{cc}
$-1$ & $0$ \\ $0$ & $1$
\end{tabular} 
\right) 
\\ \nonumber
P^{23} &=& \left( 
\begin{tabular}{cc}
$1/2$ & $\sqrt{3}/2$ \\ $\sqrt{3}/2$  & $-1/2$
\end{tabular} 
\right)
\label{eq:perm_matrices}
\eea
that have simple interpretations in the $(\rho,\lambda)$-plane as 
a reflection and a $60^{\rm o}$ rotation. Other $P_\sigma$ can be constructed from these matrices to complete the permutation group, but this will not be necessary.

Now, since our functions $\phi$ are by construction eigenfunctions of $S_{12}$ and $I_{12}$ and $l_\rho$ is fixed, they are already eigenstates of $P^{12}$ without further work,
\be 
P^{12} \phi = (-1)^{l_\rho + S_{12} + I_{12}} \phi = \pm \phi
\ee
and the only task is to discard those that are antisymmetric
$P^{12} \phi = - \phi$ from further consideration, which requires minimum bookkeeping.
This yields a new basis of functions $\phi_i'$ similar to the initial one but with $P^{12}$ antisymmetric elements filtered out.

In this new basis, we construct the matrix with elements
\be
\Pi_{ij} = \langle \phi_i' | P_{23} | \phi_j' \rangle
\ee
 and diagonalize it
\be
\Pi_{ij} c^k_j = \lambda^k c^k_j 
\ee
only the eigenstates with $\lambda^k = 1$ are physically relevant, the others should be ignored.
The new basis is given by
\be
\Phi_i = \sum_{j} c^i_j \phi_j'
\ee 

The matrix elements of $\Pi$ can be analytically calculated with the help of
the Moshinsky-Beveren-Ribeiro coefficients, but this last tedious evaluation is 
detailed on the appendix.

The resulting final basis satisfies all necessary properties (good $J$, $M_J$, symmetry and orthonormalization) and is represented on the computer in terms of the factorized HO-spin wavefunctions. The coefficients of expansion must be analogous to the fractional parentage coefficients used in nuclear structure calculations, but we have not followed the analogy.

\subsection{Control of the numerical precision}

Because we are aiming at small mass splittings within chiral multiplets,
we need to decrease as much as possible the numerical errors in diagonalizing our Hamiltonian.

There are two difficult numerical errors to control. The first is the convergence of the nine-dimensional  integrals to compute each matrix element of the potential in Eq.~(\ref{modelHelement}). The second is the truncation of the wavefunction basis composed of a three-dimensional harmonic oscillator basis per Jacobi coordinate. Once this is achieved, the Hamiltonian is a matrix of the order of $10^3 \times 10^3$, and its diagonalization is standard. 

The complexity of our numerical task, starting from the construction of an wavefunction basis with the correct symmetry, led us to develop a full C++ project, developing parallel codes for CPU clusters.

In what concerns the numerical integration, though we need up to nine dimensional integrations, our past experience~\cite{Bicudo:2009cr} suggests that Monte Carlo integration is less accurate than we need, since in Ref.  \cite{Bicudo:2009cr} we were able to achieve only a precision of $O(100)$ MeV.

Therefore we have settled for numerical Gaussian integration.
We have 9-dimensional momenta integrals, with three radial-like integrals and six angular integrals.
For the radial integrals we perform the change of coordinate for the $q$ coordinate and use Gauss-Chebyshev quadrature, for
$p_\rho$ and $p_\lambda$, we use Gauss-Laguerre integration, all with up to 12 points.
For the angular integrals, we group them in three two-dimensional solid angle integrals, and use Lebedev integration with up to 74 points. This part, was made translating the code from
\cite{Lebedev_Laikov} to modern C++.
In total, per matrix element in our variational basis, we have up to 700 million integration points. Thus we reduce the numerical integration error down to a negligible $O(2)$ MeV.

Once the numerical integration error is under control, the most important source of uncertainty in this computation is the need to truncate
at a finite shell number $N$. As discussed below Eq.~(\ref{eq:HO_energy}), the total number of basis states grows with $N_{\rm max}^4$, and the number of matrix elements of $H$ to be computed with the square of that. Thus, we have limited $N_{\rm max}=10$ or 11, depending on $J^P$, for most computations with the exception of $J=13/2$ where we find necessary to go up to 12. 

No further effort in this direction is possible to us, as the final runs for the present computations, in PC clusters with 48 cores, lasted for six months.

Since we are limited in the maximum number of harmonic oscillator shells $N_{\rm max}$, we adopt the extrapolation of our results to $N_{\rm max}$, corresponding in a sense to the extrapolation to the infinite volume limit of lattice QCD. Notice the space extent of the harmonic oscillator functions scale like $\langle r^2 \rangle \propto N_{\rm max}$, thus the volume scales like ${N_{\rm max}}^{3/2}$.   Since there is no theoretical reason to believe that there is no term with $1/N_{\rm max}$ dependence, and since this dependence is aparently followed by the four higher shells we compute, we have used this last extrapolation for our results to follow.

To quantify the uncertainty of the neglected upper shells we have looked at three errors. The first is the simple difference of the last computed shells, for example $E(N_{\rm max}=10)-
E(N_{\rm max}=8)$. Second, we extrapolate to $N_{\rm max}\to \infty$ by means of  
$E(N_{\rm max})=E_\infty + \frac{\delta}{N_{\rm max}^2}$, and finally by means of 
$E(N_{\rm max})=E_\infty + \frac{\delta}{N_{\rm max}}$.  Nevertheless we have tabulated and inspected all three for comparison.

\begin{figure}[t]
\includegraphics*[width=0.9\columnwidth]{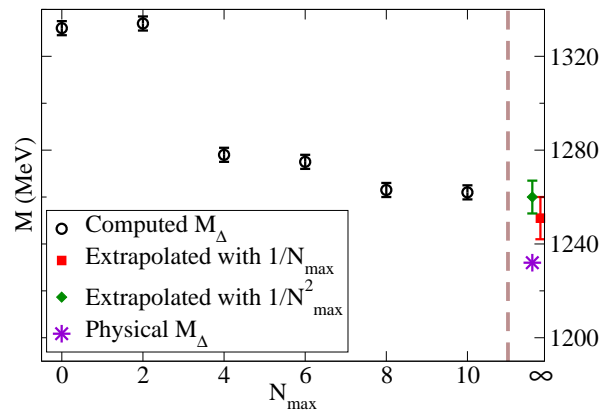}
\caption{\label{fig:DeltaExtrapolation} Black empty circles: computed $M_\Delta$ for each $N_{\rm max}$ maximum shell number as shown, with the error bar representing the numerical integration error. Filled symbols with error bars: extrapolations to $N_{\rm max}=\infty$. The error bars now include the extrapolation error.  Star: physical mass at 1232 MeV.}
\end{figure}

Fig. ~\ref{fig:DeltaExtrapolation} shows an example with the direct computations for different $N_{\rm max}$, and both extrapolations, in this case for the $J=3/2$ family of $\Delta$ baryons. The agreement is satisfactory, and the remaining difference with the experimental point must be adscribed to the model Hamiltonian.

In comparing with other works, we have noted that our prefered extrapolation to large shell-number, $E_\infty + \frac{\delta}{N_{\rm max}}$ converges more slowly than the exponential one proposed for conventional three-body nuclear calculations in~\cite{Perez:2015bqa}. 
We take the regression error in the fit as the extrapolation uncertainty.

\begin{figure}[t]
\includegraphics*[width=0.9\columnwidth]{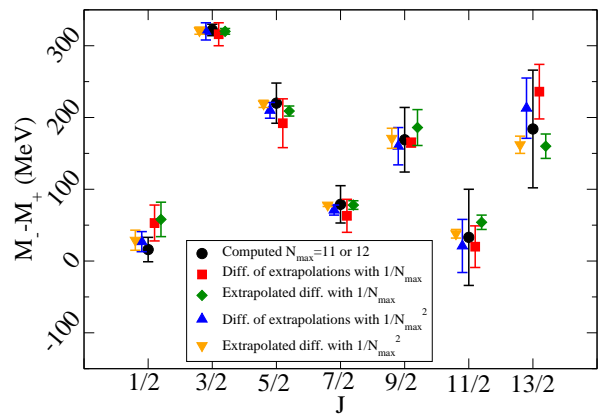}
\caption{\label{fig:Deltasplittings}  Comparison of the five methods of estimating the splitting from the variationally truncated Hamiltonian matrix elements, here for the $\Delta$ baryon family.}
\end{figure}

Proceeding beyond the absolute mass values, we next look at the parity splittings. One can adopt two extrapolation strategies: to take the difference of the extrapolations in the earlier step, or to extrapolate the difference of the computed mass values. While we choose the later option for subsequent plots, both methods are compared  
in Fig. ~\ref{fig:Deltasplittings} and we have tabulated them for all computations as an additional check.

Once we are convinced that we understand the uncertainty in our 
mass computations, adding the numerical integration error to the extrapolation error,
we proceed to report the outcome of the mass splittings in the proposed chiral multiplets.

\section{Numerical results}\label{sec:numerics}

We now proceed to systematically examine our numerical results for the parity splittings $M_+-M_-$ for increasing energy in the baryon spectrum. 

\begin{figure}[t]
\includegraphics*[width=0.9\columnwidth]{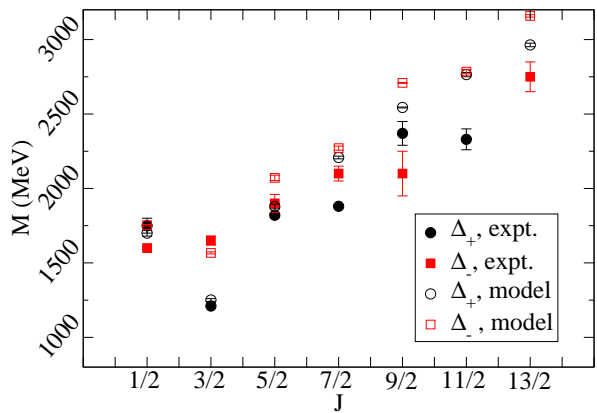}
\caption{\label{fig:thexpD} Comparison of the calculated isospin $3/2$  eigenvalues ( with positive and negative parity) of the $\Delta$ mass spectrum to the experimental data as collected by the PDG.}
\end{figure}

\begin{figure}[t]
\includegraphics*[width=0.9\columnwidth]{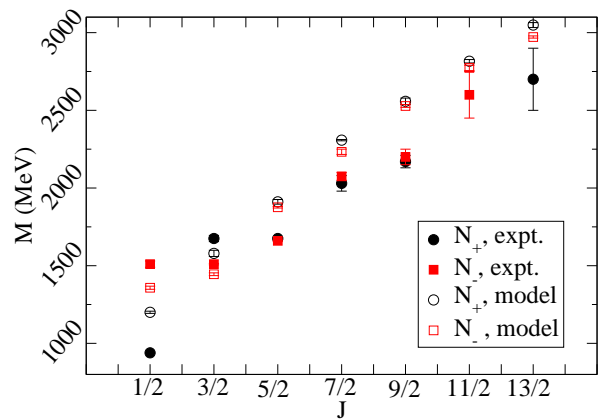}
\caption{\label{fig:thexpN}  Same as Fig. ~\ref{fig:thexpD} but for isospin 1/2 Nucleon excitations.}
\end{figure}

We first compare our spectra with the experimental ones. In Fig. \ref{fig:thexpD} we compare our results with the experimental $\Delta^*$ spectrum, and in Fig. \ref{fig:thexpN} we compare our results with the experimental $N^*$ spectrum, as a function of the total angular momentum $J$. For clarity, we only show the groundstates for each $J$ and parity $P$.

Notice our model has only one parameter in the interaction, the string tension $\sigma$. In this sense it is comparable to lattice QCD, which although it has no dimensional  parameter in the Lagrangian, upon quantization a scale appears and lattice QCD gets a dimensional scale, for instance the string tension. Quantitatively, our one-parameter model is not able to get the exact hyperfine splittings of the spectrum, but qualitatively our spectrum, adjusted to have a similar Regge slope, has a good mass splitting between the $\Delta_+$ and the $\Delta_-$ and follows the same general trend of the experimental data.

\begin{figure}[t]
\includegraphics*[width=0.9\columnwidth]{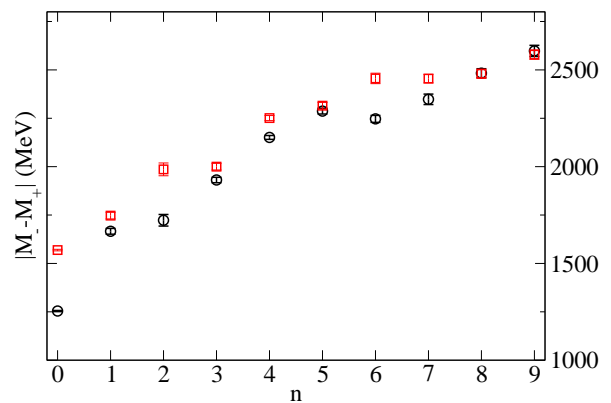}
\caption{\label{fig:radialmasses} Mass of the first ten radial excited states for the $I=J=3/2$ $\Delta$ with positive and negative parity, extrapolated from numerical data with up to 
$N_{\rm max}=10$.}
\end{figure}

We also illustrate the mass of the first ten radial excitations of a given state, The $\Delta$ with $I= 3/2$and $J=3/2$ in Fig. \ref{fig:radialmasses}. In the case of radial excitations, fewer experimental excitations are known, and we only plot the results of our model.

These Figs.  \ref{fig:thexpD}, \ref{fig:thexpN} and \ref{fig:radialmasses} illustrate the results we are able to compute in our model. For the first time, we are able to compute the theoretical baryon spectrum in a framework with chiral symmetry, with more states than the experimental data.

\subsection{Parity splittings for the $\Delta$, $I=3/2$ case }

We first study in detail the $\Delta$, $I=3/2$ spectrum. This case has the simpler symmetry; because color is antisymmetric and isospin is symmetric, the space $\times$ spin is symmetric.

\subsubsection{Splitting of the radial $\Delta$ excitations}

Let us concentrate first on a fixed channel, conveniently chosen to be the  $I=J=3/2$ $\Delta$ one. 

To assess this, we plot the actual mass of the first ten states with positive parity and also that for the negative parity ones in Fig. \ref{fig:radialmasses}.

\begin{figure}[t]
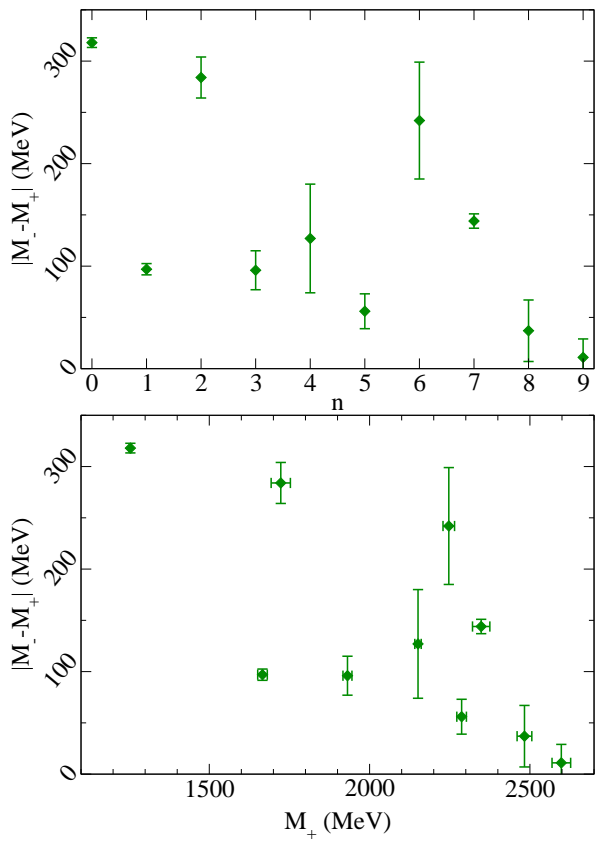

\includegraphics*[width=0.9\columnwidth]{FIGS_DIR/SplitRadialDelta.eps}
\includegraphics*[width=0.9\columnwidth]{FIGS_DIR/SplitRadialDelta_2.eps}
\caption{\label{fig:DeltaRadial}  Mass splittings in the radial excitations of the $I=J=3/2$ $\Delta$ channel. In the $OX$ axis, the top plot shows the number of the excitation and the bottom plot shows the actual mass of the positive-parity member of the doublet. In both cases the $OY$ axis provides the parity mass splittings.}
\end{figure}

Then one can compare any given state not only with the opposite parity one of the same $n$ but also with the neighbooring ones $n\pm 1$, for example.  
Our results for the parity splittings are shown in Fig. \ref{fig:DeltaRadial}.
The top plot shows the sequence $|M^{(n)}_+-M^{(n)}_-|$ against the excitation number $n$, while the top plot uses the mass of the positive parity partner as the $OX$ axis. 
There is a visible anticorrelation between the mass of the excited baryon and the parity splitting, but not a discernible power law fall. But we should not take for granted that, say the 7th positive parity baryon needs to match the 7th negative parity one (which is what is plotted). It might well be that there is an unequal number of different parity states in the lower spectrum and that they start partnering up only at quite high masses.

There is one way that, within the model calculation, we could enforce the partnering of the states, by employing the quartet states in Eq.~(\ref{quartetstates}). For example, we could compute the positive parity states up to a given shell and then apply to it the chiral charge $Q_5$ as needed to construct its chiral partner. 
This is man and computing power intensive (because part of the wavefunction corresponding to the upper shells will be out of the truncated variational space) so we have not pursued it further. 

Also it becomes a pure theoretical exercise, as experiment will not have such a handle and will be limited to analyze its results as in Fig. ~\ref{fig:radialmasses}, by giving the sequence of increasing masses for a given channel. Moreover, because the experimental states are broad and overlapping, it is unclear whether any excitations beyond the 3rd or 4th will be achievable. In all, we find that the most promising alley of experimental investigation is by studying increasing angular momentum, and now we turn our attention to it.

\subsubsection{Parity splittings in angular  $\Delta$ excitations}

\begin{figure}[t]
\includegraphics*[width=0.9\columnwidth]{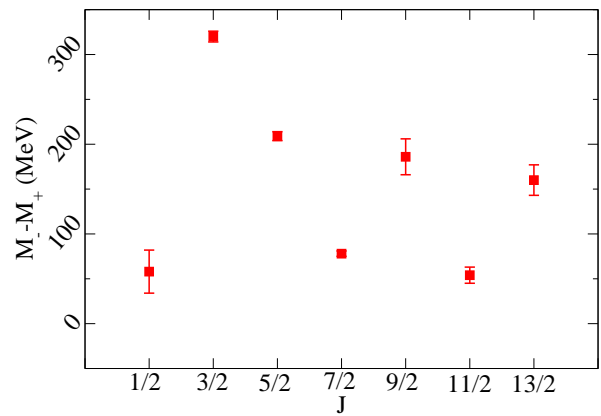}
\caption{\label{fig:DeltaJsplittings} Parity splittings for $\Delta \ I=3/2$ baryons, as a function of $J$, up  to total $J=13/2$.}
\end{figure}

We show our results for the splittings for the highly angular excited $\Delta$ baryons in Fig. \ref{fig:DeltaJsplittings}.
We remark the $\Delta$ parity splittings show a decreasing trend for high $J$, but not in a monotonous manner. The decrease is clear for the sub-spectrum even in $J- 3/2$, but for the odd $J-3/2$ sub-spectrum we cannot exclude the splitting is converging to a finite limit.


\begin{figure}[t]
\includegraphics*[width=0.9\columnwidth]{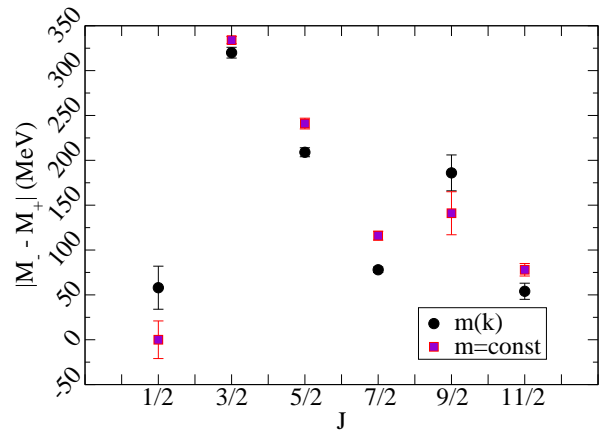}
\caption{\label{fig:constqm} Comparing the parity splittings for $\Delta \ I=3/2$ baryons with running quark mass and with constant quark mass up to total $J=11/2$.}
\end{figure}

Moreover we compare, in Fig. \ref{fig:constqm}, the model calculation for these $\Delta$ splittings with those obtained with the same code but fixing the quark mass by hand to $M=M(0)=\rm constant$. 
This allows to estimate the effect of the running mass in the spectrum, as discussed in sub-section \ref{subsec:expansion}.
Moreover this spoils the Ward identity linking the quark propagator and the pion wavefunction, and the chiral charge is no more even approximately conserved. In consequence one expects the splittings in this second calculation to lie above those in the chiral computation.
Indeed, for $J=3/2$, $5/2$, $7/2$ and $11/2$ our appreciation is correct and the parity splitting falls faster within the Cornell model with running quark mass than than in the constituent-like quark model with fixed mass.

And again, interestingly, for $J=9/2$ the {\it a priori} expected respective sizes are exchanged.


\subsection{Parity splittings in the possible $\Delta$ Yrast - Yrare quadruplet }

\begin{figure}[t]
\includegraphics*[width=0.9\columnwidth]{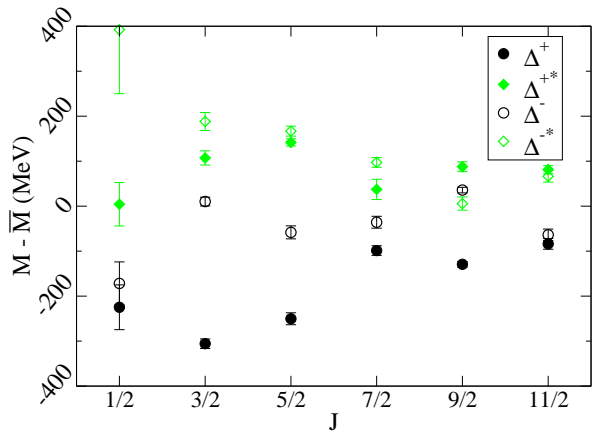}
\caption{\label{fig:Deltaexc} Parity splittings for the quadruplets including the Yrast and Yrare states with both positive and negative parity as a function of $J$. We show the extrapolation of the groundstates and the extrapolation of the first excitations in the $\Delta \ I=3/2$ channel.}
\end{figure}

We also study the splittings in the possible chiral quadruplet including the groundstate Yrast $n=0$ and the first radial excitation Yrare $n=1$ of the $\Delta \ I=3/2$ as a function of $J$. We compute the splittings up to $J=13/2$, as shown in Fig. ~\ref{fig:Deltaexc}. All states have been computed for different numbers of shells and the splittings extrapolated with $1/N_{\rm max}$. In Ref \cite{LlanesEstrada:2011jd} we discussed the possible subdivision of the quadruplet in two doublets, after diagonalizing Eq. (\ref{quartetstates}). Again, this doubling seems to be setting in for $J- 3/2$ even, not but for $J-3/2$ odd.

\subsection{Parity splittings in the possible $\Delta-N$ quaduplet }

\begin{figure}[t]
\includegraphics*[width=0.9\columnwidth]{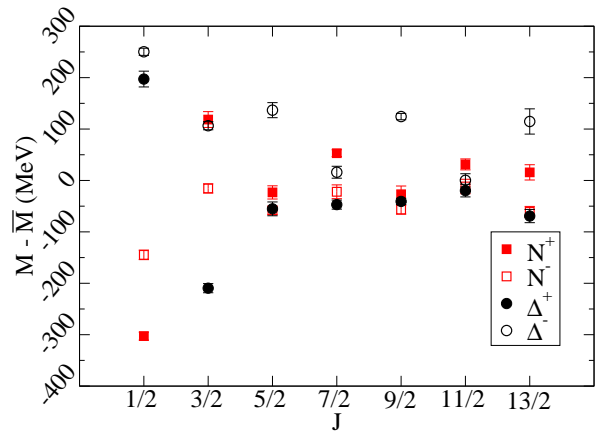}
\caption{\label{fig:NDofE} Yrast Nucleon and $\Delta$ parity splittings in the model. The nucleon ones already follow the expected decreasing trend for these  moderate (and experimentally accessible to partial wave analysis) angular momenta, while the $\Delta$ splittings still show oscillations and are inconclusive.}
\end{figure}

Fig. ~\ref{fig:NDofE} shows the resulting parity splittings for the quadruplet including both $I=3/2$ $\Delta$ and $I=1/2$ $N$ baryons with $P=\pm 1$, as function of angular momentum $J$. This is the central outcome of our work. 

Again, we observe in our results some level of even-odd staggering. The even $J-3/2$ Yrast nucleon ( corresponding to $3/2$, $7/2$ and $11/2$) parity splittings decrease very satisfactorily within the quadruplet, and show both the 
parity and isospin degeneracy expected from insensitivity to chiral symmetry breaking in the high spectrum. However, in  the odd $J-3/2$ Yrast nucleon (corresponding to  $J=1/2$, $5/2$, $9/2$ and $13/2$) the splittings are apparently not converging to zero for infinitely large $J$.

\subsection{Quark momentum distributions} \label{subsec:momdist}

\begin{figure}
\includegraphics*[width=0.9\columnwidth]{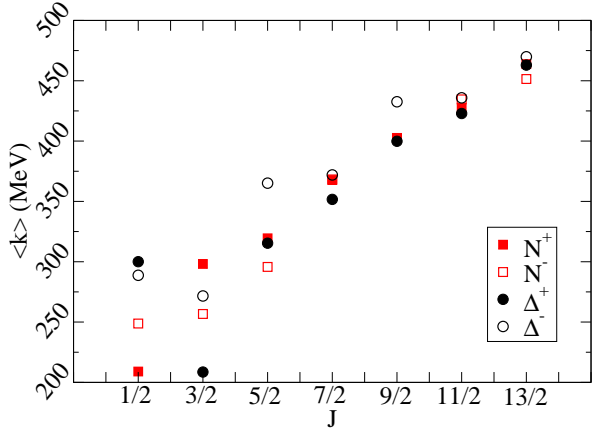}
\caption{\label{fig:AverageK} Average quark momentum $\langle k \rangle$ for each of the positive and negative parity N and $\Delta$ resonances.}
\end{figure}

Since the key to dropping parity splittings is in the quark running mass and in the quark momentum distribution,
we now analyze them in detail.
In Fig. ~\ref{fig:AverageK} we examine the average momentum $\langle k \rangle$ for a quark in each of the states.
The average momentum clearly does increase with $J$. This confirms our use of the virial theorem.

\begin{figure}
\includegraphics*[width=0.9\columnwidth]{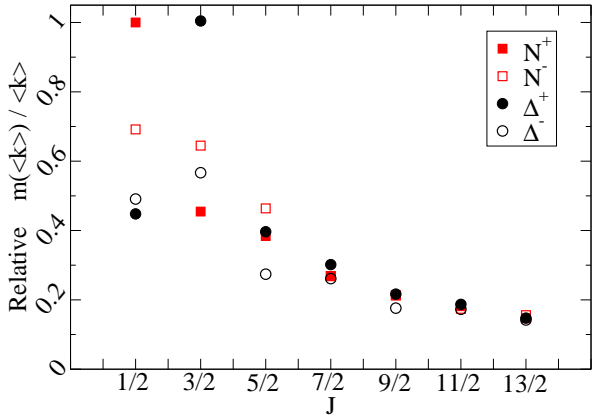}
\caption{\label{fig:averageM} Drop of the average (running) quark mass across the spectrum due to increasing average momentum. This is fully extracted from the variational computation.}
\end{figure}

Then this has to be compared to the decrease in the quark mass for that average momentum.
Fig. ~\ref{fig:averageM} shows $m(\langle k\rangle)/\langle k\rangle$ for the same states. Since the quark mass in this $\gamma_0\gamma_0$ model is rather small, we have preferred to divide by the same quantity for the lightest state of each isospin-parity combination, so that we only plot the drop relative to that reference point. 
Clearly, the average of the quark mass decreases with momentum, as we anticipated in our theorem.

\begin{figure}[t]
\includegraphics*[width=0.9\columnwidth]{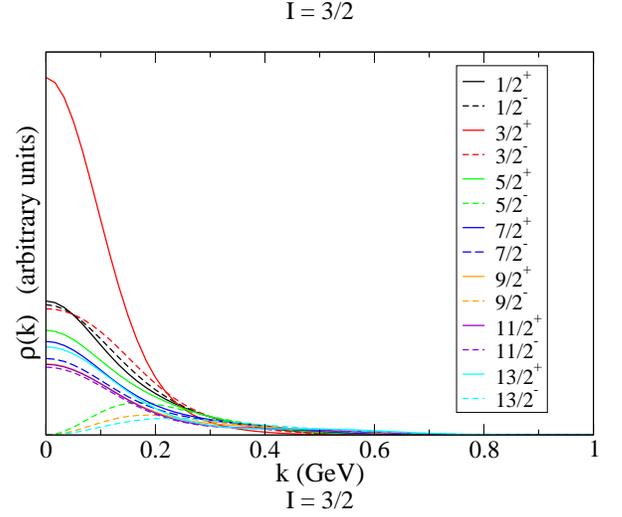}
\includegraphics*[width=0.9\columnwidth]{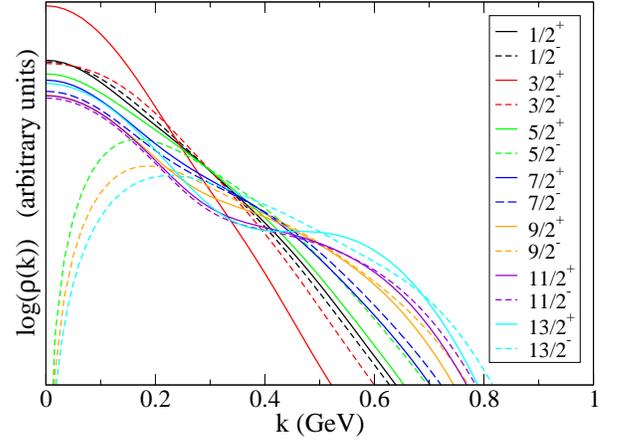}
\caption{\label{fig:momdists} Momentum wavefunction squared (top) and its logarithm (bottom) for various $I=3/2$ $\Delta^J$ combinations. (Color online)}
\end{figure}

Then, for more detail, a different important vista is shown in Fig. ~\ref{fig:momdists}. The squared wavefunction in arbitrary units is shown for most of the $\Delta$ states discussed in this work.The top plot displays $|\Psi|^2$ in linear scale, the bottom one in logarithmic scale, for best visibility.

It is clearly seen that increasing the angular momentum quantum number $J$ pushes the linear momentum distribution to higher values (towards the right of the plots). 

There are three states that deserve special mention by their behaviour at low momentum. The top line there (corresponding to the ground state $3/2^+$ baryons) is much more peaked at low momentum than the others.
But there are three other states, with $J^P=5/2^-$, $J^P=9/2^-$ and  $J^P=13/2^-$ that distinguish themselves by having vanishing wavefunction for zero average quark momentum. 

The reason for this behaviour is found examining the wavefunction basis, and is due to specific quantum number combinatorics. One finds that only in these three cases we do not have a $l_\lambda=0$ function on the lowest shell.
This means that their splittings to the (differently behaved) positive parity states is larger than otherwise expected, as can be seen in Fig. ~\ref{fig:constqm} below.
Notice we would expect, with high $J$, this component with $l_\lambda=0$ and with $\rho(k)=0$, should eventually vanish for all states. Thus the states $J^P=5/2^-$, $J^P=9/2^-$ and  $J^P=13/2^-$ are the ones with the expected behaviour.

\begin{figure*}
\includegraphics[width=0.45\columnwidth]{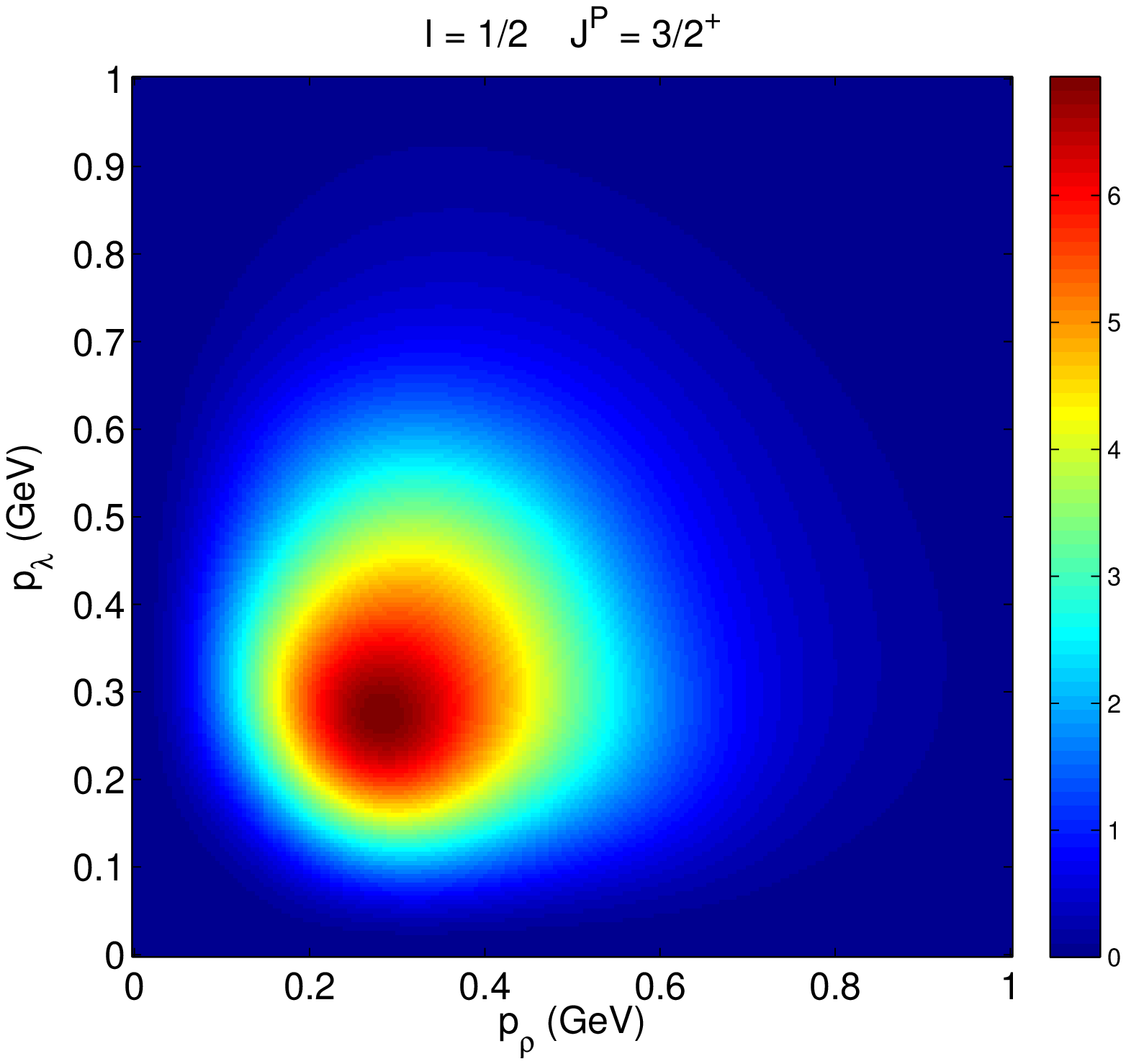} \quad \includegraphics[width=0.45\columnwidth]{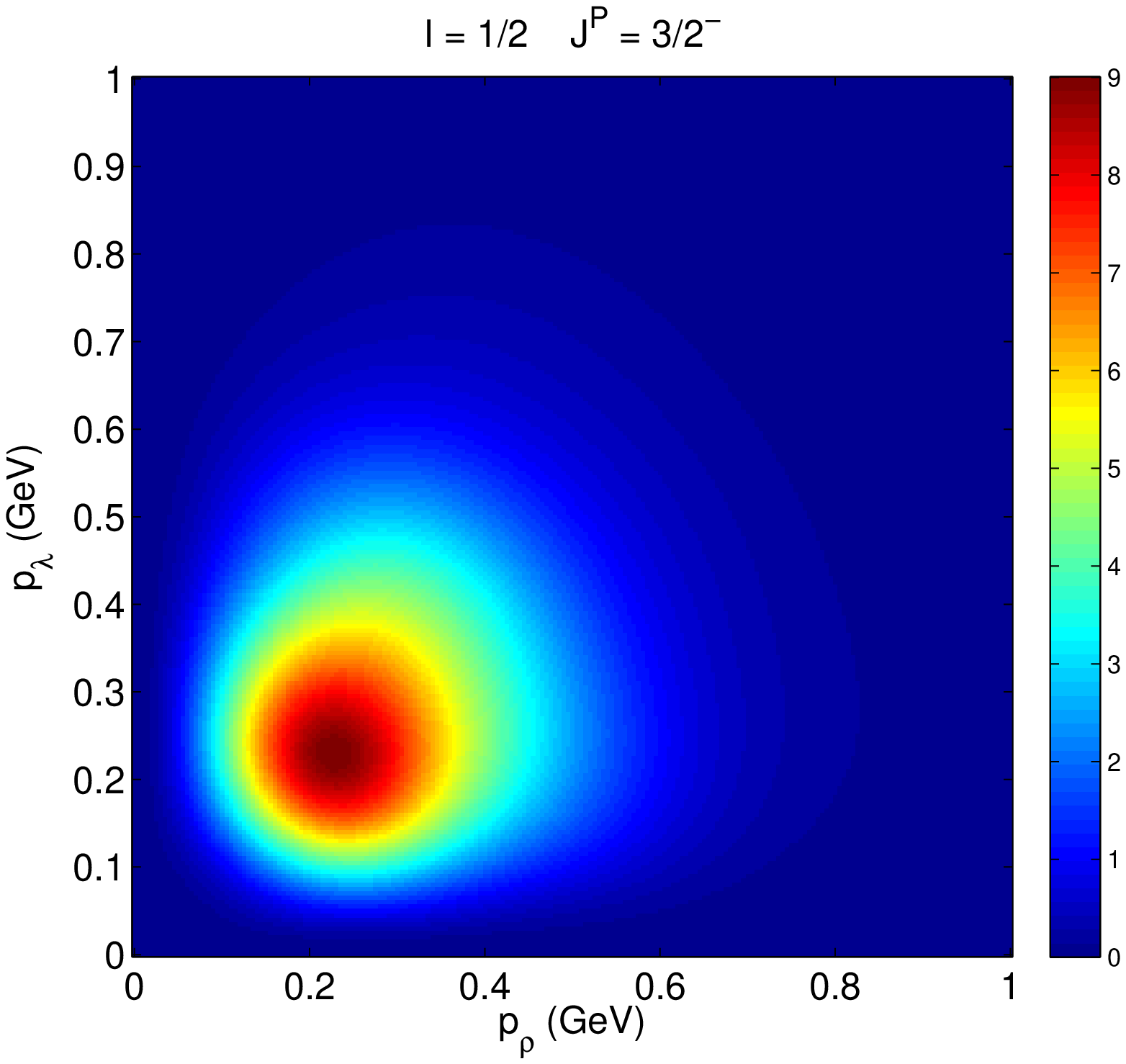}  \quad \includegraphics[width=0.45\columnwidth]{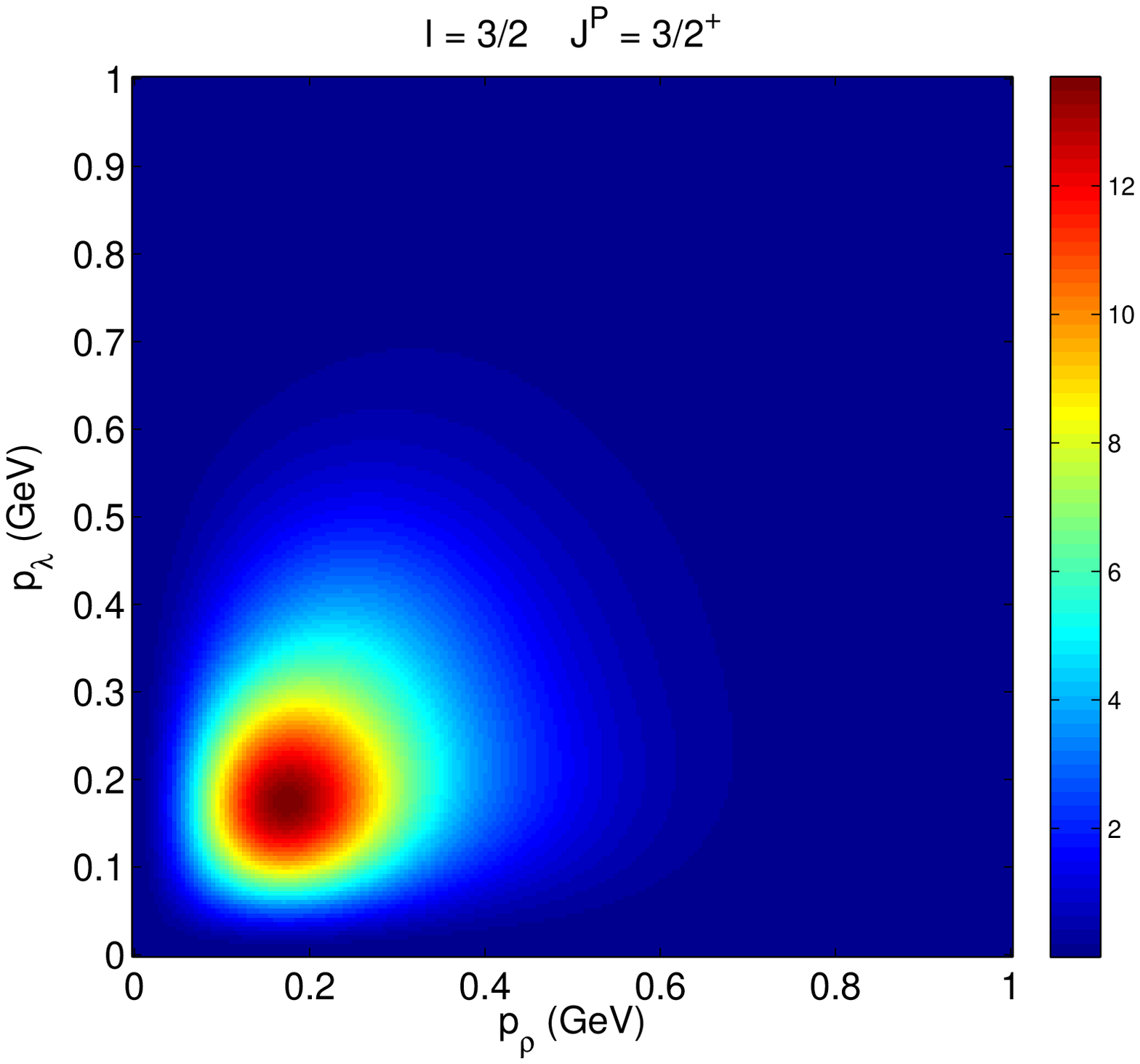}  \quad \includegraphics[width=0.45\columnwidth]{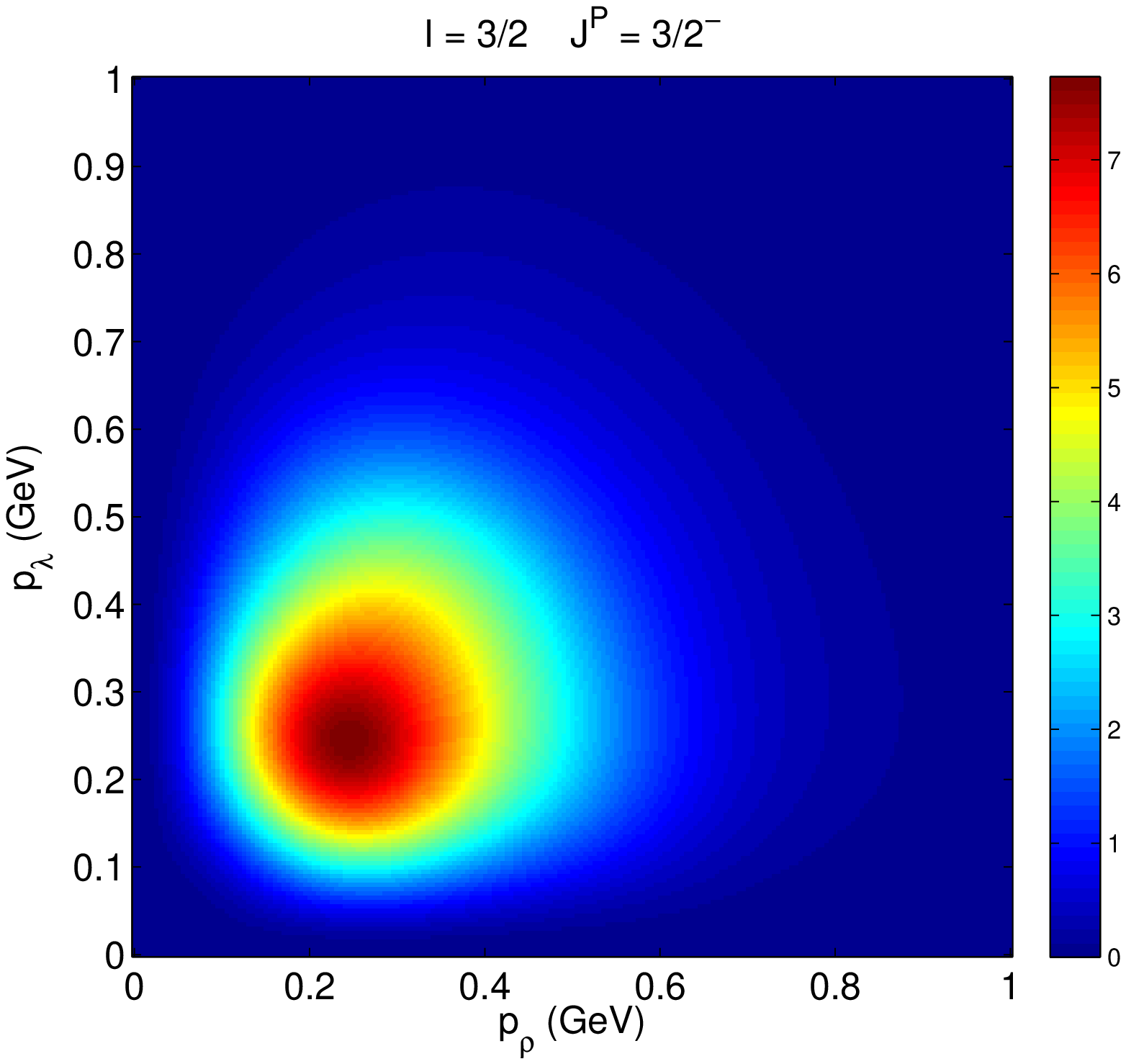}
\\
\includegraphics[width=0.45\columnwidth]{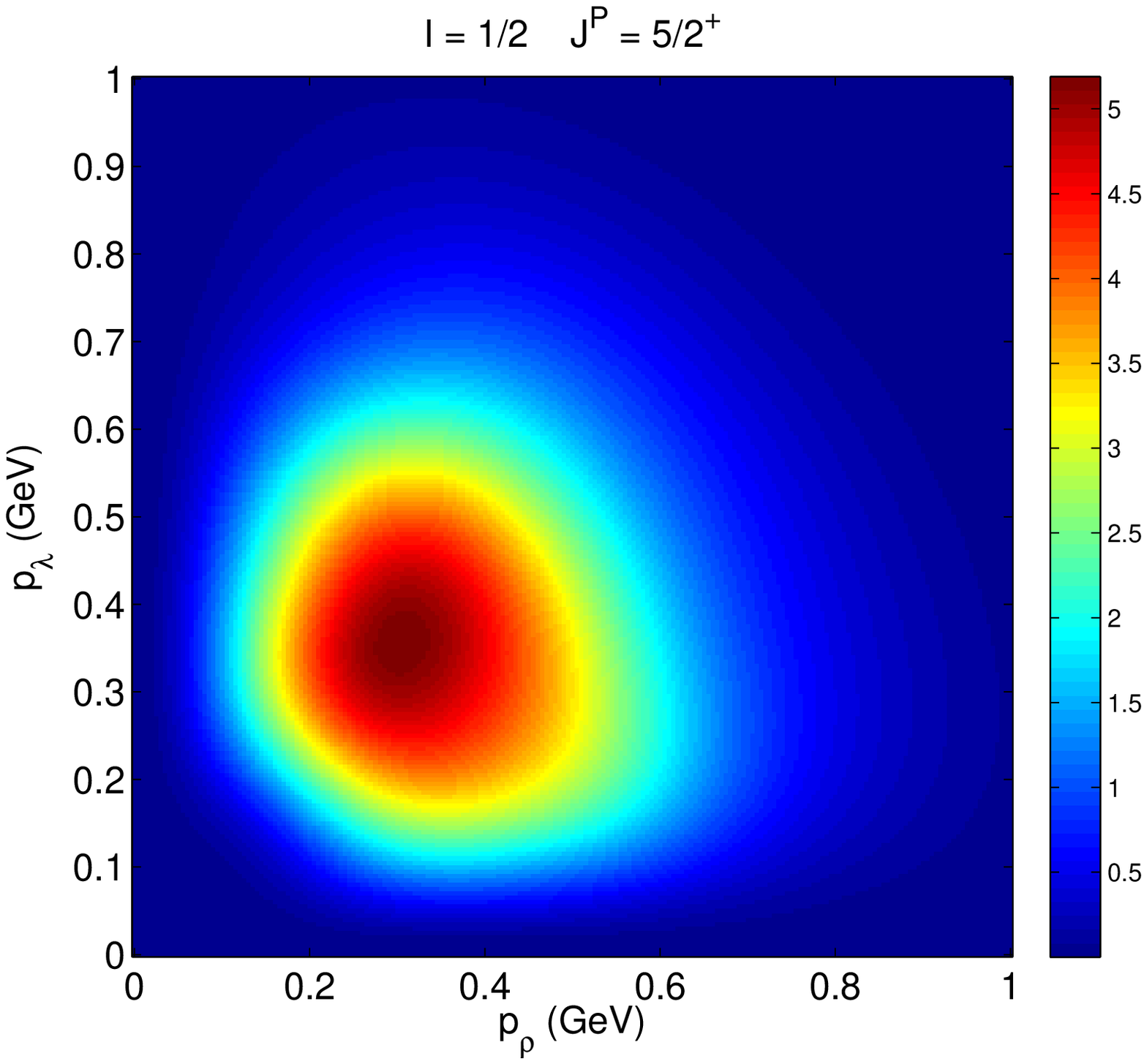}  \quad \includegraphics[width=0.45\columnwidth]{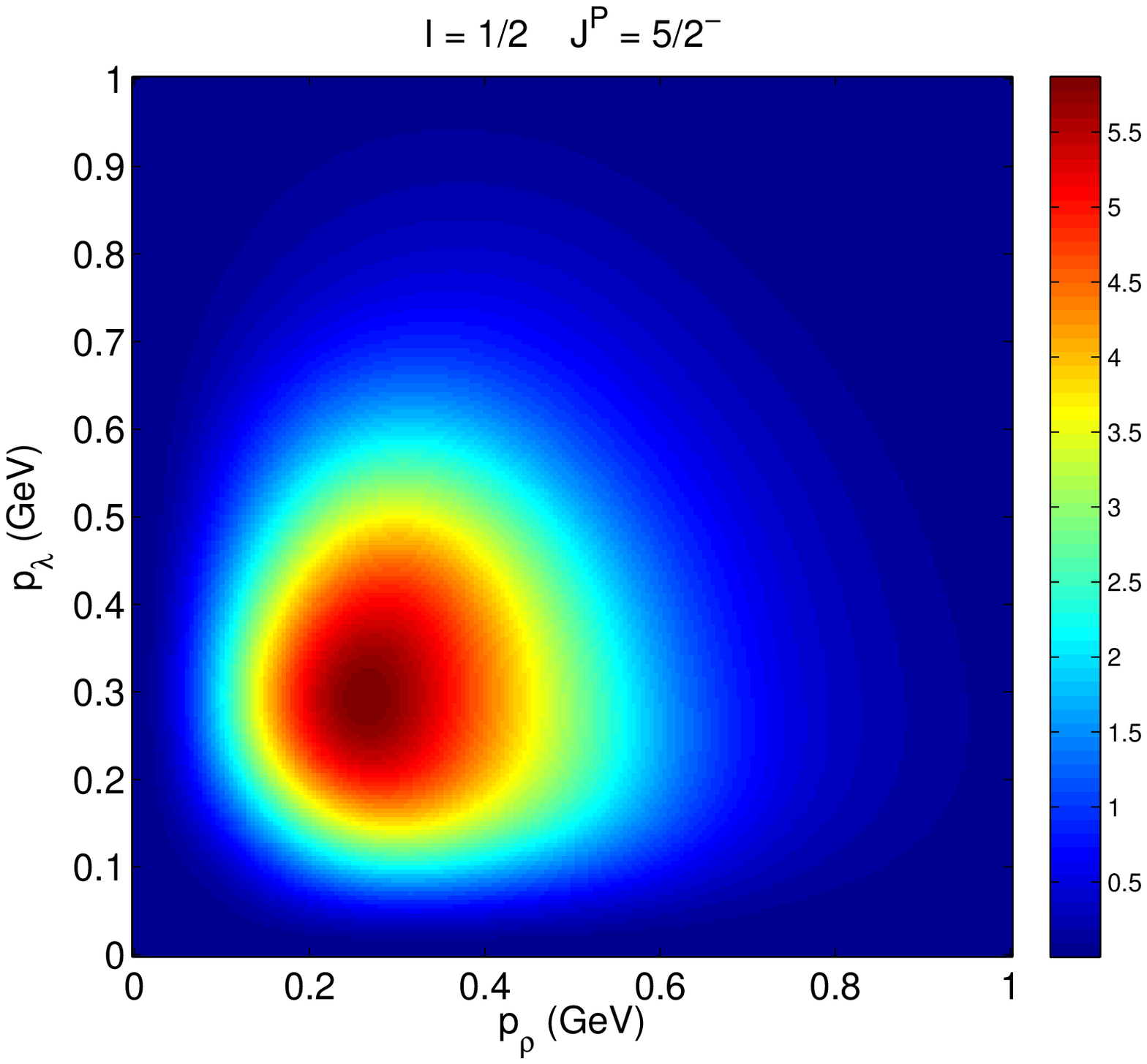}  \quad \includegraphics[width=0.45\columnwidth]{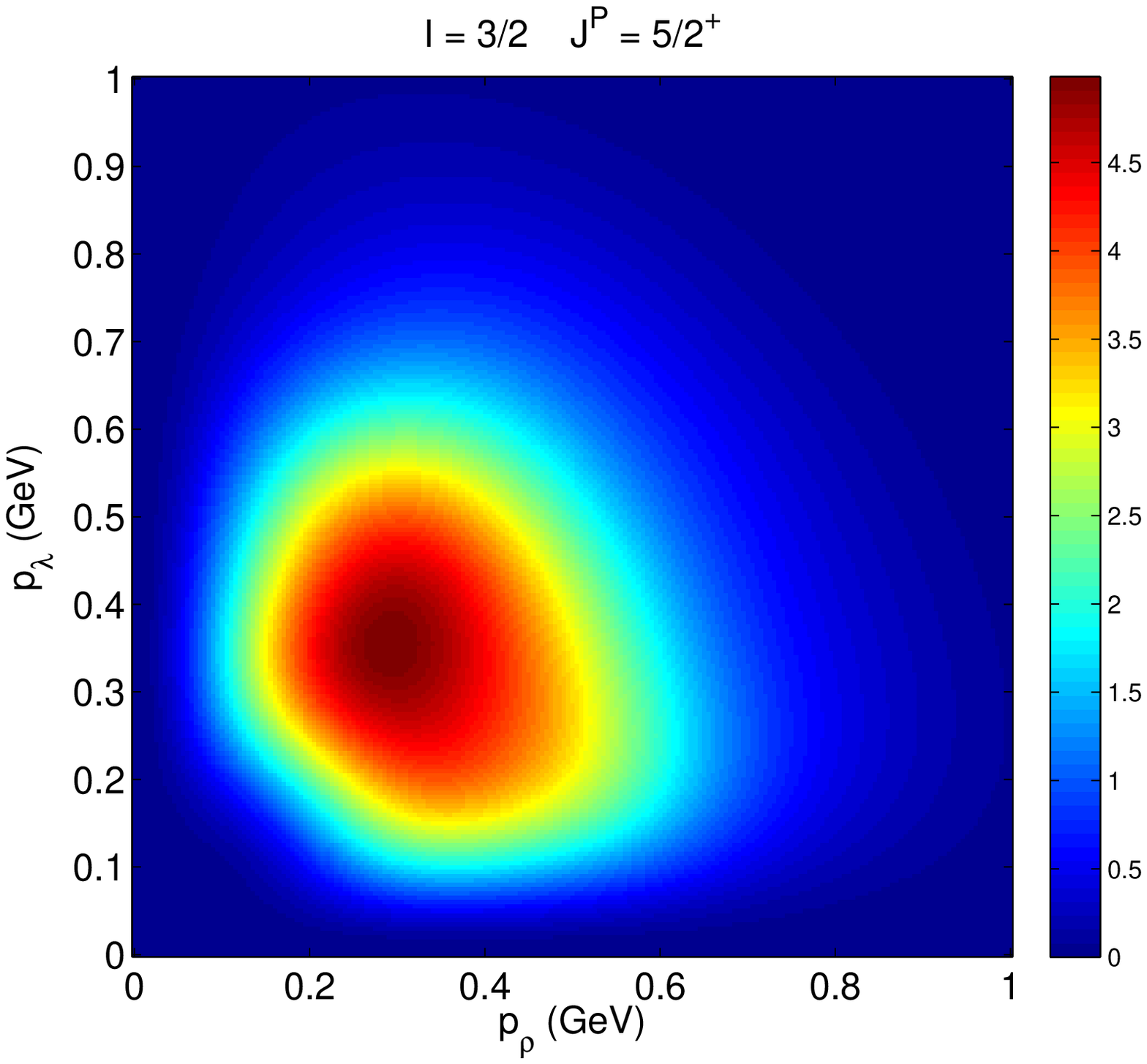}  \quad \includegraphics[width=0.45\columnwidth]{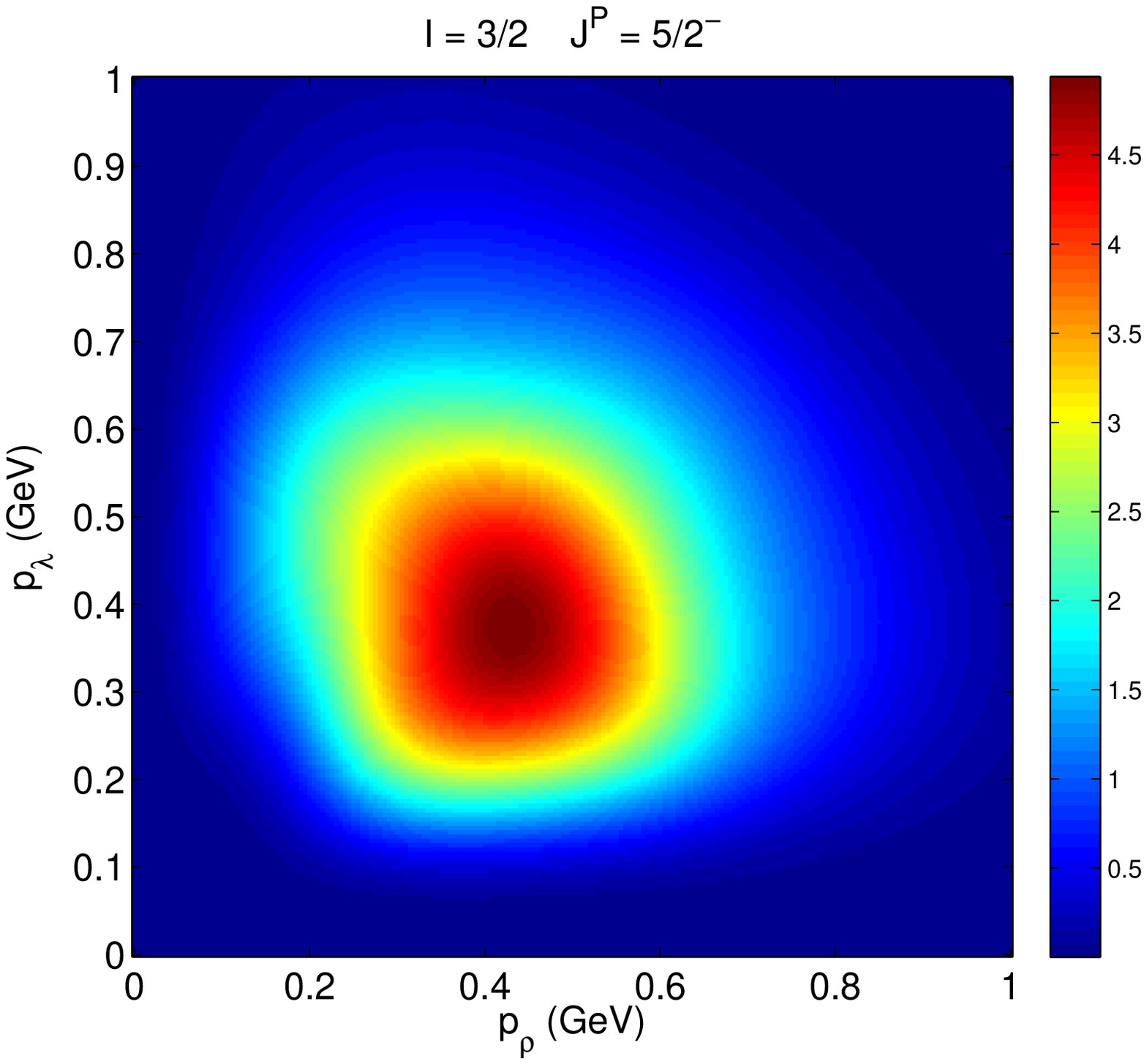}
\\
\includegraphics[width=0.45\columnwidth]{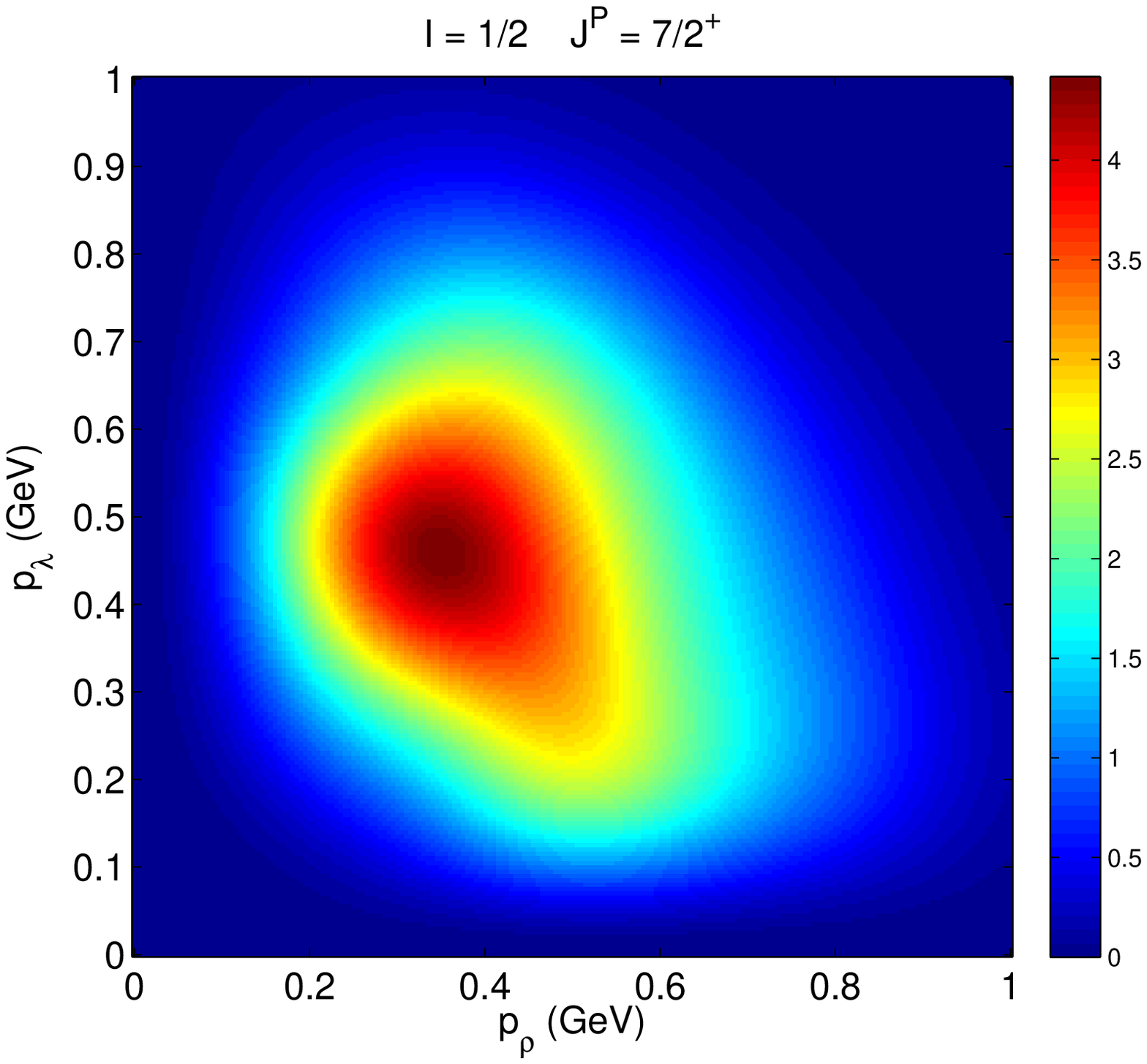}  \quad \includegraphics[width=0.45\columnwidth]{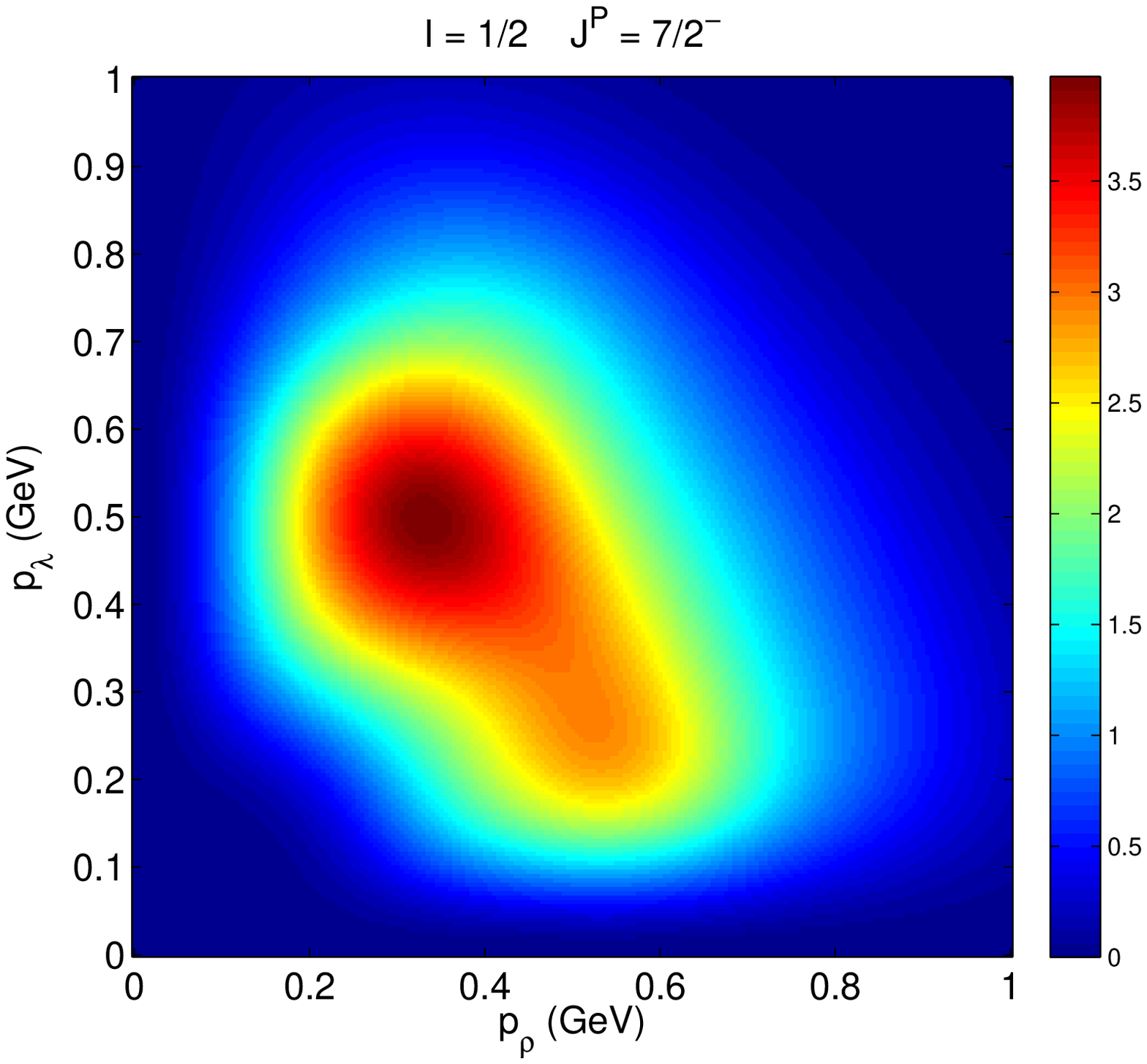}  \quad \includegraphics[width=0.45\columnwidth]{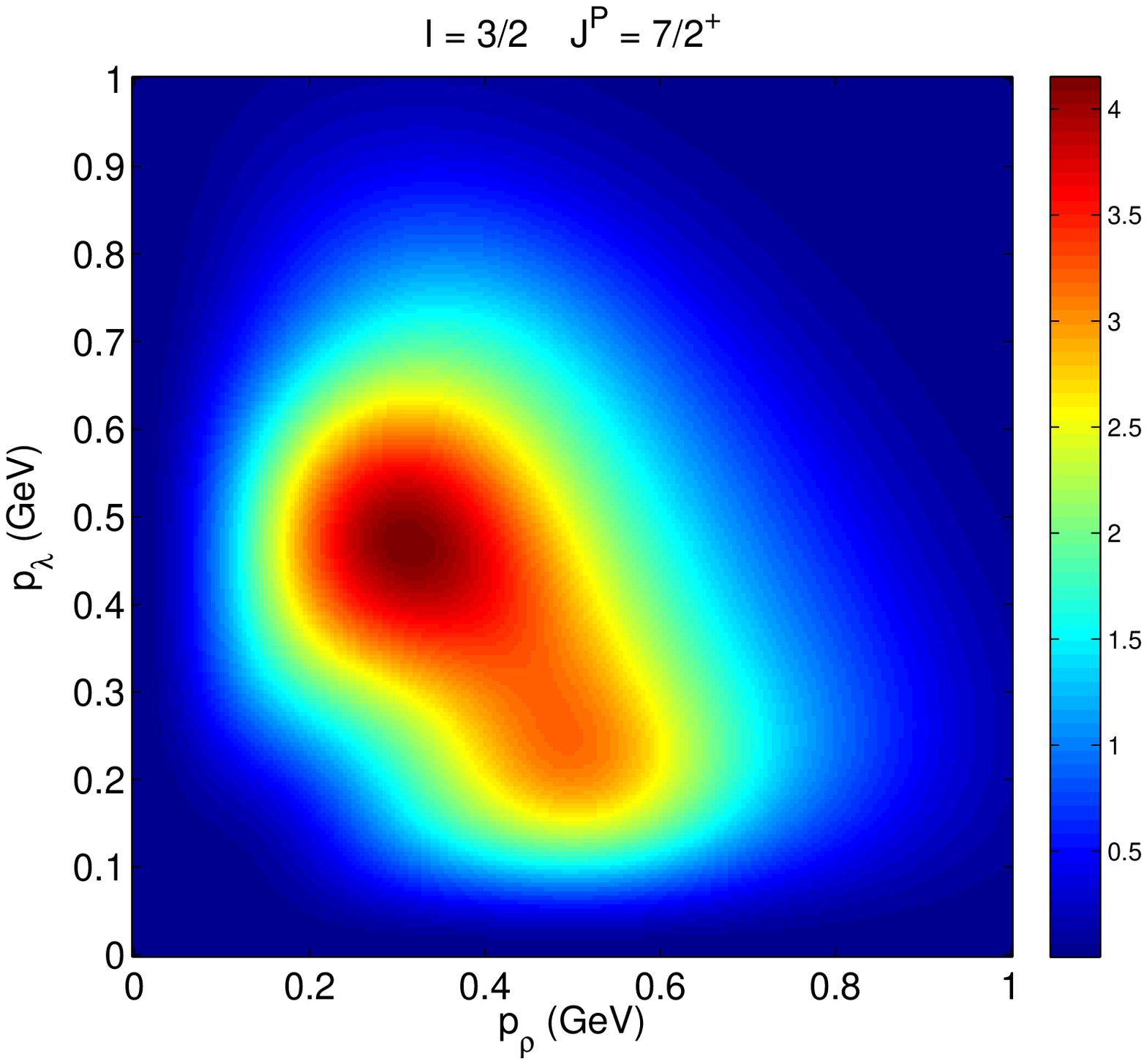}  \quad \includegraphics[width=0.45\columnwidth]{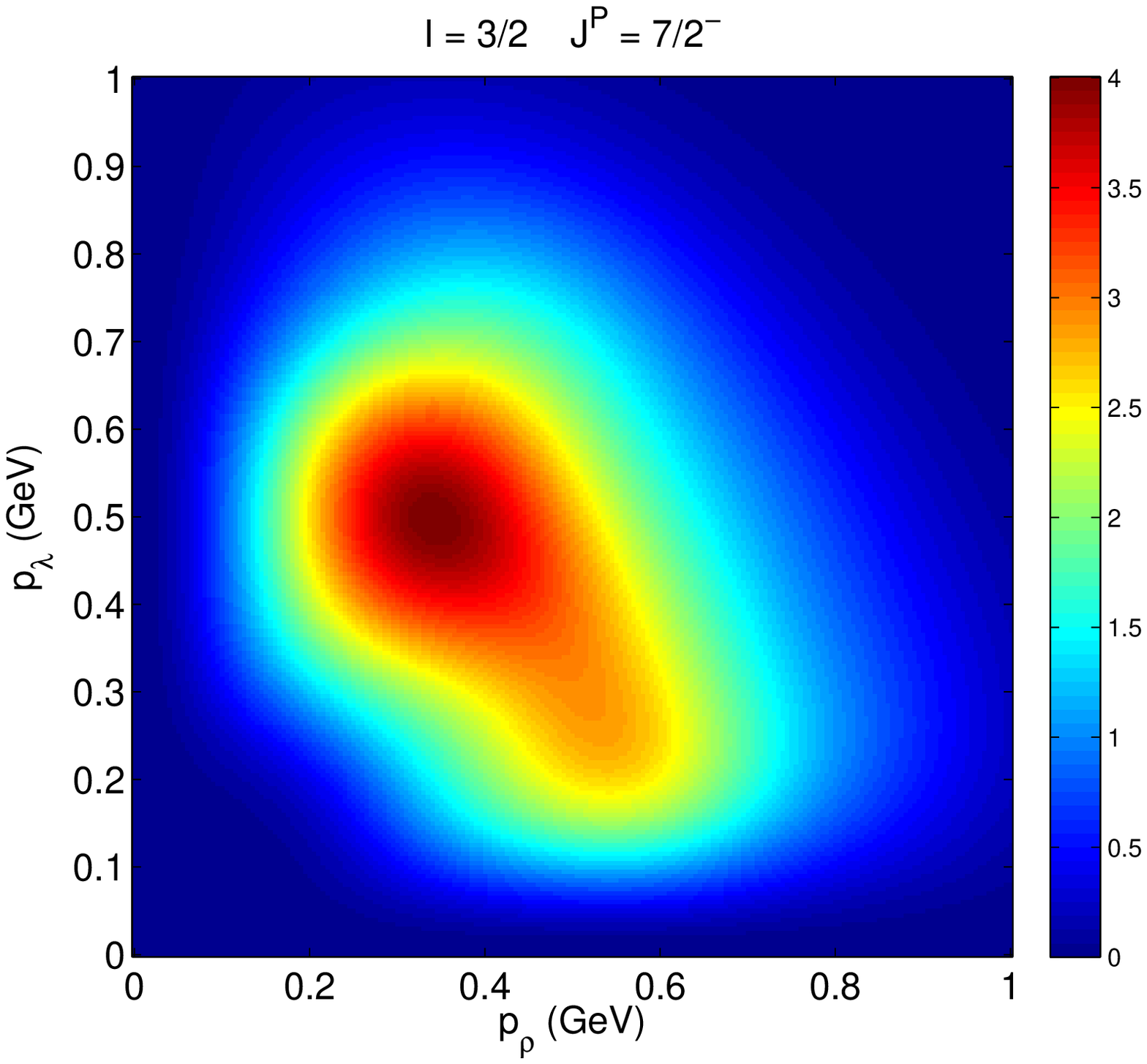}
\\
\includegraphics[width=0.45\columnwidth]{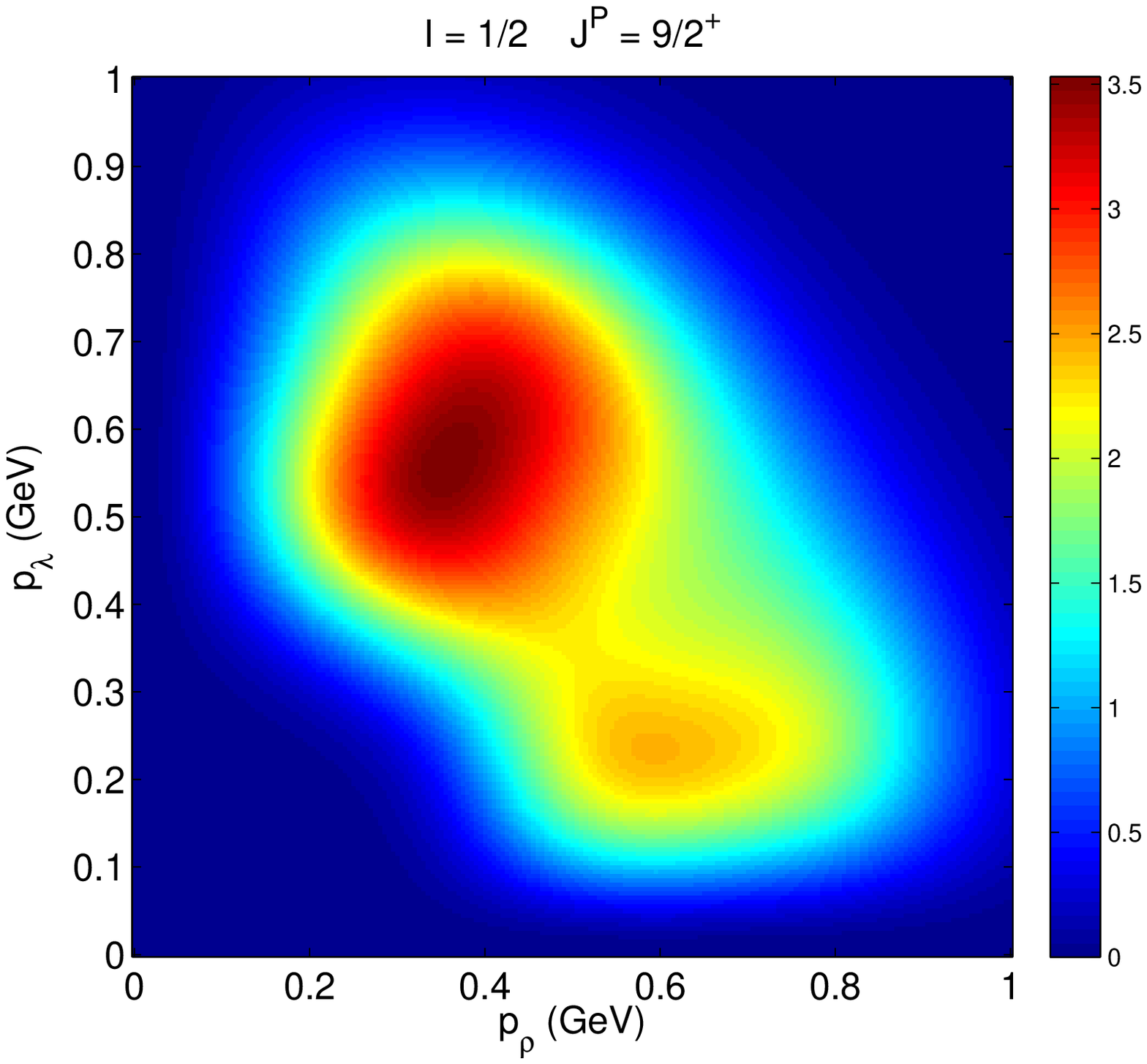}  \quad \includegraphics[width=0.45\columnwidth]{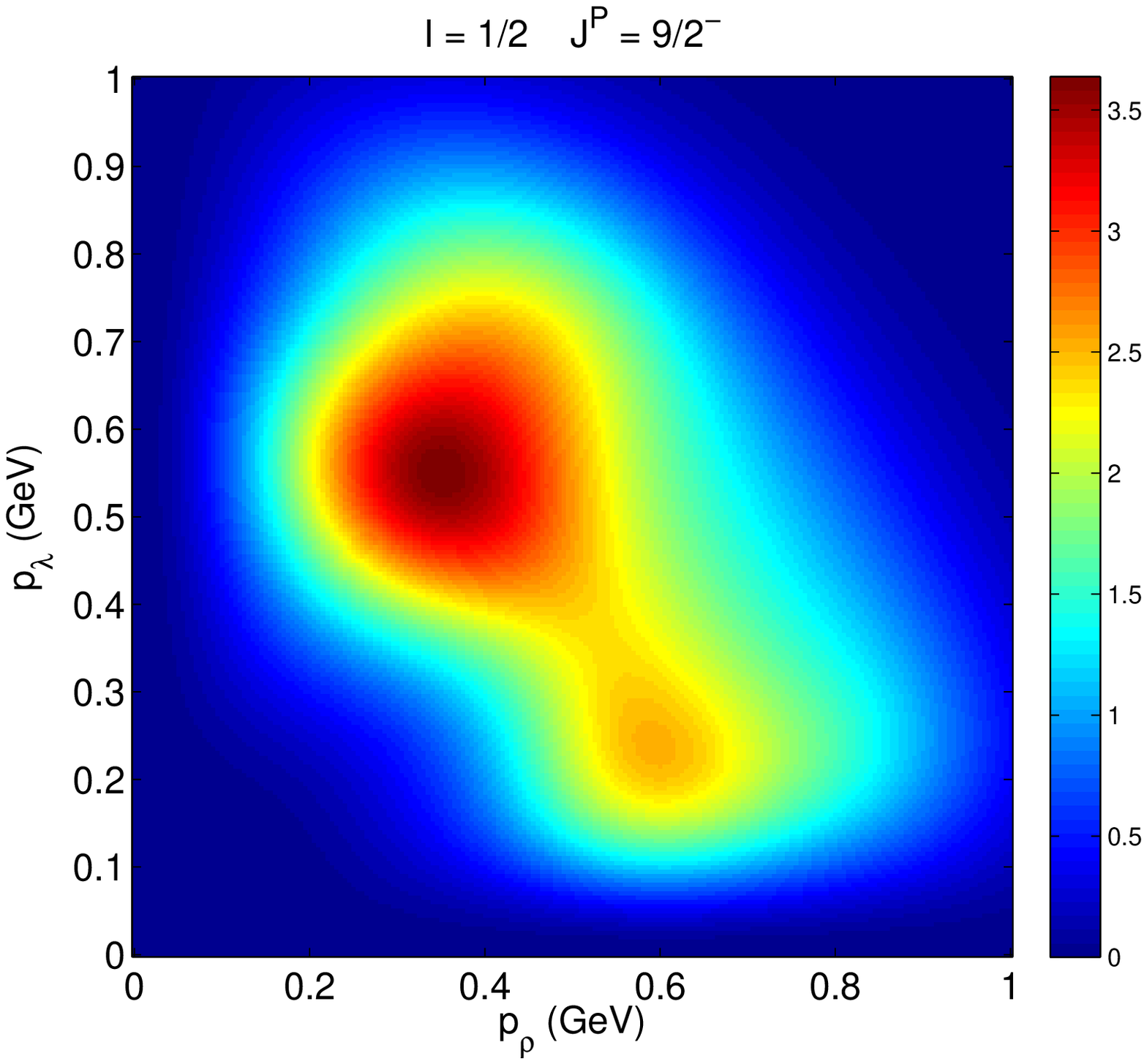}  \quad \includegraphics[width=0.45\columnwidth]{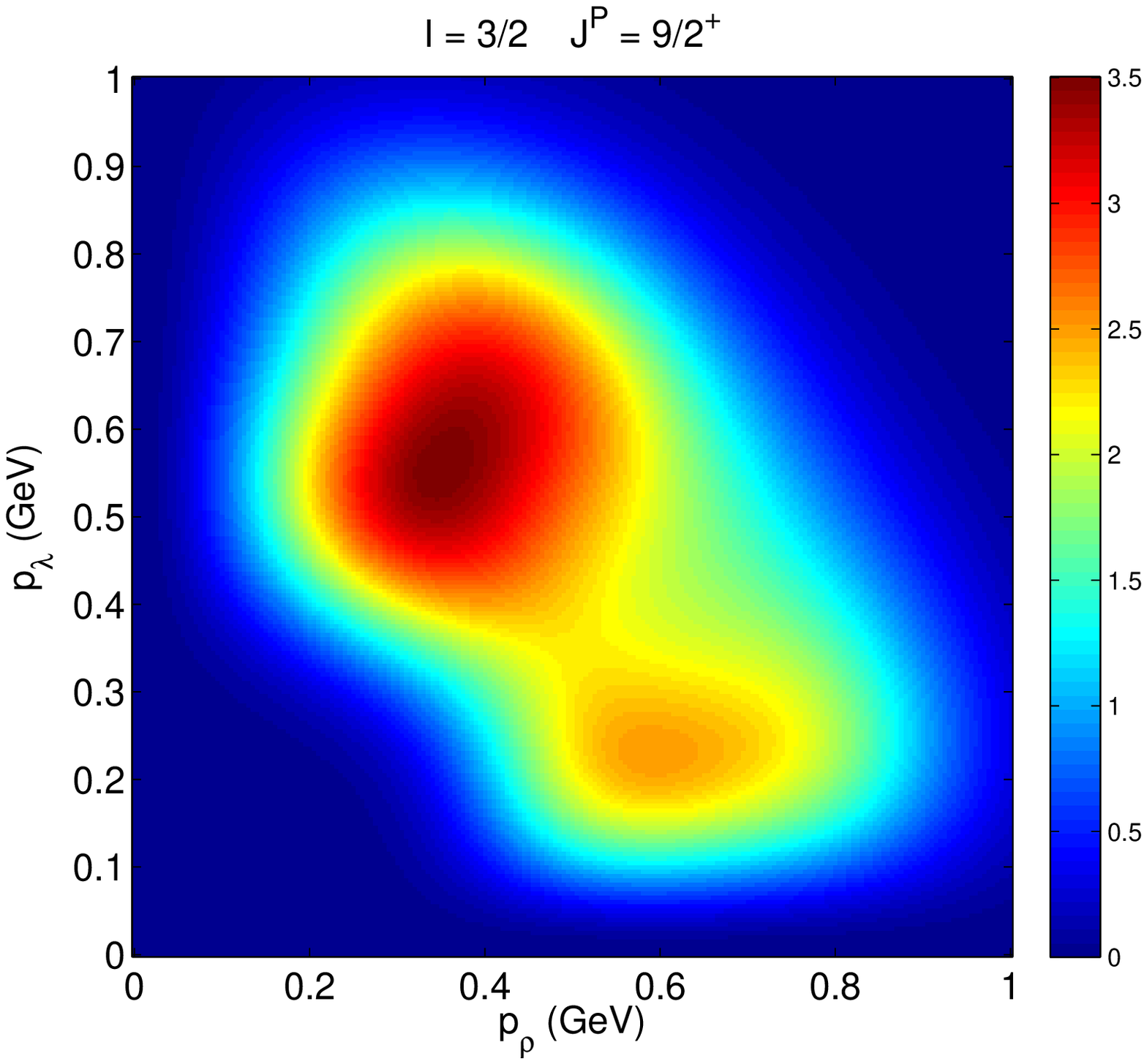}  \quad \includegraphics[width=0.45\columnwidth]{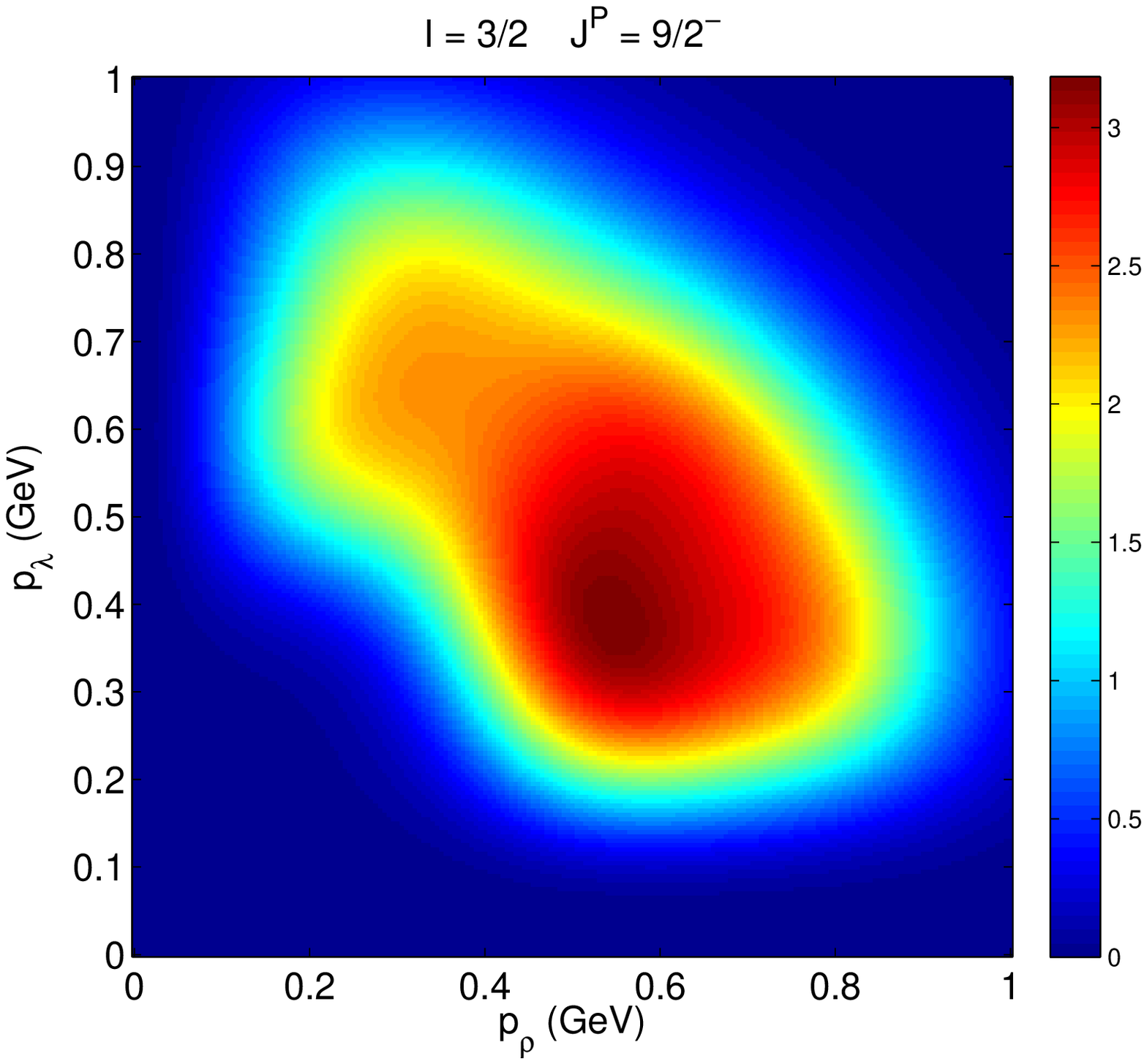}
\\
\includegraphics[width=0.45\columnwidth]{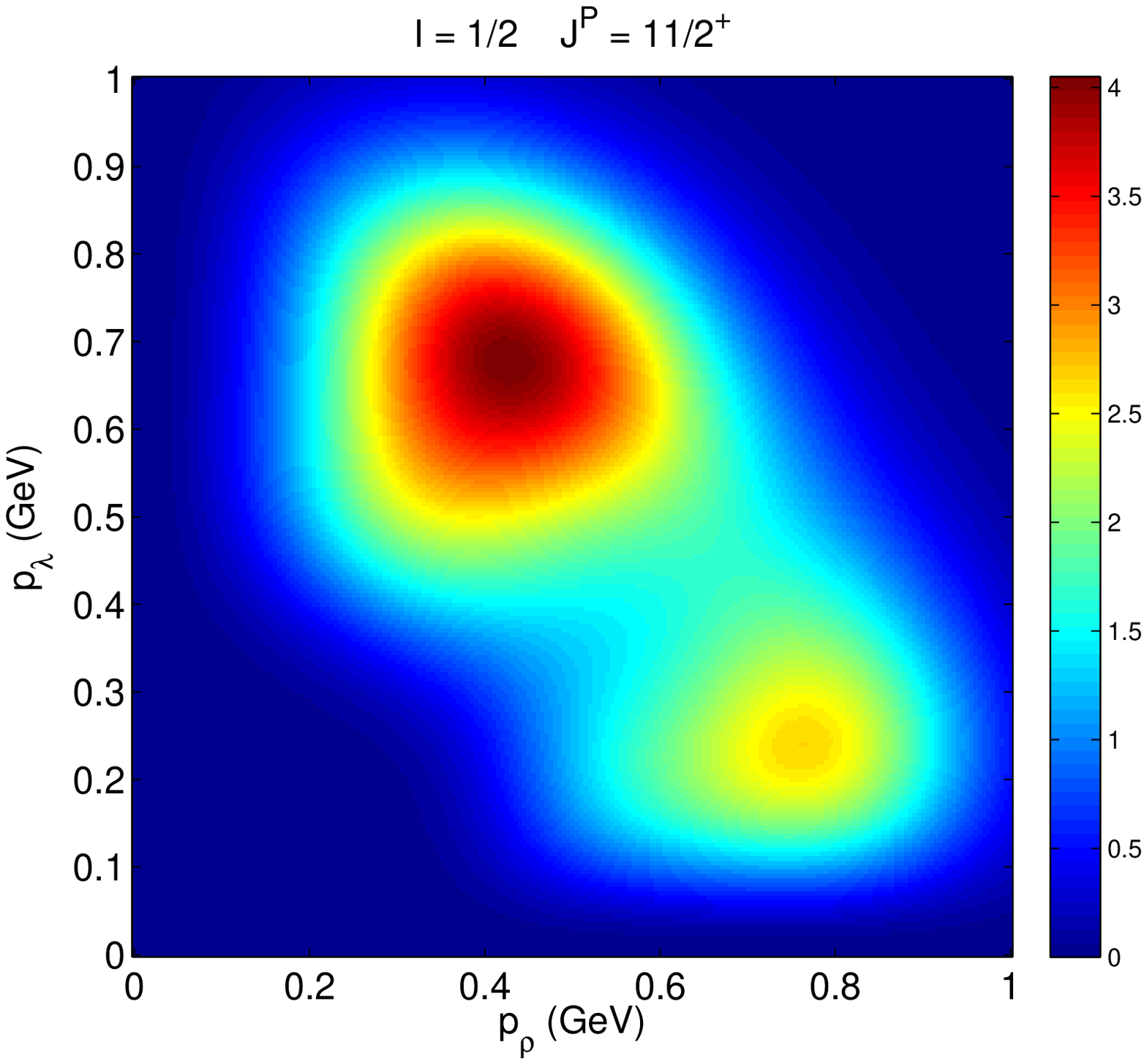}  \quad \includegraphics[width=0.45\columnwidth]{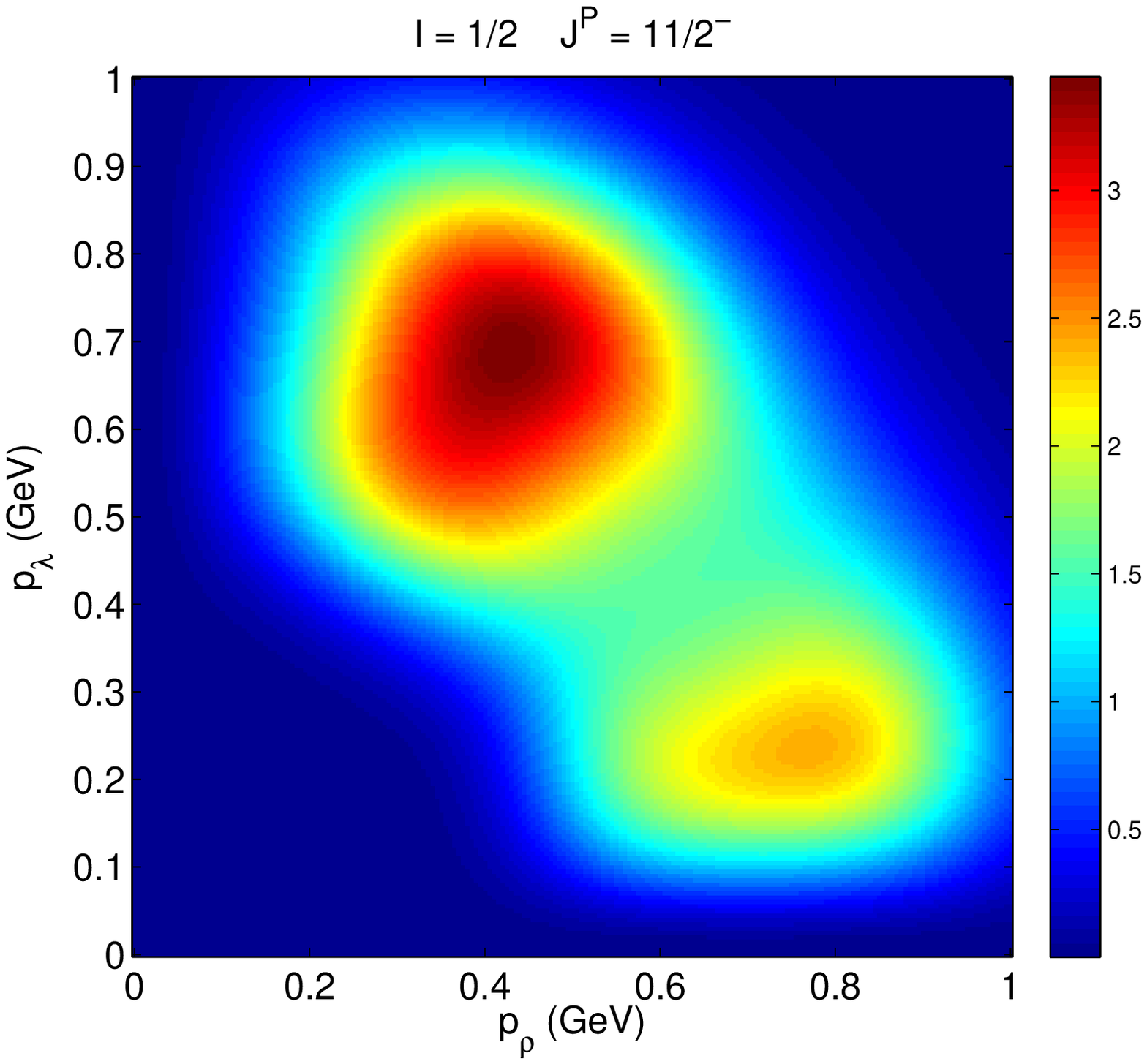}  \quad \includegraphics[width=0.45\columnwidth]{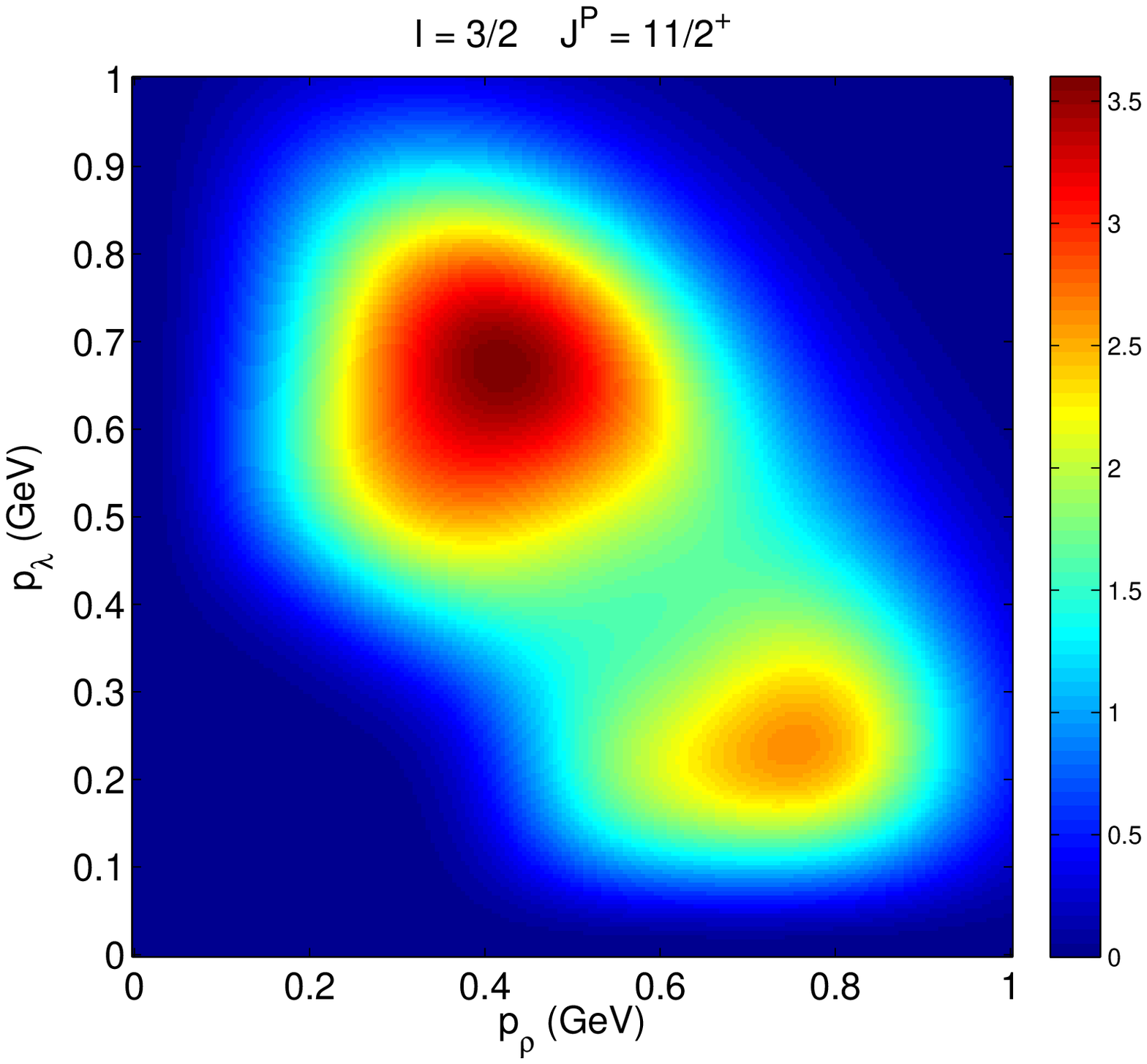}  \quad \includegraphics[width=0.45\columnwidth]{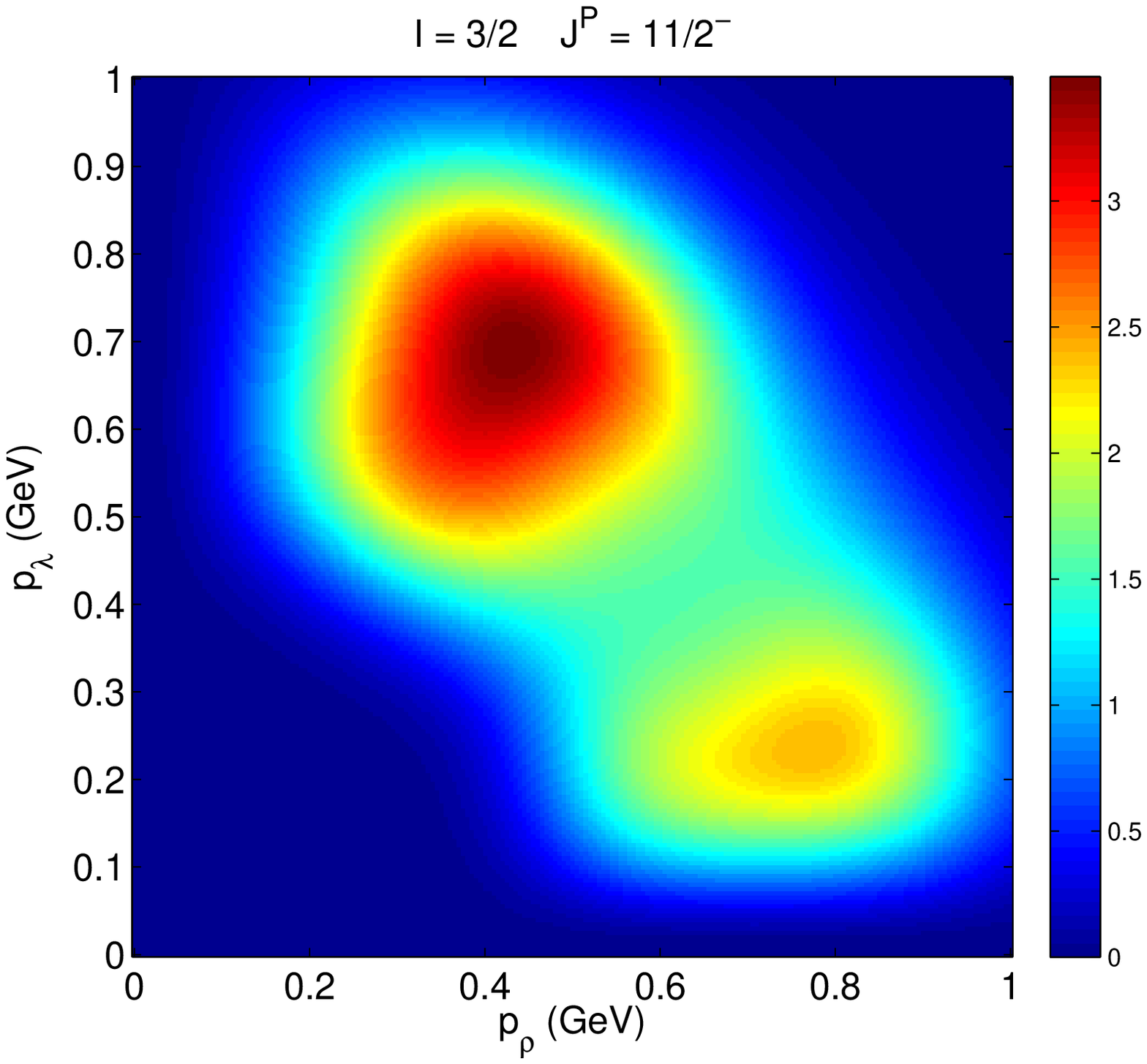}
\\
\includegraphics[width=0.45\columnwidth]{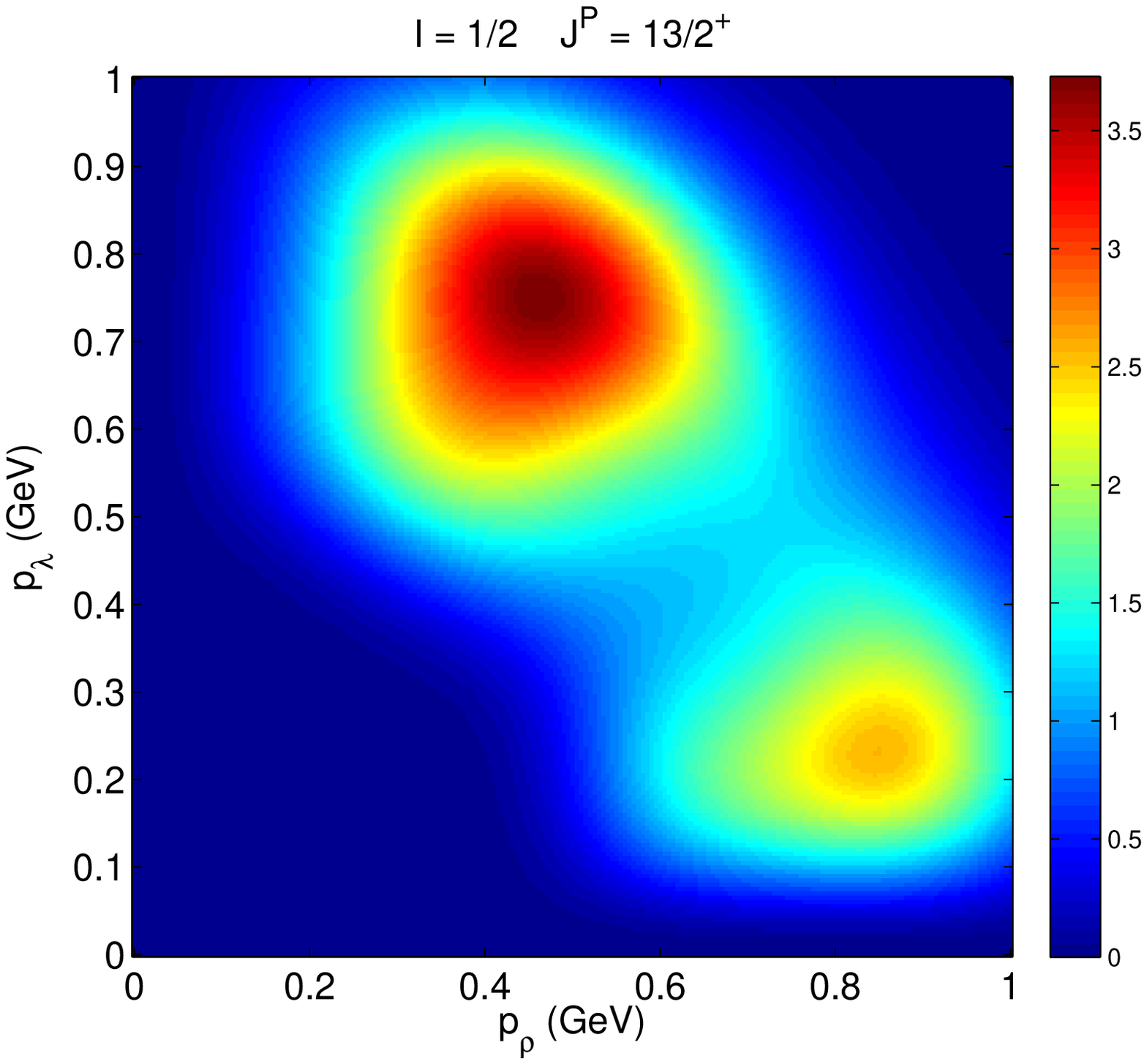}  \quad \includegraphics[width=0.45\columnwidth]{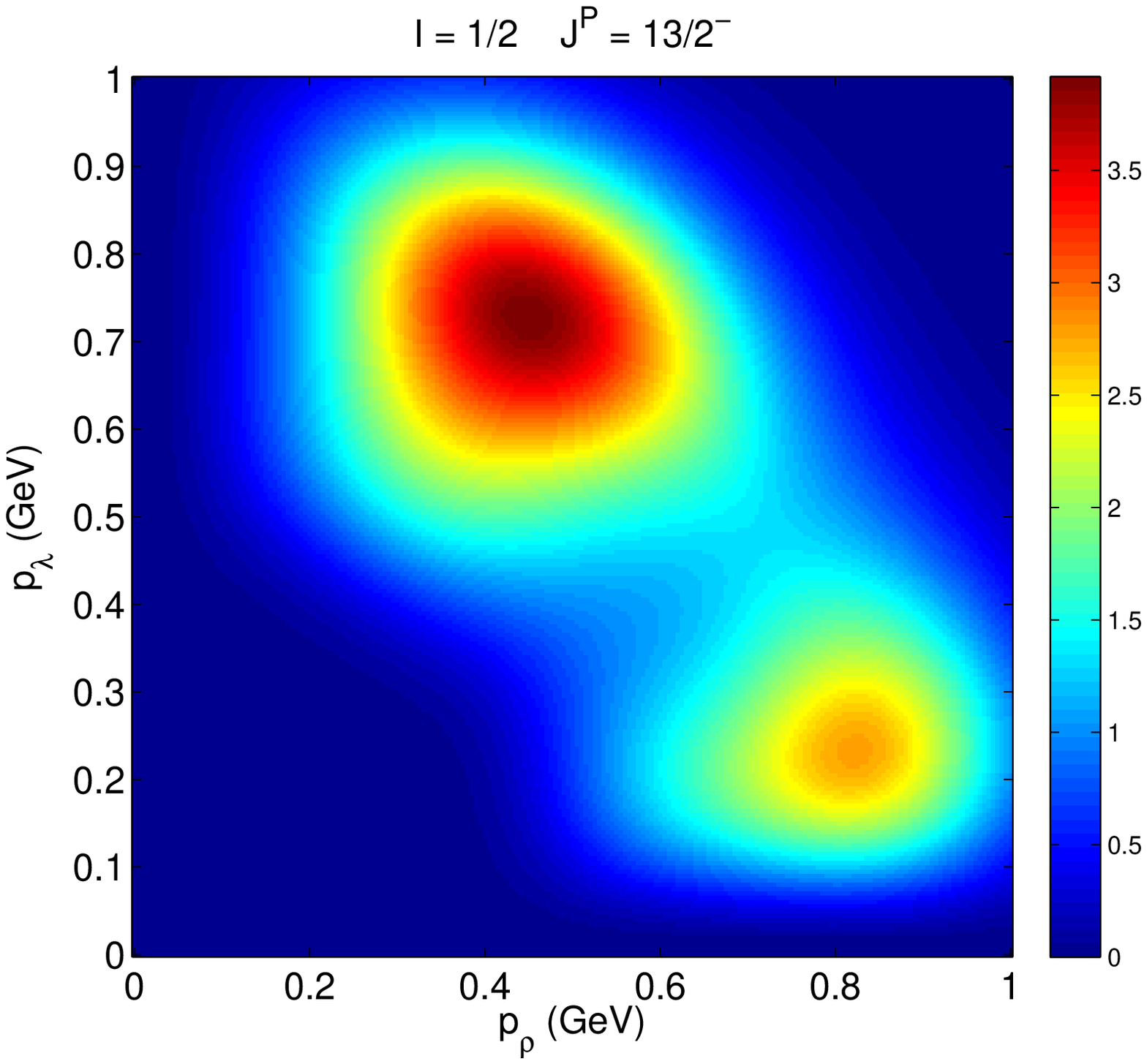}  \quad \includegraphics[width=0.45\columnwidth]{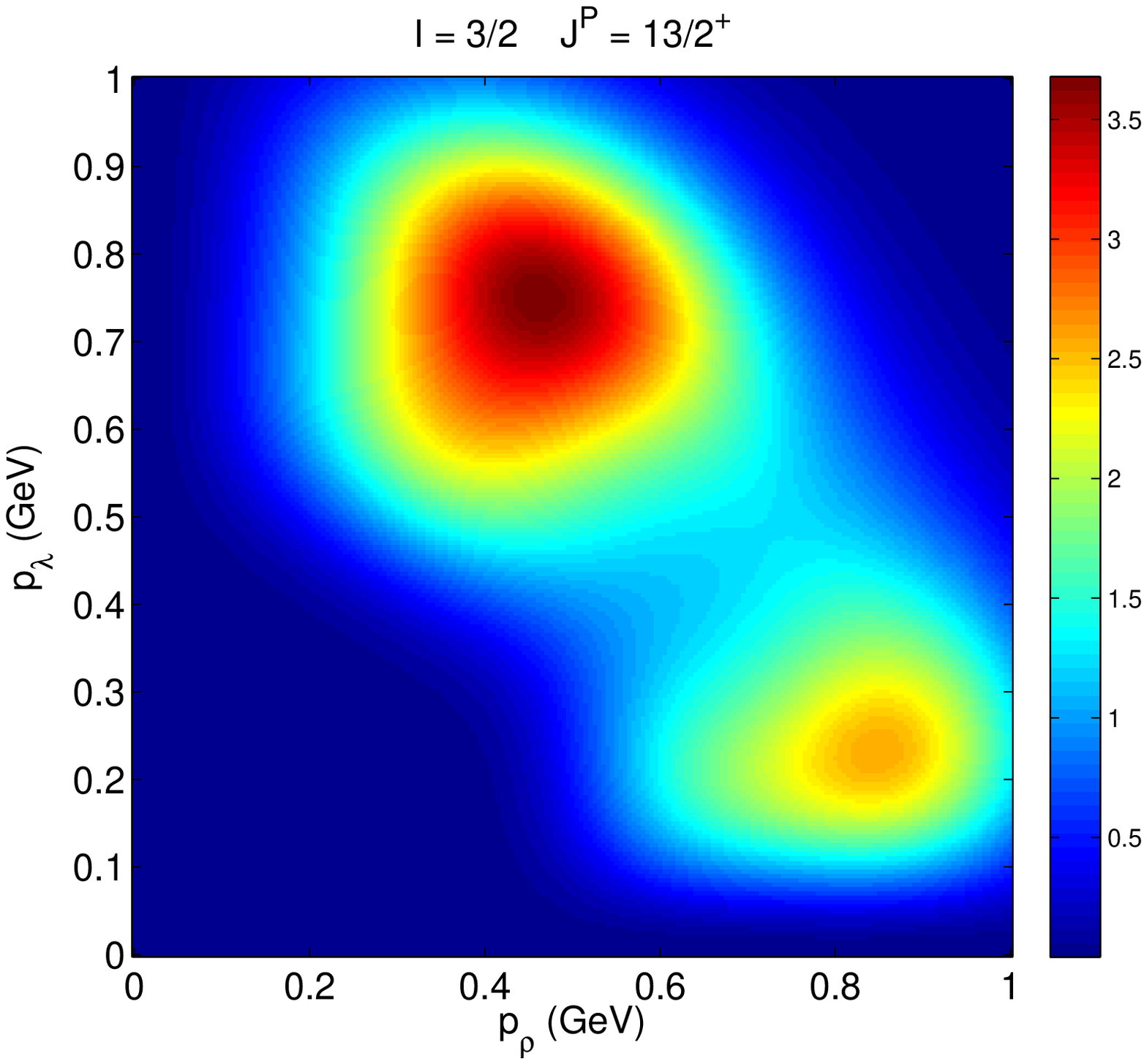}  \quad \includegraphics[width=0.45\columnwidth]{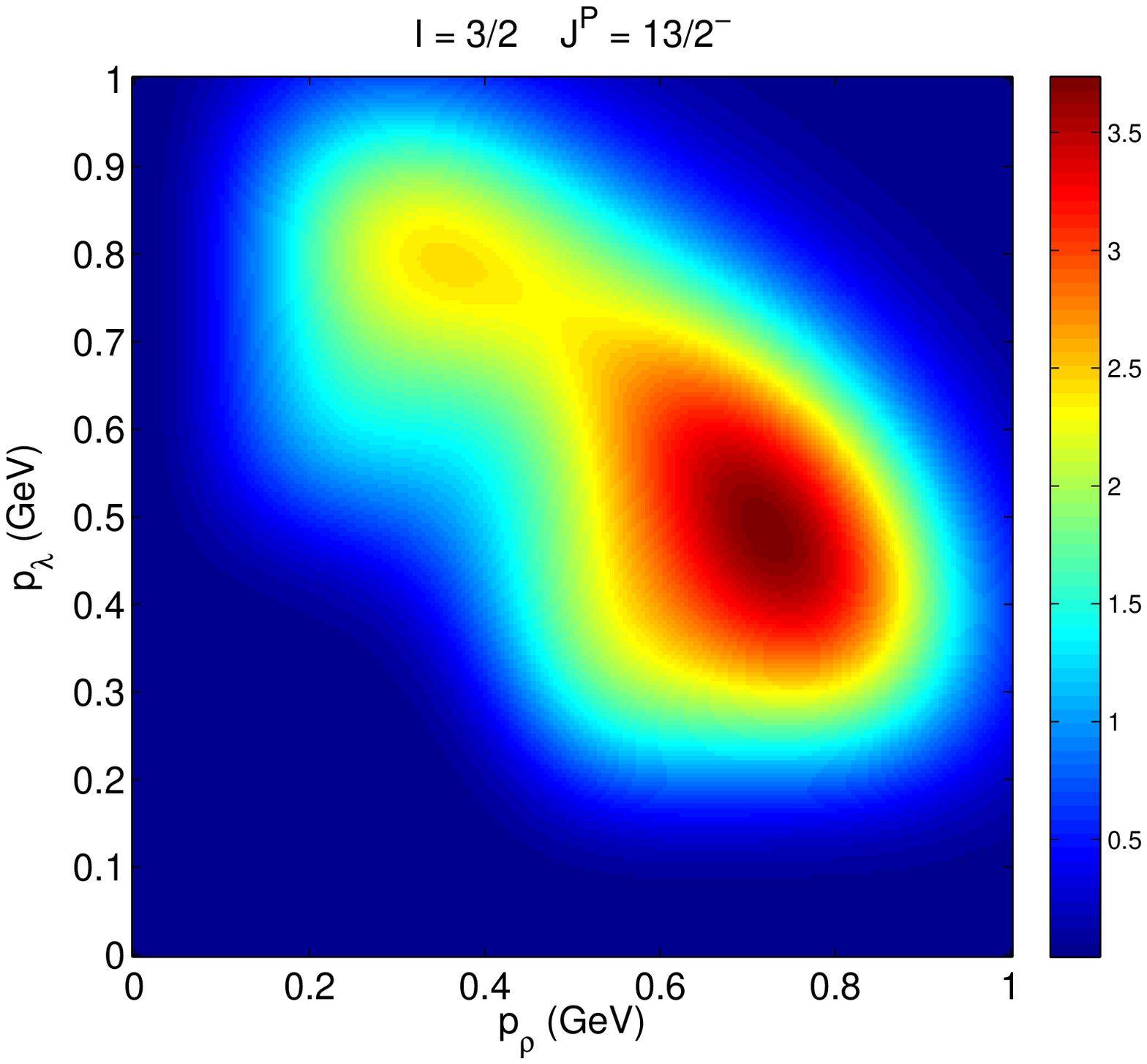}
\caption{\label{fig:allcontours} 
Momenta density distributions of the Yrast light baryons; from left to right: $N^+, N^-, \Delta^+$ and $\Delta^-$;  from top to bottom: $J = 3/2$ to $J=13/2$.  Starting from $J=7/2$, there is a clear general trend  as a function of $J$, with the increase of $p_\lambda$, except for the $\Delta_-$ at $J=9/2$ and $J=13/2$ which follow a different behaviour, with the increase of $p_\rho$. }
\end{figure*}

We can gain even one more insight from the further  $p_\rho-p_\lambda$ plane density plots in Fig. ~\ref{fig:allcontours}. For both spins and both parities we see that there are two areas of larger density, one with $|p_\rho|>|p_\lambda|$ and another with $|p_\lambda|>|p_\rho|$.
There is a difference though, distinguishing states with 
$J-3/2$ odd from states
with $J-3/2$ even. 
As representative of the second class of states we have spin $J=11/2$ (Fig. ~\ref{fig:allcontours}), and similarly spin $7/2$: the two regions in the $p_\rho-p_\lambda$ plane are asymmetrically populated, but this asymmetry seems to be equal for both parities. In both it is more likely to have $|p_\lambda|$ larger than $|p_\rho|$.
On the contrary, for spin $J=9/2$ (Fig. ~\ref{fig:allcontours}) and also for $13/2$
we find that the asymmetry in the population of the two regions is not equal for positive and negative parity: in the first case, $|p_\lambda|$ is larger, whereas in the second one, $|p_\rho|$ is larger.

\section{Discussion}\label{sec:conclusion}

\begin{table*}
\caption{
\label{tab:LScomp}
Angular momentum $(L,S)$ composition extracted from the variational computation, 
in percentage ({\it e.g.}, 47(2,3) means that 47\% of that state's wavefunction has $L=2$,
$S=3$). Fractions smaller than 4\% are omitted.}				
\begin{ruledtabular}
\begin{tabular}{|c|c|c|c|c|}
$J$  &  $\left(\frac{1}{2}\right)^+$  & $\left(\frac{1}{2}\right)^-$         &$\left(\frac{3}{2}\right)^+$& $\left(\frac{3}{2}\right)^-$     \\
\hline
 1/2 &100(0,1/2)	                     & 78(1,3/2) 22(1,1/2)            & 97(2,3/2)               & 97(1,1/2)                     \\
 3/2 &42(2,3/2) 38(2,1/2) 16(1,1/2) & 69(1,3/2) 31(1,1/2)            & 99(0,3/2)               & 97(1,1/2)                     \\
 5/2 &52(2,1/2) 47(2,3/2)	             & 98(1,3/2)                         & 71(2,3/2) 29(2,1/2)  & 83(3,3/2) 13(2,1/2)        \\
 7/2 &71(2,3/2) 16(4,1/2) 10(4,3/2) & 60(3,3/2) 38(3,1/2)            & 98(2,3/2)               & 93(3,1/2) 6(3,3/2)         \\
 9/2 &62(4,1/2) 36(4,3/2)	             & 97(3,3/2)                         & 70(4,3/2) 29(4,1/2)  & 72(5,3/2) 12(3,3/2) 10(4, 1/2) \\
 11/2&49(4,3/2) 30(6,1/2) 18(6,3/2) & 59(5,3/2) 39(5,1/2)            & 98(4,3/2)               & 96(5,1/2)                     \\
 13/2&64(6,1/2) 34(6,3/2)	             & 80(5,3/2) 9(7,3/2) 8(7,1/2) & 71(6,3/2) 29(6,1/2)	 & 93(5,3/2) 4(7,3/2)         \\ 
\end{tabular}
\end{ruledtabular}
\end{table*}

We develop numerical techniques to compute with high precision, for the first time in the framework of chiral invariant quark models, the excited light baryon spectra. We analyse in detail the possible parity doublets both in radial and angular states, both for the $\Delta$ $I=3/2$ and Nucleon $I=1/2$ isospins. While we were not expecting parity doublets in the radial excited spectrum, the setting of parity quadruplets was previously conjectured for high $J$. Notice, for static-light and light-light mesons parity doublets have already been clearly demonstrated in the framework used here. 

However we find an interesting and puzzling result: the spectrum of high $J$ Yrast baryons is not as simple as anticipated, because we have two independent Jacobi coordinates, in momentum space $p_\rho$ and $p_\lambda$. This is scrutinized in subsection~\ref{subsec:momdist}, clearly seen in Fig. ~\ref{fig:allcontours}, where we show the density plots for the momenta density distributions of the Yrast light baryons; from left to right: $N^+, N^-, \Delta^+$ and $\Delta^-$;  from top to bottom: $J = 3/2$ to $J=13/2$.  
To summarize our analysis of the puzzle, 
\begin{itemize}
\item
It appears that the wavefunctions are in general concentrated in the same region of the $(p_\rho,p_\lambda)$ plane (with a corresponding approximate degeneracy in the spectrum) except for the $\Delta_-$ at $J=9/2$ and $J=13/2$ which follow a different behaviour.
The general trend, as a function of $J$, is the increase of $p_\lambda$, except for the $\Delta_-$ at $J=9/2$ and $J=13/2$ where it is $p_\rho$ who increases.
\item
Moreover, most wavefunctions, except for  the $\Delta_-$ at $J=9/2$ and $J=13/2$, maintain a small but non-vanishing component at $p \sim 0$.
\end{itemize}

A further hint in this direction is presented in table~\ref{tab:LScomp}.
There we have extracted the percentage of various Russell-Saunders coupling $(L,S)$ 
wavefunction for each of the variationally computed states, to a precision of 4\%.
The effect is clearest for the $\Delta$ baryons that are flavor symmetric, so there is less
entanglement. Looking at $\Delta(7/2)$ and $\Delta(11/2)$ with both parities, we see that
the chiral partner wavefunctions are stepping in $L$ and $S$: the two percentage distributions are consistent with $(L,S)\to (L-1,S+1)$ as befits the spin-orbit coupling chiral charge. However, if one focuses instead on  $\Delta(9/2)$ and $\Delta(13/2)$, one sees that there is no simple relation among the wavefunctions of the would-be parity partners. 

Nevertheless, having an exchange in the role of the variables $(p_\rho,p_\lambda)$ or having a non vanishing density at $p \sim 0$ is in principle possible in baryons, even though the wavefunction in isospin $\times$ spin $\times$ space/momentum is symmetric. Clearly, this is possible only in baryons, it certainly does not occur when there is only one Jacobi coordinate as in mesons. Thus our unexpected result makes the nucleon and $\Delta$ spectra particularly special within the different hadron spectra.

As an outlook, to further understand the symmetries of the light excited baryons, an even more intensive computational effort will be necessary. It will be important to study even higher $J$ to clarify the trends in the spectrum. It will also be interesting to develop other chiral invariant quark models, possibly using Coulomb gauge with a lesser degree of truncation or using a Cornell-like potential \cite{LlanesEstrada:2001kr,TorresRincon:2010fu,LlanesEstrada:2010bs,Eichten:1978tg}, to check if the puzzle is general or if it is an artefact of the utilized model. Since this will require new C++ codes and even more efficient supercomputers, 
we leave further investigation of this issue to future work.

With our results at hand, the baryon spectrum is less promising than the meson spectrum for studying chiral symmetry in Wigner mode and for extracting the quark mass running in the transition between Goldstone and Wigner modes: theory is needed to guide experiment in the selection of quantum numbers. We propose to look at the sequence of even $J-3/2$ of baryon spins $3/2$, $7/2$, $11/2$ and give up on $5/2$, $9/2$, $13/2$, nevertheless model dependence remains to be addressed and additional work would appear necessary.

A further prediction of the possible Wigner-Weyl realization of chiral symmetry in the high spectrum is that, because the second term in Eq.~(\ref{EffectQ5}) would be small, the pion would decouple from the baryon, so that processes such as $N^*\to N\pi$ become
rarer. However, in view of our mass spectra, in-depth study of this phenomenon is also better deferred to future work.

\acknowledgments

PB and MC acknowledge the use of the CPU and GPU servers of PtQCD, supported by CFTP, FCT and
NVIDIA and grant FCT UID/FIS/00777/2013.  
MC is supported by FCT under the contract SFRH/BPD/73140/2010.
The  work  of  FJLE  relied  on  the  Spanish  Excellence
Network on Hadronic Physics FIS2014-57026-REDT, and
grants  UCM:910309,  MINECO:FPA2014-53375-C2-1-P.
He  also  thanks  the  Department  of  Energy's  Institute
for Nuclear Theory at the University of Washington for
its  partial  support  and  hospitality.

\appendix
\section{Analytical computation of exchange-operators matrix elements}\label{sec:trocas}
We have relegated to this appendix the analytical evaluation of the
$P^{23}$ operator in Eq.~(\ref{eq:perm_matrices}) within states of the unsymmetrized basis.
These are given by
\begin{widetext} 
\be
|\phi_k \rangle = | I I_{12} J M L S S_{12} n_\rho l_\rho n_\lambda n_\lambda \rangle =
| I I_{12} \rangle \langle J M | L M_l S M_s \rangle | S M_s S_{12} \rangle
\langle L M_l | l_\rho m_\rho l_\lambda m_\lambda \rangle | n_\rho l_\rho m_\rho \rangle
| n_\lambda l_\lambda m_\lambda \rangle  
\ee
and so, the matrix elements are
\bea \label{P23expanded}
\langle \phi_k | P^{23} | \phi_{k'} \rangle &=& \langle I I_{12} | P^{23} | I I_{12}' \rangle \,
\langle S S_{12} | P^{23} | S S_{12}' \rangle \,
\langle l_\rho m_\rho l_\lambda m_\lambda | L M_l \rangle \, \langle L M_l | l_\rho' m_\rho' l_\lambda' m_\lambda' \rangle \\ \nonumber && \langle n_\rho l_\rho m_\rho n_\lambda l_\lambda m_\lambda |
P^{23} | n_\rho' l_\rho' m_\rho' n_\lambda' l_\lambda' m_\lambda' \rangle
\eea
\end{widetext}

\subsection{Momentum-modulus dependent part}

The effect of exchange operators on the momentum, spin and flavor parts of
the unsymmetrized basis wavefunctions can be separately calculated because of the factorization in Eq.~(\ref{P23expanded}).
The momentum-modulus dependent braket, the last term in that Eq.~(\ref{P23expanded}), is a known exchange matrix because of the Gaussian times polynomial structure of the HO basis,
\be
\langle n_\rho l_\rho m_\rho n_\lambda l_\lambda m_\lambda | P^{23}
| n_\rho' l_\rho' m_\rho' n_\lambda' l_\lambda' m_\lambda' \rangle = 
\mD^{[N][l][m_\rho][m_\lambda]}_{[l'][m'_\rho][m'_\lambda][\sigma]} \; .
\ee
Indeed, the $\mD$ numbers are the MBRB coefficients for the specific case of transforming a product of two harmonic oscillator wavefunctions. The coefficients are needed when expressing the product of harmonic oscillator wavefunctions within a shell $N$, with quantum numbers
$\left\{\begin{array}{ccc} n^{N,l}_\rho l^{N,l}_\rho m_\rho
\\ n^{N,l}_\lambda l^{N,l}_\lambda m_\lambda \end{array} \right\}$,
and with permuted variables, as a linear combination of products of
harmonic oscillator wavefunctions with quantum numbers
$\left\{\begin{array}{ccc} n^{N,l'}_\rho l^{N,l'}_\rho m'_\rho
\\ n^{N,l'}_\lambda l^{N,l'}_\lambda m'_\lambda \end{array} \right\}$ and
with the original variables. An explicit expression for the $\mD$
coefficients is given by (we drop the index N for the orbital quantum numbers)
\begin{widetext}
\begin{multline} 
\mD^{[N][l][m_\rho][m_\lambda]}_{[l'][m'_\rho][m'_\lambda][\sigma]} =
\delta(2n^{l}_\rho + l^{l}_\rho + 2n^{l}_\lambda + l^{l}_\lambda,
2n^{l'}_\rho + l^{l'}_\rho + 2n^{l'}_\lambda + l^{l'}_\lambda) \,
\frac{\pi}{4} \,
(-1)^{n^{l'}_\rho+n^{l'}_\lambda+n^{l}_\rho+n^{l}_\lambda} \\ \times
\sqrt{\frac{n^{l'}_\rho! \, n^{l'}_\lambda! \, n^{l}_\rho! \,
    n^{l}_\lambda!  \, \Gamma(n^{l'}_\rho + l^{l'}_\rho + 3/2) \,
    \Gamma(n^{l'}_\lambda + l^{l'}_\lambda + 3/2) \, \Gamma(n^{l}_\rho
    + l^{l}_\rho + 3/2) \, \Gamma(n^{l}_\lambda + l^{l}_\lambda +
    3/2)} {(2l^{l'}_\rho + 1) \, (2l^{l'}_\lambda + 1) \, (2l^{l}_\rho
    + 1) \, (2l^{l}_\lambda + 1)} } \\ \times \sum_{n_{\rho\rho}}
\sum_{n_{\rho\lambda}} \sum_{n_{\lambda\rho}}
\sum_{n_{\lambda\lambda}} \sum_{l_{\rho\rho}}
\sum^{l_{\rho\rho}}_{m_{\rho\rho} = -l_{\rho\rho}}
\sum^{l_{\rho\lambda}}_{m_{\rho\lambda} = -l_{\rho\lambda}}
\sum^{l_{\lambda\rho}}_{m_{\lambda\rho} = -l_{\lambda\rho}}
\sum^{l_{\lambda\lambda}}_{m_{\lambda\lambda} = -l_{\lambda\lambda}}
\la l_{\rho\rho} \, m_{\rho\rho} \, l_{\rho\lambda} \, m_{\rho\lambda} \ar
l^{l'}_\rho \,
m'_\rho \ra \, \la l_{\rho\rho} \, 0 \, l_{\rho\lambda} \, 0 \ar
l^{l'}_\rho \, 0 \ra \\ \times \la l_{\lambda\rho} \, m_{\lambda\rho} \,
l_{\lambda\lambda} \, m_{\lambda\lambda} \ar l^{l'}_\lambda \,
m'_\lambda \ra \, \la l_{\lambda\rho} \, 0 \, l_{\lambda\lambda} \, 0 \ar
l^{l'}_\lambda \, 0 \ra \, \la l_{\rho\rho} \, m_{\rho\rho} \,
l_{\lambda\rho} \, m_{\lambda\rho} \ar l^{l}_\rho \, m_\rho \ra \, \la
l_{\rho\rho} \, 0 \, l_{\lambda\rho} \, 0 \ar l^{l}_\rho \, 0 \ra \, \la
l_{\rho\lambda} \, m_{\rho\lambda} \, l_{\lambda\lambda} \,
m_{\lambda\lambda} \ar l^{l}_\lambda \, m_\lambda \ra \, \la
l_{\rho\lambda} \, 0 \, l_{\lambda\lambda} \, 0 \ar l^{l}_\lambda \,
0 \ra \\ \times
\frac{(\mP^{\rho\rho}_\sigma)^{2n_{\rho\rho}+l_{\rho\rho}} \,
  (\mP^{\rho\lambda}_\sigma)^{2n_{\rho\lambda}+l_{\rho\lambda}} \,
  (\mP^{\lambda\rho}_\sigma)^{2n_{\lambda\rho}+l_{\lambda\rho}} \,
  (\mP^{\lambda\lambda}_\sigma)^{2n_{\lambda\lambda}+l_{\lambda\lambda}}
  \, (2l_{\rho\rho}+1) \, (2l_{\rho\lambda}+1) \, (2l_{\lambda\rho}+1)
  \, (2l_{\lambda\lambda}+1)} {n_{\rho\rho}! \, n_{\rho\lambda}!
    \, n_{\lambda\rho}! \, n_{\lambda\lambda}! \, \Gamma(n_{\rho\rho}
    + l_{\rho\rho} + 3/2) \, \Gamma(n_{\rho\lambda} + l_{\rho\lambda}
    + 3/2) \, \Gamma(n_{\lambda\rho} + l_{\lambda\rho} + 3/2) \,
    \Gamma(n_{\lambda\lambda} + l_{\lambda\lambda} + 3/2)} \; .
\label{eq:beverencoef}
\end{multline}
\end{widetext}
Such an expression deserves quite some explanations! 
First, the Kronecker $\delta$ ensures that the harmonic oscillator shell number
$N$ remains the same and that there is no mixing between
HO-wavefunctions of different HO-shells. The $n$'s, $l$'s and $m$'s
with a single $\rho$/$\lambda$ subscript are orbital quantum numbers
which are not summed over (=external quantum numbers), but are indices to the
$\mD$-coefficient. The $n$'s, $l$'s and $m$'s with a double
$\rho$/$\lambda$ subscript are summed over for every $\mD$
coefficient. The values that these indices may take depend on the external
quantum numbers. More specifically, the following conditions need to
be fulfilled:
\begin{subequations}
\be
2 n_{\rho\rho} + l_{\rho\rho} + 2n_{\rho\lambda} + l_{\rho\lambda} = 2
n^{l'}_\rho + l^{l'}_\rho
\ee
\be
2 n_{\lambda\rho} + l_{\lambda\rho} + 2n_{\lambda\lambda} + l_{\lambda\lambda} = 2
n^{l'}_\lambda + l^{l'}_\lambda
\ee
\be
2 n_{\rho\rho} + l_{\rho\rho} + 2 n_{\lambda\rho} + l_{\lambda\rho} =
2 n^{l}_\rho + l^{l}_\rho
\ee
\be
2n_{\rho\lambda} + l_{\rho\lambda} + 2n_{\lambda\lambda} +
l_{\lambda\lambda} = 2 n^{l}_\lambda + l^{l}_\lambda \; .
\ee
\label{eq:cond_beveren}
\end{subequations}
These conditions give rise to the $\delta$-function in
Eq.~(\ref{eq:beverencoef}) and fix the values of $l_{\rho\lambda}$,
$l_{\lambda\rho}$ and $l_{\lambda\lambda}$. Moreover, they also
determine the maximum value of the internal $n$-quantum numbers and of
$l_{\rho\rho}$:
\begin{subequations}
\be
n_{\rho\rho} : 0 \to \textrm{Min} \left( \frac{2 n^{l'}_\rho +
  l^{l'}_\rho} {2}, \frac{2 n^{l}_\rho + l^{l}_\rho} {2} \right)
\ee
\be
n_{\rho\lambda} : 0 \to \textrm{Min} \left( \frac{2 n^{l'}_\rho +
  l^{l'}_\rho} {2}, \frac{2 n^{l}_\lambda + l^{l}_\lambda} {2} \right)
\ee
\be
n_{\lambda\rho} : 0 \to \textrm{Min} \left( \frac{2 n^{l'}_\lambda +
  l^{l'}_\lambda} {2}, \frac{2 n^{l}_\rho + l^{l}_\rho} {2} \right)
\ee
\be
n_{\lambda\lambda} : 0 \to \textrm{Min} \left( \frac{2 n^{l'}_\lambda +
  l^{l'}_\lambda} {2}, \frac{2 n^{l}_\lambda + l^{l}_\lambda} {2} \right)
\ee
\be
l_{\rho\rho} : 0 \to \textrm{Min} \left( 2 n^{l'}_\rho + l^{l'}_\rho, 2
n^{l}_\rho + l^{l}_\rho \right)
\ee
\be
l_{\rho\lambda} = 2 n^{l'}_\rho + l^{l'}_\rho - 2 n_{\rho\rho} - 2
n_{\rho\lambda} - l_{\rho\rho}
\ee
\be
l_{\lambda\rho} = 2 n^{l}_\rho + l^{l}_\rho - 2 n_{\rho\rho} - 2
n_{\lambda\rho} - l_{\rho\rho}
\ee
\be
l_{\lambda\lambda} = 2 n^{l}_\lambda + l^{l}_\lambda - 2 n^{l'}_\rho
- l^{l'}_\rho + 2 n_{\rho\rho} - 2 n_{\lambda\lambda} + l_{\rho\rho} \; .
\ee
\label{eq:lim_beveren}
\end{subequations}
The numbers $\mP^{ij}_\sigma$ in Eq.~(\ref{eq:beverencoef}) with
$i,j=\rho,\lambda$ denote the matrix elements of the permutation
matrix with index $\sigma$ in Eq.~(\ref{eq:perm_matrices}), with $\rho
\to 1$ and $\lambda \to 2$ (\textit{e.g.}
$\mP^{\lambda\rho}_{\sigma=132} = \sqrt{3}/2$).

One more remark is in order. Inside the large expression for the MBRB
coefficients in Eq.~(\ref{eq:beverencoef}), there are four CG-coefficients for which the magnetic quantum numbers are zero. These four coefficients conspire so that the MBRB coefficients are only non-zero when ($l^{l}_\rho$,$l^{l}_\lambda$) and
($l^{l'}_\rho$,$l^{l'}_\lambda$) couple to the same $L$~ \cite{Varshalovich:1988}.

In the end, the availability of the analytical result in Eq.~(\ref{eq:beverencoef}) 
is what motivates our use of the HO basis to variationally expand all wavefunctions.

\subsection{Spin-dependent part}
We now turn to the spin terms in the first line of Eq.~(\ref{P23expanded}).
The spin state is described by three quantum numbers $S$, $m_S$ and $S_{12}$, given in the order $ | S \, m_S \, S_{12} \rangle $.

If the spin $S$ takes the value $S = 3/2$, the $S_{12}$ quantum number becomes redundant as it must necessarily be $S_{12} = 1$.
The action of the exchange operators is also trivial, because the states are completely symmetric in spin and the matrix element needed for Eq.~(\ref{P23expanded}) is given by
\be
\langle \frac{3}{2} \, m_S \, 1 | P^{23} | \frac{3}{2} \, m_S' \, 1 \rangle = \delta_{m_S m_S'}
\ee

For $S = 1/2$, we can instead have both $S_{12} = 0$ and $S_{12} = 1$, so for each $m_S$ we have two states to consider. We give them for $m_S=+1/2$,
\bea
| \frac{1}{2} \, \frac{1}{2} \, 0 \rangle &=&
\frac{ | \uparrow \downarrow \uparrow \rangle - | \downarrow \uparrow \uparrow \rangle }{ \sqrt{2} } \\
| \frac{1}{2} \, \frac{1}{2} \, 1 \rangle &=&
\frac{ 2 | \uparrow \uparrow \downarrow \rangle - | \uparrow \downarrow \uparrow \rangle - | \downarrow \uparrow \uparrow \rangle }{ \sqrt{6} } \ .
\eea
Applying $P^{23}$ to these states results in
\be
P^{23}\left(\begin{array}{c}
|\frac{1}{2}\, \frac{1}{2} \, 0 \rangle\\
|\frac{1}{2}\, \frac{1}{2}\, 1  \rangle
\end{array}\right)=\left(\begin{array}{cc}
\frac{1}{2} & \frac{\sqrt{3}}{2}\\
\frac{\sqrt{3}}{2} & -\frac{1}{2}
\end{array}\right)\left(\begin{array}{c}
|\frac{1}{2}\, \frac{1}{2}\, 0\rangle\\
|\frac{1}{2}\, \frac{1}{2}\, 1\rangle
\end{array}\right)\ .
\ee
Curiously, this is the same matrix $\mathcal{P}_{23}$ that represents $P^{23}$ in $\rho$--$\lambda$ momentum space.

As the result is also valid for $m_S = - 1/2$, we can directly write the matrix element for $S = 1/2$
\be
\langle \frac{1}{2} \, i \, m_S | P^{23} | \frac{1}{2} \, j \, m_S' \rangle =
\mathcal{P}_{23}^{ij} \, \delta_{m_S m_S'}
\ee
 where the indices $i, j = 1, 2$ correspond to $S_{12}, S_{12}' = 0, 1$.

Finally, the computation of the $P^{23}$ exchange matrix elements in isospin space is similar to the one in spin space and we do not repeat the discussion for the sake of brevity.


\end{document}